# In-situ process monitoring and adaptive quality enhancement in laser additive manufacturing: a critical review

Lequn Chen [a, b, *], Guijun Bi [c], Xiling Yao [d, **], Jinlong Su [a], Chaolin Tan [a], Wenhe Feng [a], Michalis Benakis [a], Youxiang Chew [a,***], Seung Ki Moon [b, ****]

[a] Advanced Remanufacturing and Technology Centre (ARTC), Agency for Science, Technology and Research (A*STAR), 3 Cleantech Loop, 637143, Singapore

[b] School of Mechanical and Aerospace Engineering, Nanyang Technological University, 639798, Singapore

[c] Institute of Intelligent Manufacturing, Guangdong Academy of Sciences, Guangzhou, 510070, China

[d] Singapore Institute of Manufacturing Technology (SIMTech), Agency for Science, Technology and Research (A*STAR), 5 Cleantech Loop, 636732, Singapore

***Corresponding authors**:
E-mail addresses: chen1470@e.ntu.edu.sg (L. Chen), yaox@outlook.com (X. Yao), skmoon@ntu.edu.sg (S.K. Moon), chewyx@artc.a-star.edu.sg (Y. Chew)

**Abstract:** Laser Additive Manufacturing (LAM) presents unparalleled opportunities for fabricating complex, high-performance structures and components with unique material properties. Despite these advancements, achieving consistent part quality and process repeatability remains challenging. This paper provides a comprehensive review of various state-of-the-art in-situ process monitoring techniques, including optical-based monitoring, acoustic-based sensing, laser line scanning, and operando X-ray monitoring. These techniques are evaluated for their capabilities and limitations in detecting defects within Laser Powder Bed Fusion (LPBF) and Laser Directed Energy Deposition (LDED) processes. Furthermore, the review discusses emerging multisensor monitoring and machine learning (ML)-assisted defect detection methods, benchmarking ML models tailored for in-situ defect detection. The paper also discusses in-situ adaptive defect remediation strategies that advance LAM towards zero-defect autonomous operations, focusing on real-time closed-loop feedback control and defect correction methods. Research gaps such as the need for standardization, improved reliability and sensitivity, and decision-making strategies beyond early stopping are highlighted. Future directions are proposed, with an emphasis on multimodal sensor fusion for multiscale defect prediction and fault diagnosis, ultimately enabling self-adaptation in LAM processes. This paper aims to equip researchers and industry professionals with a holistic understanding of the current capabilities, limitations, and future directions in in-situ process monitoring and adaptive quality enhancement in LAM.

**Keywords:** Additive manufacturing, laser directed energy deposition, laser powder bed fusion, in-situ monitoring, multisensor data fusion, defect detection, closed-loop control, digital twin, machine learning

# Contents

## Nomenclature

Acoustic emission (AE)
Artificial Intelligence (AI)
Additive Manufacturing (AM)
Artificial Neural Network (ANN)
Back Propagating Neural Network (BPNN)
Complementary Metal Oxide Semiconductor (CMOS)
Charge-coupled device (CCD)
Convolutional Neural Networks (CNN)
Cyber-Physical Production Systems (CPPS)
Cyber-Physical Systems (CPS)
Design for Additive Manufacturing (DfAM)
Direct metal printing (DMP)
Decision Tree (DT)
Dynamic X-ray Radiography (DXR)
Fibre Bragg grating (FBG)
Finite element method (FEM)
Functionally Graded Materials (FGM)
Field of View (FoV)
Frame per second (fps)
General Adversarial Network (GAN)
Gaussian process regression (GPR)
Internet of Things (IoT)
Inline coherent imaging (ICI)
K Nearest Neighbour (KNN)
Laser additive manufacturing (LAM)
Laser Directed Energy deposition (LDED)
Lack of Fusion (LoF)
Laser powder bed fusion (LPBF)
Laser-Powder Directed Energy deposition (LP-DED)
Laser-Wire Directed Energy deposition (LW-DED)
Mel-frequency cepstrum coefficient (MFCC)
Multiple-Input Multiple-Output (MIMO)
Multi-scale Convolutional Neural Network (MsCNN)
Medium Wavelength Infrared (MWIR)

Non-destructive testing (NDT)
Optical Microscope (OM)
Optical Coherence Tomography (OCT)
Principal component analysis (PCA)
Physics-informed machine learning (PIML)
Random Forrest (RF)
Reinforcement learning (RL)
Root mean square error (RMSE)
Robot Operating System (ROS)
Spectral convolutional neural networks (SCNN)
Standard deviation (SD)
Scanning Electron Microscope (SEM)
Single-Input Single-Output (SISO)
Selective laser melting (SLM)
Selective laser sintering (SLS)
Self-organizing Maps (SOM)
Short-time Fourier transform (STFT)
Support vector machine (SVM)
Short Wavelength Infrared (SWIR)
Tool-Centre-Point (TCP)
Time-frequency representations (TFR)
Ultimate tensile strength (UTS)
Variational Auto-Encoder (VAE)
Variable polarity plasma arc welding (VPPA)
Wire and arc additive manufacturing (WAAM)
Wavelet transform (WT)
X-ray Computed Tomography (X-CT)
X-ray diffraction (XRD)
Yield strength (YS)

# Graphical Abstract

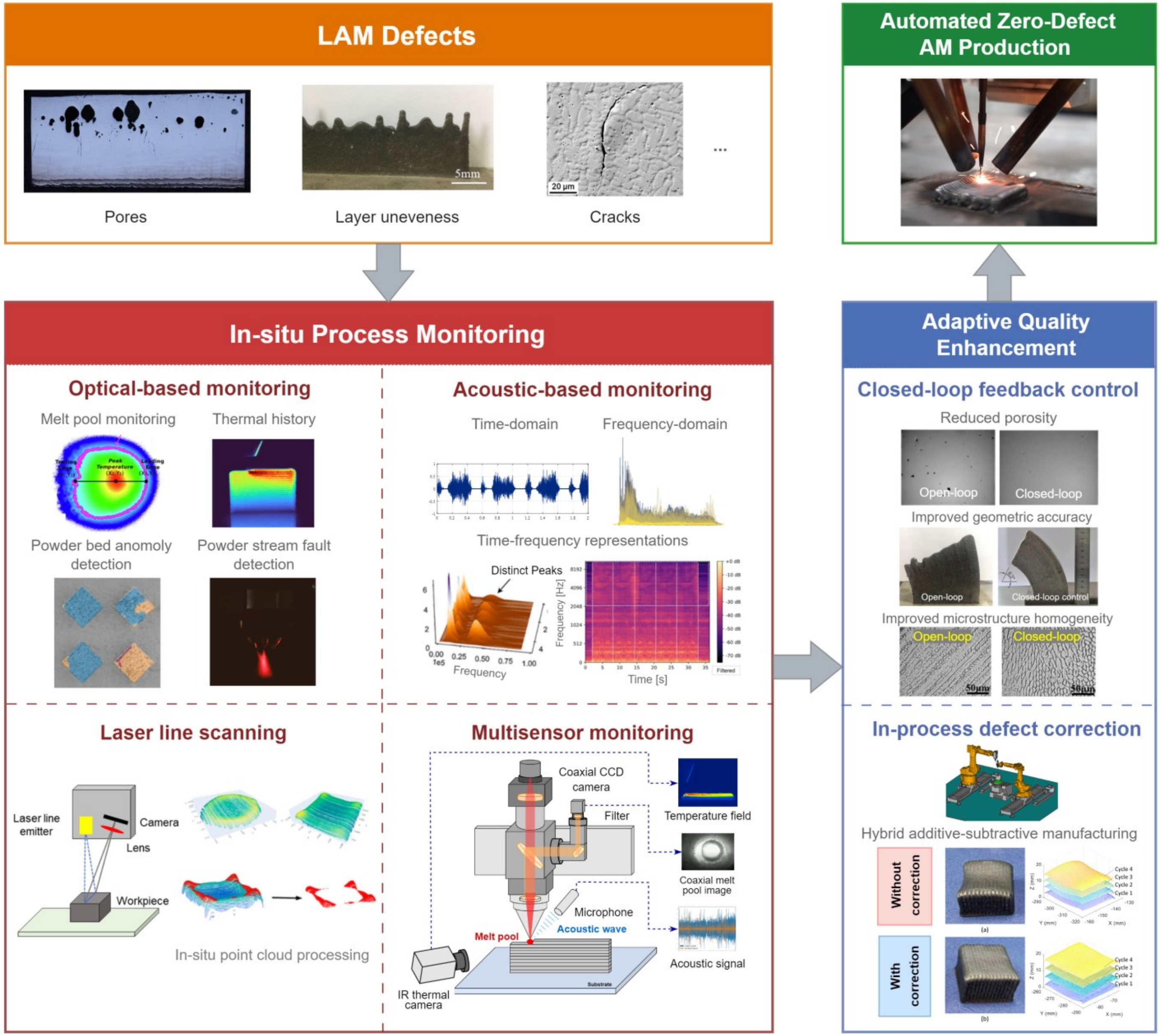

## 1. Introduction

As defined by the International Organization for Standardization (ISO)/American Society for Testing and Materials (ASTM) F2792-12a [1], Laser Additive Manufacturing (LAM) includes two primary processes, namely, Laser Powder Bed Fusion (LPBF) [2] and Laser Directed Energy Deposition (LDED) [3]. LAM has witnessed tremendous development over the past several decades, transforming from a tool used for rapid prototyping to a method for direct manufacturing of functional, high-performance components [4–6]. LAM has attracted considerable interests from aerospace, automotive, biomedical, marine, and offshore industries [7–16].

The unprecedented advantages of LAM stem from its layer-by-layer material addition methodology, offering unparalleled design flexibility [17–19], enhanced mechanical properties [20–22], improved energy efficiency [23–25], the capability to fabricate topologically optimized complex structures and lightweight components [26–31], and the integration of multiple functionalities within a single component for enhanced performance [32–35]. Furthermore, the fully digitized and automated process chain of LAM supports manufacturing-on-demand for highly customized, low-volume production, establishing new business models that reduce dependency on logistics, supply chains, and spare component storage while shortening production lifecycles [36–38].

However, despite these significant advancements, the inherent stochastic nature of the LAM process can lead to the formation of defects such as porosity, cracks, and distortions, which can substantially degrade the mechanical properties of as-fabricated parts. These defects occur due to complex thermal dynamics and intricate interactions between the laser beam, feedstock and base materials, which can be influenced by many factors such as unstable printing speed and dynamic heat accumulation. Defects could occur even with pre-optimized process parameters [39–42]. This challenge not only impedes further industrial adoption of LAM but also poses failure risks in critical applications where the reliability of the part is paramount. It emphasizes the importance of early defect detection and correction in ensuring as-built part quality while also improving the reliability and reproducibility of LAM processes.

Recognizing this critical need, the field has seen a surge in research activities focused on in-situ monitoring and closed-loop quality enhancement. Data from the Scopus database indicates a substantial increase in LAM monitoring research publications over the past decade (**Figure 1**(a)), with US, China, and Germany at the forefront. The trend in various sensing approaches is also illuminating, with optical-based monitoring leading the way, and a growing interest in acoustic, and multisensor monitoring (**Figure 1**(b)). A considerable rise in the integration of artificial intelligence (AI) in LAM monitoring signifies the fusion of traditional sensing approaches with advanced data analysis methods, indicating a multidisciplinary evolution of this field (**Figure 1**(c)).

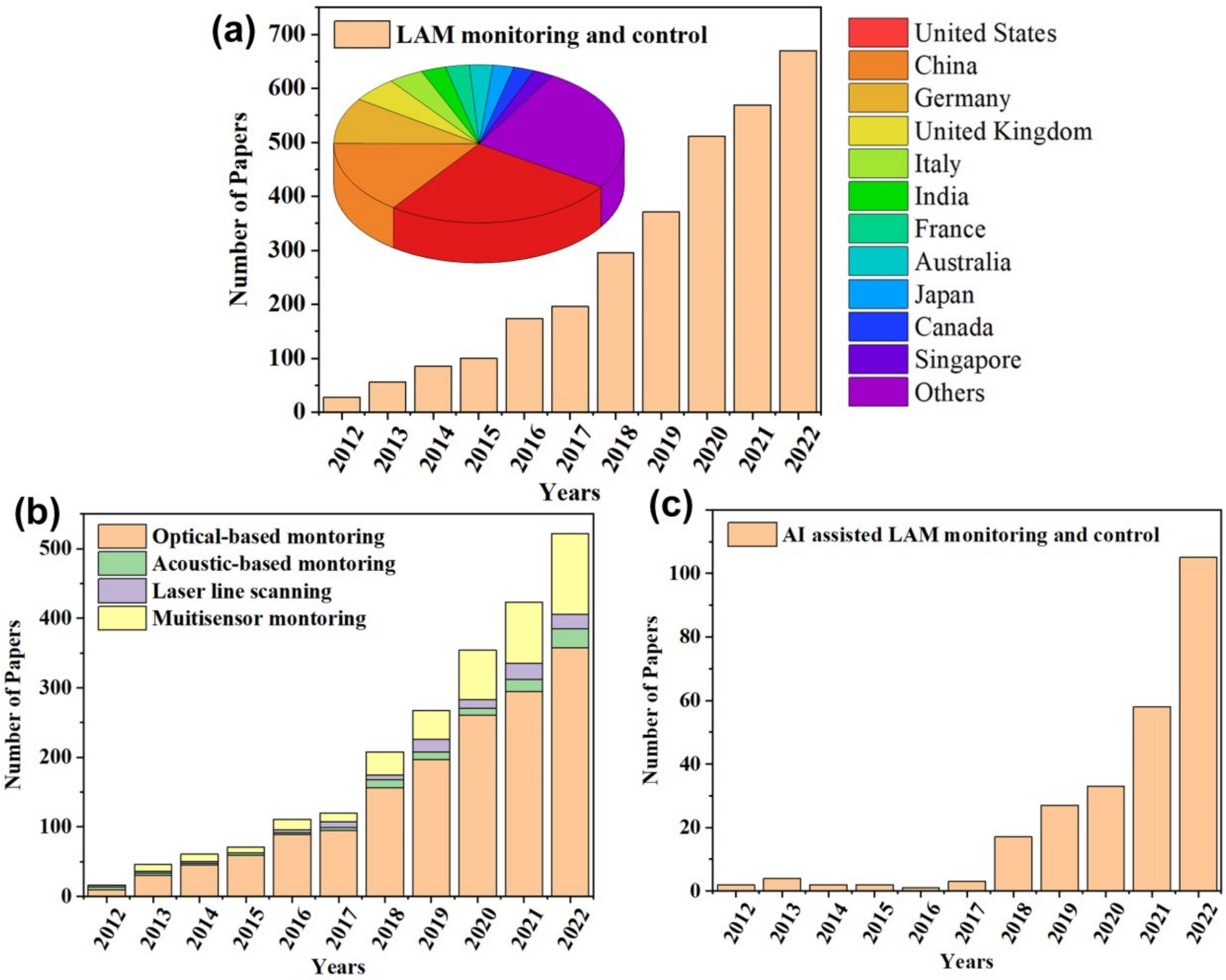

**Figure 1.** A decade of LAM monitoring research publications - a statistical overview from the Scopus database: (a) Publication statistics on LAM monitoring research, with a focus on leading countries. (b) Breakdown of publication statistics for different sensing approaches used in LAM monitoring research. (c) Trends in AI-assisted LAM monitoring publications.

Previous reviews in in-situ process monitoring and control in LAM have made significant contributions to the field [43–60]. However, many of these reviews exhibit limitations that our work aims to address, as summarized in **Table 1**. A common theme across previous reviews, such as those by Herzog et al. [52] and Grasso et al. [50], is a narrowed focus on specific aspects of LAM, such as ML for defect detection or sensor signals. Other reviews, such as Everton et al. [45] and Chua et al. [61], while providing valuable insights, do not fully incorporate the latest advancements in monitoring technologies or real-time process control, thereby leaving a gap in terms of comprehensive and up-to-date coverage. Furthermore, several reviews including Xia et al. [49] and Tang et al. [48] focus exclusively on particular AM methods, such as WAAM and DED. This specialization results in a lack of a holistic view that encompasses the breadth of LAM techniques. Similarly, works like Segovia

Ramírez et al. [56] and Lu et al. [47] focus predominantly on non-destructive testing, offering less insight into the in-situ process monitoring and control aspects that are crucial for the advancement of LAM.

In contrast, the proposed review provides a comprehensive and critical examination of various in-situ sensing methodologies, including but not limited to optical, acoustic, and infrared thermography monitoring, alongside the integration of advanced ML-assisted defect detection methods. We address the existing research gaps identified in previous studies by emphasizing the need for standardization, improved reliability, and sensitivity, as well as enhanced data interpretation and decision-making strategies. Additionally, this review explores the state-of-the-art adaptive quality enhancement methods, positioning it as a pathway towards achieving zero-defect autonomous manufacturing. This approach not only bridges the gaps observed in earlier reviews but also sets a foundation for future research directions in the field.

**Table 1.** Previous review on process monitoring and control in LAM.

| Review paper | Key contributions | Limitations | Year |
|---|---|---|---|
| Gunasegaram [62] | • ML strategies for adaptive control in metal AM.<br>• Proposes a framework for ML-assisted closed-loop control (CLC) focusing on defect avoidance, mitigation, and repair. | • Limited focus on in-situ process monitoring, multisensor fusion and sensing technologies<br>• Lacks implementation challenges of ML-assisted CLC | 2024 |
| Herzog et al. [52] | • ML approaches and data structures in defect detection for LAM<br>• Monitoring technology trends comparison | • Limited focus on process control and adaptive quality enhancement methods | 2023 |
| Cai et al. [58] | • Detailed survey of in-situ process sensing and control in metal AM<br>• Review of various signal monitoring methods and closed-loop control strategies | • Lacks coverage on the very latest developments of vision and acoustic-based monitoring<br>• Lacks coverage on multisensor monitoring and data fusion.<br>• Limited focus on in-process adaptive defect correction. | 2023 |
| Segovia Ramírez et al. [56] | • Review of NDT methods in AM. | • Focused on non-destructive testing, less on in-situ process monitoring<br>• Limited focus on process control | 2023 |
| Qin et al. [54] | • Systematic review of ML in AM | • Broad focus on AM, not specific to Laser-based AM or in-situ monitoring. | 2022 |
| | • Cluster analysis in AM literature, including DfAM, material analysis, monitoring, and sustainability. | • Limited focus on process control | |
| AbouelNour and Gupta [51] | • In-situ monitoring for subsurface and internal defects in AM<br>• Focus on imaging and acoustic methods review | • Focus primarily on imaging and acoustic methods, lacks surface defect detection.<br>• Limited review on process control and quality enhancement | 2022 |
| Grasso et al. [50] | • Sensor signals, in-situ sensing and monitoring in metal PBF | • Focus on PBF, lacks broader LAM coverage<br>• Limited focus on in-process quality enhancement | 2021 |
| Xia et al. [49] | • Monitoring and control in WAAM<br>• Sensor-based feedback control | • Specific to WAAM, not applicable to broader LAM | 2020 |
| Tang et al. [48] | • In-situ monitoring in metal DED | • Specific to DED, lacks coverage of other LAM techniques<br>• Lacks review on process control | 2020 |
| Lu and Wong [47] | • Review of NDT in AM<br>• In-process inspection NDT methods | • Focus on NDT, less on in-situ process monitoring and control | 2018 |
| Chua et al. [61] | • Quality control in metal AM<br>• Real-time inspection methods | • Lacks coverage of recent developments | 2017 |
| Everton et al. [45] | • Quality assurance in AM<br>• Developments in process control for AM | • Lacks coverage of recent developments | 2016 |
| Tapia and Elwany [43] | • Overview of process monitoring and control in metal-based AM. | • Lacks coverage of recent developments | 2014 |

The structure of this review is designed to provide a comprehensive understanding of in-situ process monitoring and adaptive quality enhancement in LAM. The graphical abstract shown in **Figure 2** visually outlines the scope and interconnectedness of the various sections. **Section 2** lays the foundational knowledge of LPBF and LDED techniques, and the common defects encountered in LAM. Building upon this, **Section 3** provides a critical examination of in-situ monitoring methodologies, highlighting its capability in early defect detection. This includes a focused discussion on ML-assisted defect identification. **Section 4** advances the discussion by reviewing state-of-the-art closed-loop adaptive quality enhancement techniques, which ensures the quality consistency of the printed products. The review concludes with **Section 5**, which summarizes the key findings and identifies crucial research gaps and future perspectives. A road map towards fully autonomous, zero-defect LAM production is proposed. By integrating these distinct yet interconnected aspects, this review seeks to stimulate further research in this field, guiding the development of industrial-grade in-situ process monitoring and control applications for LAM technologies.

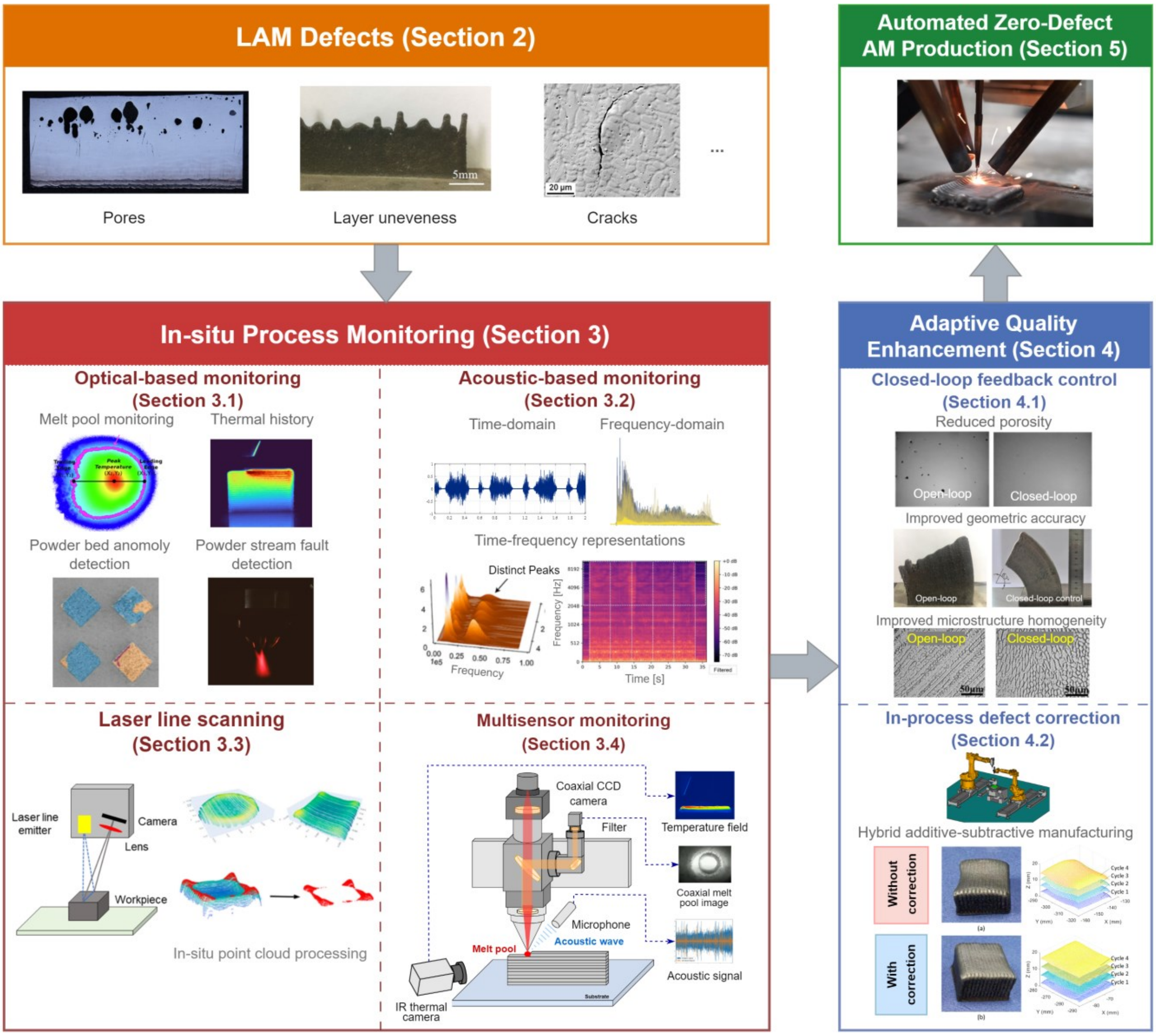

**Figure 2**. Graphical abstract: a review of in-situ process monitoring and adaptive quality enhancement in LAM. Image source from Refs. [63–74].

## 2. Laser Additive Manufacturing and Defects

This section investigates the complexities of Laser Additive Manufacturing (LAM) and its associated defects. The discussion begins with the analysis of Laser Powder Bed Fusion (LPBF) in **Section 2.1**, highlighting its importance in creating complex geometries with precision. This is followed by a review of Laser Directed Energy Deposition (LDED) in **Section 2.2**, which is known for its abilities in material addition and component repair. The discussion continues with an in-depth examination of common LAM defects in **Section 2.3**, establishing a connection to the specific characteristics of LPBF and LDED. Lastly, **Section 2.4** conducts a comparative analysis between LPBF and LDED dynamics, and their implications for defect formation and in-situ monitoring. This systematic review is essential for understanding the challenges in achieving defect-free LAM processes and sets the context for the subsequent sections on in-situ monitoring and adaptive quality enhancement.

### *2.1. Laser Powder Bed Fusion*

Originating in the late 1980s, Laser Powder Bed Fusion (LPBF) has evolved into a critical pillar of the AM industry, attracting widespread interest across various sectors due to its capability to produce complex geometries with high precision [10]. As depicted in **Figure 3**(a), the LPBF process works on the principle of selective point-by-point irradiation scanning to build three-dimensional parts. The process begins with a recoater blade spreading a thin layer of fine metal powder over the build platform, and a laser beam selectively scans and melts the powder according to the part's cross-sectional geometry. As the laser beam relocates, the molten powders undergo a rapid solidification process. The build platform then descends by a height equivalent to the predetermined layer thickness, marking the commencement of another cycle of this layer-by-layer procedure until the part's complete fabrication. During scanning, a melt pool is formed, which solidifies upon cooling, forming a fusion bond with the preceding layer and adjacent tracks. The complex interplay between the laser beam and the metallic powder leads to the melting, solidification, and densification of the powder particles. The behaviour of these melt pool dynamics is significantly influenced by key process parameters, including laser power, scan speed, layer thickness, and hatch spacing.

LPBF's primary advantage lies in its capacity to fabricate parts with complex geometries that traditional manufacturing methods find challenging or impossible to produce. The rapid cooling and solidification processes produce incredibly fine microstructures with grain sizes as small as hundreds of nanometres, enabling LPBF-fabricated parts with higher mechanical strength than cast and forged products. LPBF can handle a wide array of materials, ranging from various types of metals to alloys. Furthermore, due to the smaller laser beam size, LPBF can achieve enhanced dimensional accuracy (up to ±0.05 mm) and superior surface quality (Ra≤10 μm), surpassing other metal AM methods such as LDED and WAAM [10].

However, LPBF has the following limitations: (1) Achieving fully dense parts can be challenging due to the process's stochastic nature and the potential for defects such as porosity and cracks. (2) The process is highly sensitive to powder characteristics and process parameters, necessitating careful and time-consuming process parameter optimization for attaining the desired part quality. (3) It can only produce relatively small-sized parts due to the low build efficiency and the restricted build capacity within the enclosed chamber. (4) The high thermal gradients induced by the rapid heating and cooling cycles can lead to substantial residual stresses and distortions in the fabricated parts [75].

## *2.2. Laser Directed Energy Deposition*

Laser Directed Energy Deposition (LDED) is another key LAM process that emerged in the early 1990s. It is widely used for repairing, coating, and fabricating complex structures directly from a CAD model. As illustrated in **Figure 3**(b), LDED utilizes a focused laser source to melt metallic materials as they are deposited onto a substrate. LDED can use either powder or wire (or both) as feedstock materials. The melt pool dynamics in LDED involve complex interactions between the laser beam, the powder/wire material, and the substrate. The key process parameters governing this interaction include laser power, scan speed, hatch space, laser spot size, and the powder/wire feed rate.

One of the key advantages of LDED is its ability to add material to existing parts, making it an excellent choice for repair and coating applications [76]. Moreover, LDED has unique advantages in manufacturing multi-material components [77,78] and functionally graded materials (FGMs) [79,80]. It also utilizes higher laser power and larger laser spot size to achieve faster build rate than LPBF process, making it suitable for the fabrication of larger parts such as propellers and rocket engine nozzles [81]. LDED also outperforms other DED methods like WAAM and EBAM in mechanical performance (e.g., strength, fatigue life, etc.) and surface roughness. Lastly, LDED can handle a diverse range of materials, including metals, ceramics, and composites.

However, LDED also has several limitations. The process can result in a relatively coarse microstructure due to the high thermal input and large melt pool, which weaken the mechanical strength of the fabricated parts. Like LPBF, the process is sensitive to various processing parameters, requiring careful fine-tuning and optimization. Furthermore, creating highly complex geometries with high precision using LDED is challenging. Increased build rates, while boosting productivity, compromise surface roughness and dimensional accuracy. Addressing issues like geometric distortions, defect occurrences, and microstructure inhomogeneity often proves difficult due to localized heat accumulation and residual stress.

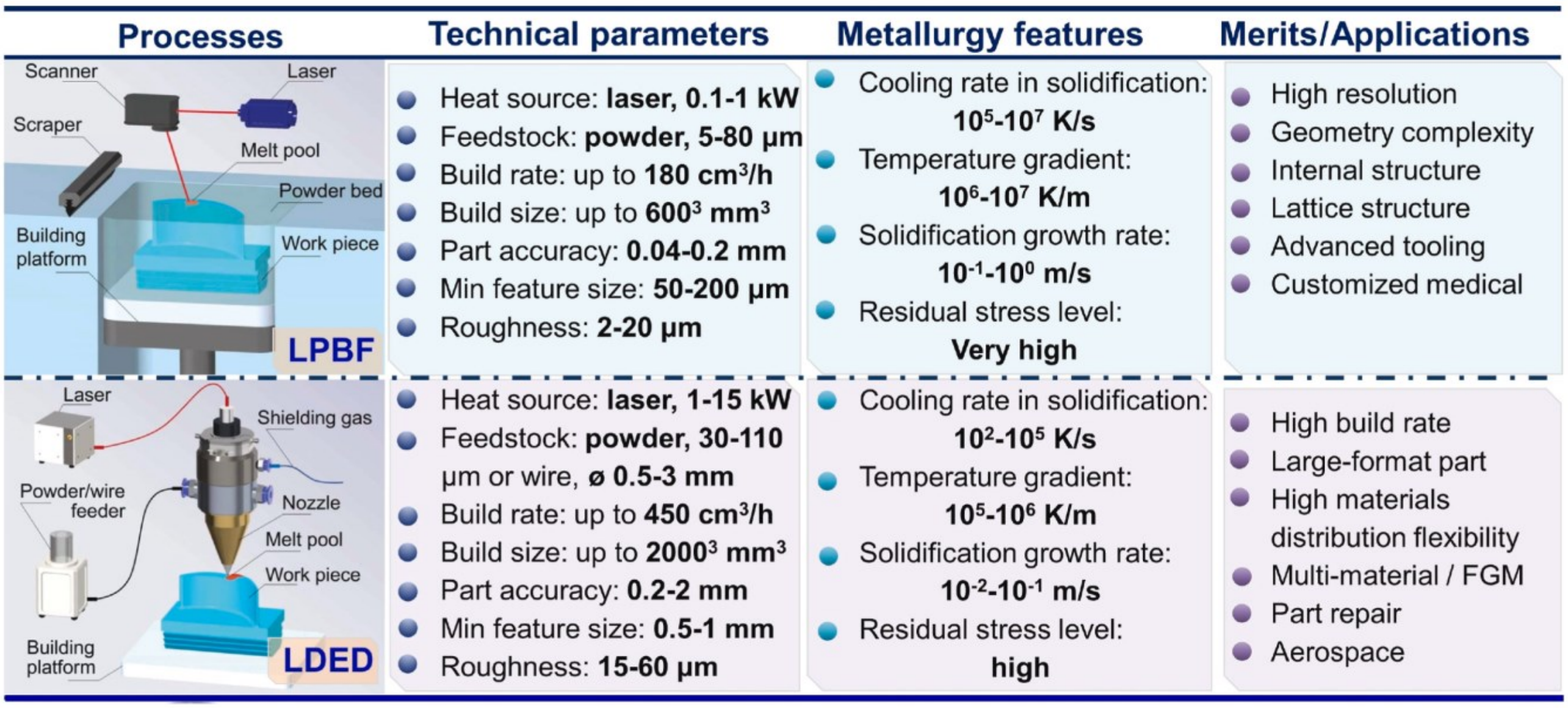

**Figure 3**. Summary of key parameters, features, and applications of laser powder bed fusion (LPBF) and laser directed energy deposition (LDED) [41,82].

### *2.3. Defects in LAM*

Achieving high quality consistency and process repeatability in LAM remains a challenging task. LAM is sensitive to various process parameters, including laser power, scanning speed, layer thickness, hatch spacing, as well as to powder particles composition [83]. Optimizing these parameters through various methods like trial-and-error [84], mechanistic modelling [85–87], or machine learning [88–90] are often both time-consuming and expensive. Yet, even with these optimized parameters, variations in part quality persist. Defects such as pores and cracks can stochastically emerge due to dynamic heat accumulation and residual stress.

The thermal dynamics during the LAM fabrication of metallic parts is a critical aspect in the defect formation mechanisms. As the substrate temperature rises, a cascade of complications can occur due to heat build-up, such as nonuniform tracks, expanded heat-affected zones, excessive dilutions, geometric distortions, and cracking [91]. The dynamic and stochastic nature of the melt pool metallurgical process in LAM contributes to the uncertainty of defects [92]. It is particularly challenging to determine the transitions from a conduction region (non-defective) to abnormal states like keyhole porosity region or lack-of-fusion (LoF) region [93–95].

Previous literature [96–98] reveals a range of defects, spanning from micro/meso scales (μm level) to macro scales (mm level), as summarized in **Table 2** and depicted in **Figure 4**. Micro/meso scale defects include porosity, cracks, discontinuity, and microstructure inhomogeneity. Porosity is a prevalent micro/meso scale defect in LAM, substantially undermining mechanical performance such as strength, ductility, and fatigue life [98]. LoF pores form when the volumetric energy density is insufficient, resulting from incomplete melting, poor bonding of the melt pool with neighbouring layers, and crack propagations [99,100]. Balling, a subtype of LoF, frequently emerges in LPBF due to poor melt pool flowability, leading to unstable, discontinuous melting tracks [101]. Conversely, excessive volumetric energy can yield keyhole pores [102–105]. These large, spherical keyhole pores are often the result of material evaporation due to high energy density and unstable melt pool dynamics. The interaction of the fusion zone with pre-existing pores can exacerbate the situation, leading to crack propagation and formation of larger pores, thereby further reducing mechanical properties such as strength, fatigue life, and corrosion resistance. Gas-induced pore formation mechanisms differ between LPBF and LDED. In LPBF, gas entrapment from surface fluctuations can cause pores [95], while in LDED, high-velocity powder feedstock with protective gas can break through and inject into the melt pool, leading to gas-induced pores [94].

Cracking and delamination, often instigated by residual stress or partially melted powder, also pose significant challenges, with rapid re-heating and cooling cycles and high temperature gradients resulting in high thermal stress within as-printed parts [106–108]. These cracks can culminate in premature failures and reduced mechanical performance. Moreover, the build-up of residual stress and localized heat accumulation can lead to microstructure inhomogeneity and anisotropic mechanical properties (**Figure 4** (h)) [63].

On the macro scale, localized heat accumulation can result in high surface roughness and geometric distortions [109]. Uneven layers and significant geometric distortions, as depicted in **Figure 4**(e) and (g), may be induced by residual stress. These macroscopic defects are intrinsically linked to the stand-off distance and dilution in the LDED process. A low dilution could result in insufficient layer bonding, while a high dilution indicates extensive heat-affected zones and a higher likelihood of thermal expansion failures [110,111]. By addressing these macro-scale defects, such as surface roughness and geometric distortions, the overall quality and performance of LAM-produced parts can be significantly enhanced. Furthermore, residual stress, the primary cause of many macroscopic and microscopic defects, could be alleviated through post-printing heat treatments or adjustments to the scanning strategies during the printing process.

However, the inherent complexity and stochastic nature of the LAM process means that defects can still occur even with optimized process parameters. Therefore, there is a growing recognition of the necessity for in-situ monitoring systems and real-time defect detection methodologies. These techniques enable the detection of defects during the manufacturing process itself, effectively reducing the reliance on post-production inspections. This shift signifies an evolutionary leap in the LAM process - from a reactive, post-production inspection approach towards a proactive, real-time quality assurance strategy. Through immediate detection and in-process remediation, the upper limits of repeatability, reliability, and consistency in LAM-produced parts are continually being expanded. This perspective sets the stage for the subsequent section, providing a comprehensive review of the most recent advancements in in-situ quality monitoring in LAM.

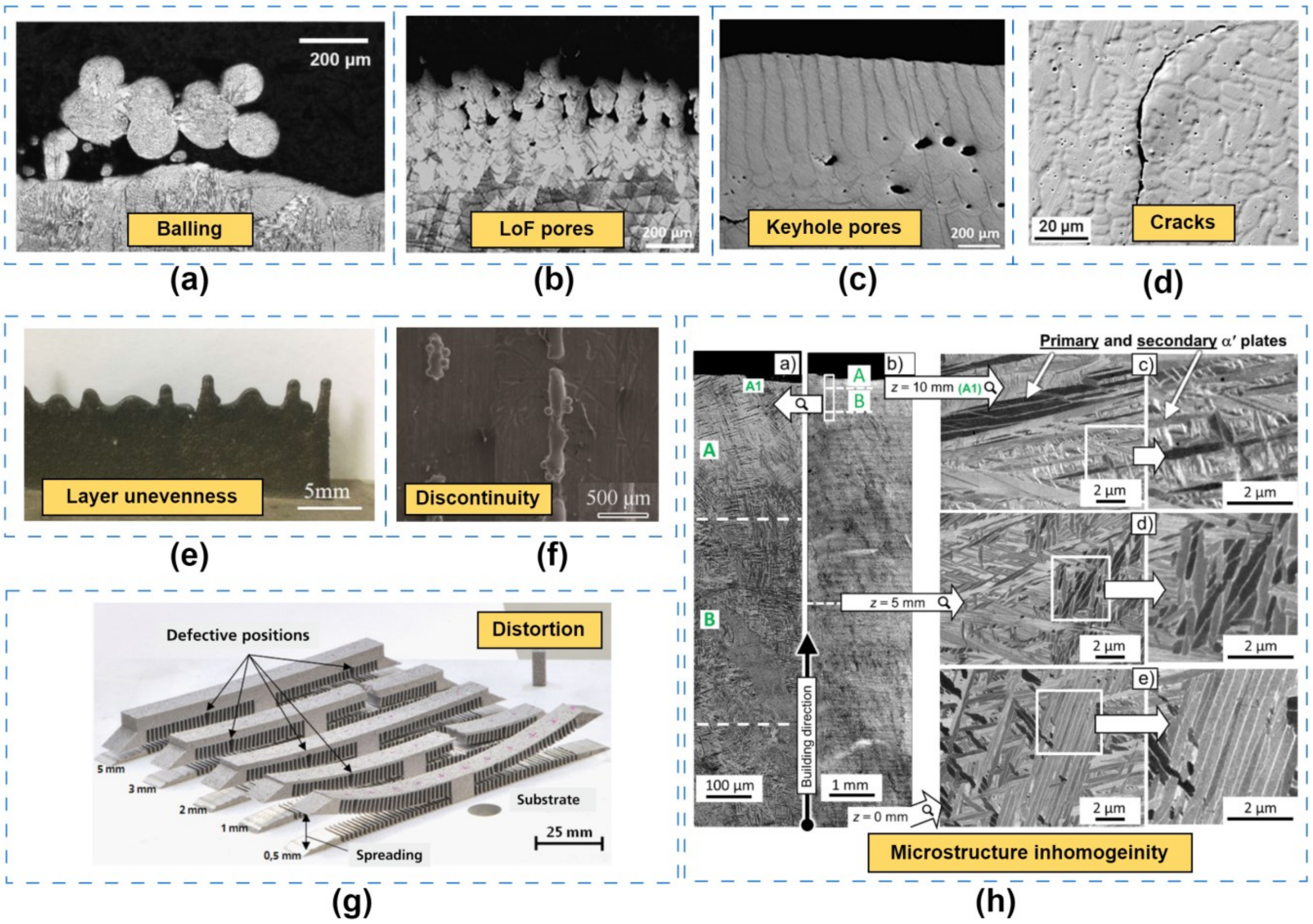

**Figure 4.** Illustrations of common defects in parts fabricated using LAM (including LPBF and LDED): (a) Balling [112] (commonly observed in LPBF); (b) Lack of Fusion (LoF) pores [66]; (c) keyhole pores [66]; (d) Cracks [74]; (e) Layer unevenness [113]; (f) Discontinuity [114]; (g) Distortions induced by thermal residual stress [115]; (h) Microstructure inhomogeneity [116].

**Table 2**. Common defects in parts fabricated by LAM: characteristics, potential causes, and consequences [96–98].

| Defects | Characteristics | Potential causes | Consequences | Ref |
|---|---|---|---|---|
| Balling (common in LPBF parts) | • micro-meter-scaled balls or highly coarsened balls. | • Low volumetric energy density due to high scanning speed or low laser power leading to higher melt pool surface tension.<br>• Poor melt pool flowability leading to limited substrate contact.<br>• Unstable, discontinuous melting tracks due to poor wetting. | • Reduced mechanical properties, especially fatigue life<br>• Facilitated crack propagation, anisotropy, corrosion resistance degradation | [117] |
| LoF pores | • Poorly bonded or incompletely melted metal particles.<br>• Irregular, elongated shapes, potentially larger than 500 μm. | • Insufficient energy input leading to incomplete melting and poor bonding with adjacent layers. | • Degradation of mechanical properties | [118] |
| Keyhole pores | • High depth-to-width ratio of the fusion zone.<br>• Large spheroidal shapes. | • Material evaporation due to high volumetric energy input.<br>• Unstable melt pool dynamics and gas entrapment due to localized heat accumulation. | • Degradation of mechanical properties<br>• Formation of larger pores due to interaction with pre-existing pores, leading to void/crack propagation | [102–104] |
| Gas induced pores | • Small and spherical shapes. | • Residual gas from feedstock powder (for LDED)<br>• Gas entrapment from powder bed surface fluctuations (for LPBF)<br>• Gas release from the liquid melt during solidification | • Degradation of mechanical properties<br>• Facilitated propagation of large pores across multiple layers | [3] |
| Cracks | • Intergranular cracks.<br>• Cracks along grain boundaries. | • *Hot cracking* due to tensile stresses on solidifying metal.<br>• *Liquid cracking* due to presence of undesirable phases, such as partially melted zone, and stress concentration during solidification.<br>• *Ductility dip cracking* in FCC alloys at elevated temperatures | • Reduced mechanical properties, especially low fatigue limit and low ductility<br>• Compromised structural integrity, and potential failure of the additively manufactured component | [108,119] |
| Distortions and delamination | • Macro-scale defects (mm).<br>• Uneven surface and high surface roughness. | • High cooling rates and thermal stresses<br>• Rapid heating and cooling cycles<br>• Residual stress | • Loss of geometric accuracy<br>• Degradation in fatigue behaviour | [120–122] |
| Microstructure inhomogeneity | • Variations in microstructure characteristics within the part. | • Localized heat accumulation<br>• Rapid heating and cooling cycles | • Variation in mechanical strength and ductility<br>• Anisotropic mechanical properties | [73,123] |

## *2.4. Comparative Analysis of LPBF and LDED Dynamics: Implications for Defect Formation and In-Situ Monitoring*

LPBF and LDED exhibit different process dynamics, as highlighted in **Figure 3**. LPBF typically operates at a lower power scale (generally not exceeding 1 kW), which is suitable for high-precision, smaller-scale parts. In contrast, LDED, which may utilize laser power of up to 15 kW, allowing for larger-scale structures and higher deposition rates. Such difference in laser power significantly affect the melt pool characteristics, thermal gradients, and solidification rates, leading to different microstructural features and defect profiles.

The solidification dynamics involved in LPBF and LDED have a significant influence on the types of defects that typically occur during the process. LPBF, with its distinctive rapid cooling rates ($10^5$ – $10^7$ K/s), is prone to the formation of microcracks caused by the stresses of rapid solidification. In contrast, the significantly slower cooling rates in LDED ($10^2$ – $10^5$ K/s) can result in gas porosity or delamination problems due to localized heat accumulation. Residual stresses also differ between the two processes. The steep thermal gradients of LPBF ($10^6$ – $10^7$ K/m) frequently result in larger amounts of residual stress within the manufactured parts [82]. Although it may be detrimental to mechanical properties, it can also lead to formation of finer microstructures. The smaller temperature gradient ($10^5$ – $10^6$ K/m) of LDED reduces residual stress levels. However, the prolonged thermal exposure may result in coarser microstructures [82].

The feedstock material also presents a notable difference: LPBF's exclusive use of powder allows for a finer resolution in part features and surface finish. In comparison, LDED's flexibility in using either powder or wire feedstock allows for a wider range of part geometries and surface roughness levels. This is reflected in the minimum achievable feature sizes: LPBF can produce detailed features down to 50-100 µm scales, while LDED is better suited to produce larger features with 0.5 – 1mm range.

These fundamental differences in process dynamics necessitates specialized in-situ monitoring strategies. For LPBF, high temporal resolution is required to capture rapid solidification processes, necessitating monitoring devices with high data acquisition rates, such as high-frequency coaxial cameras (up to 10 kHz) [39]. In contrast, the higher build rate and coarser feature resolution in LDED demand monitoring solutions that can accommodate larger field-of-views and potentially slower temporal resolutions (30 Hz – 1000 Hz) [70]. In addition, LDED's typically noisier environment and larger laser spot size present a different challenge for acoustic monitoring, necessitating robust noise-cancellation techniques and sensors that can distinguish process noise from actual defects [68]. The use of robotic arms in LDED also impacts sensor placement, offering flexibility while also requiring careful calibration to ensure accurate data capture.

ML models for each process must be developed with these factors into account. The large amounts of data produced by LPBF's high sampling rates must be efficiently processed and analysed, which poses significant challenge in handling the high data flow. On the other hand, LDED's comparatively scarcer data may result in less generalizability of the model [124]. Ultimately, the development of in-situ monitoring solutions must be tailored to each process's specific characteristics, ensuring that sensor placement, data handling, and ML models are all aligned with the distinct needs of LPBF and LDED. The following sections of this paper will investigate specific in-situ sensing technologies and their applications in each process, providing a comprehensive understanding of how these techniques can handle the challenges highlighted herein.

## 3. In-Situ Process Monitoring And Defect Detection In LAM

In this section, a critical review of the latest advancements in in-situ monitoring in LAM is presented, emphasizing its pivotal role in early defect detection to prevent quality deterioration and potential build failure. This review connects directly to previous discussion on LAM defects, underscoring in-situ monitoring as a key solution. It offers a holistic exploration of diverse in-situ sensing methodologies: optical-based monitoring (**Section 3.1**), acoustic-based monitoring (**Section 3.2**), laser line scanning (**Section 3.3**), other emerging methods including operando X-ray (**Section 3.4**), and multisensor monitoring and data fusion (**Section 3.5**). These methods contribute significantly to understanding critical process signatures like melt pool dynamics, thermal histories, and acoustic features from laser-material interactions. In addition, the increasing integration of ML in enhancing defect detection capabilities is critically evaluated, with a focus on ML models specifically tailored for in-situ defect detection. This multi-faceted evaluation provides a comprehensive picture of the current state-of-the-art and points to the potential future directions of in-situ monitoring and defect detection in LAM.

### *3.1. Optical-Based Monitoring*

This subsection investigates optical-based monitoring in LAM, a crucial technique for defecting process anomaly and defects. It begins with an examination of melt pool dynamics in **Section 3.1.1**, in which optical sensing plays a key role in capturing and interpreting melt pool visual and thermal features. The extracted melt pool visual features are critical for determining process quality. The discussions then shifts to how these features aid in detecting defects in **Section 3.1.2**, leveraging advanced ML techniques. Subsequently, this subsection expands to cover anomaly detection and fault diagnosis, illustrating the broader applications of optical monitoring in **Section 3.1.3**. Special attention is given to infrared thermal imaging in **Section 3.1.4**, which a distinct approach for localized quality assessments in LAM. Finally, this subsection discusses the challenges and limitations of optical monitoring, offering a holistic overview and future prospects for its applications in LAM.

### *3.1.1. Melt Pool Dynamics And Visual Feature Extractions*

The melt pool dynamics can be captured through optical-based monitoring [108]. Off-axis and coaxial camera setups are widely used in melt pool monitoring, with each offering specific advantages and disadvantages. **Figure 5** presents a comparison of these setups in both LDED and LPBF. Off-axis camera configurations, as depicted in **Figure 5**(a) and (c), have been extensively used in various studies. Although this setup can be simple to install into the AM machine, they also pose unique challenges, particularly the difficulties in measuring melt pool dimensions due to oblique viewing angles. To get correct melt pool features in this setting, image transformation and calibration procedures are required, which can be time-consuming [125–128]. On the other hand, coaxial camera monitoring setups, illustrated in **Figure 5**(b) and (d), have been increasingly favoured due to their direct, overhead viewing capabilities. This coaxial approach essentially eliminates the need for the tedious image transformation and calibration, enhancing operational efficiency. A notable study by Tang et al. [129] highlighted the significant advantage of coaxial setups, revealing that the visual characteristics of the coaxial melt pool directly correspond to underlying metallurgical phenomena, including the states of melting, cooling, and heat transfer.

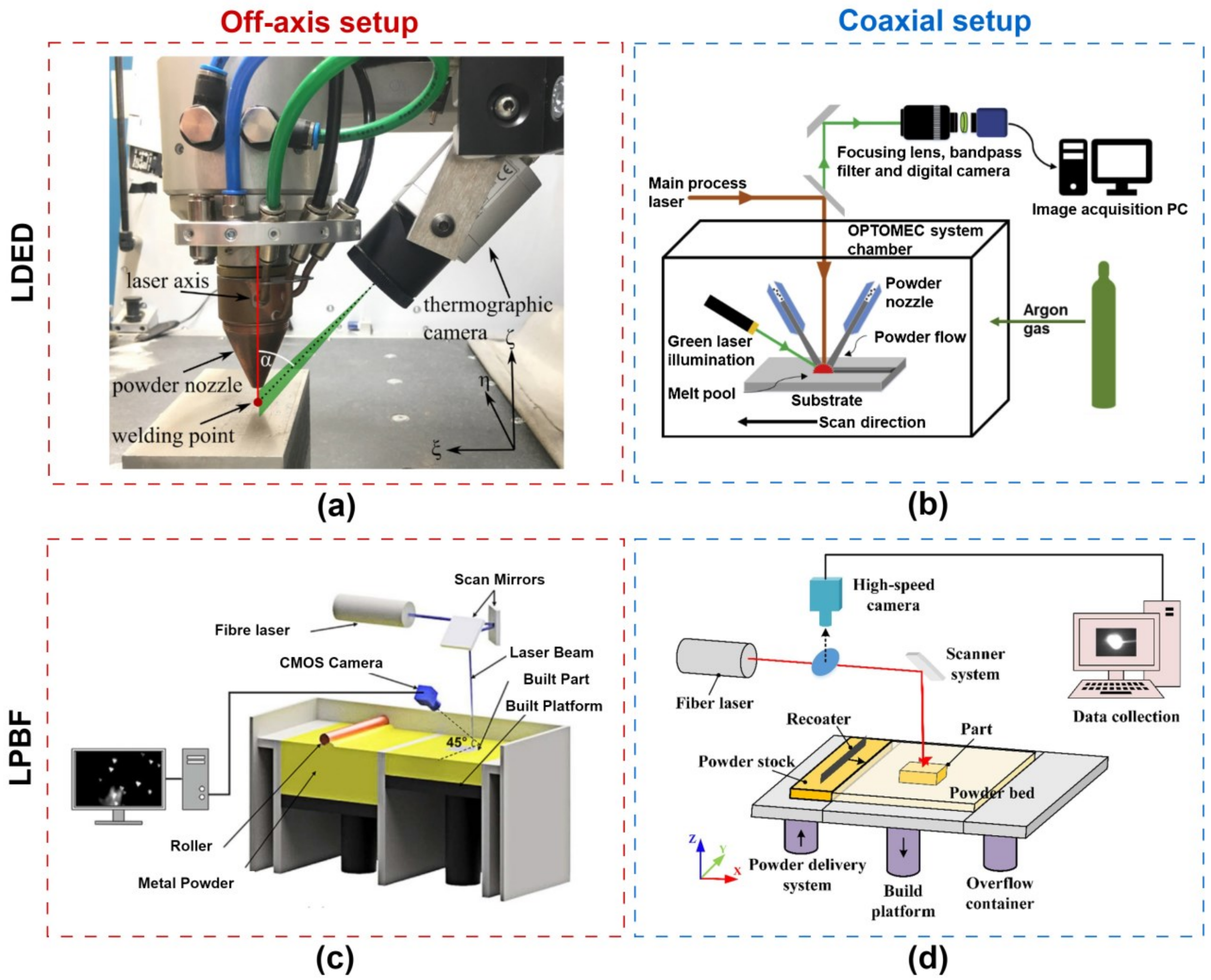

**Figure 5**. Comparison and visualization of off-axis and coaxial camera setups: (a) Off-axis camera mounting in LDED [130]. (b) Coaxial camera setup in LDED [131]. (c) Off-axis camera configuration in LPBF [132]. (d) Coaxial camera setup in LPBF [133].

Numerous studies have been dedicated to extracting physics-informed features from melt pool images as quality indicators [125,126,134–148]. The melt pool features can be used for defect detection, process-structure-property (PSP) causal analytics, and design rule constructions in LAM processes [149–151]. **Table 3** provides a summary of these features, such as melt pool width, size, moment of aera, peak temperature, etc. For example, Bi et al. [152–156] discovered that the melt pool size and temperature signals are both positively influenced by the input energy density. Ocylok et al. [157] showed that coaxial CMOS camera melt pool geometry strongly correlates with process stability. Increased input energy density yields a broader, deeper, and more asymmetrical melt pool [158]. Furthermore, Chen et al. [159] investigated melt pool evolution in the LDED process when printing on an inclined substrate with a non-vertically irradiated laser beam. The melt pool area can be linked to process stability and solidification state when depositing materials over sloped surfaces. Image processing techniques such as binary thresholding and edge extraction are commonly used to extract these melt pool morphological features.

**Figure 6** presents examples of melt pool visual characteristics and image processing techniques. A fixed threshold value can isolate the melt pool area from raw coaxial CMOS/CCD camera images, as shown in **Figure 6**(a) [70,160]. A series of image processing techniques, including binarization, elliptical fitting, and convex hull area extraction, can be used to distil melt pool geometric features that reflects the underlying physics, such as solidification, cooling, and heat conduction. **Figure 6**(b) reveals the coaxial visual characteristics in the head region and middle region, capturing vital details such as solid-liquid interface slags, boundary textures, and crucial morphological attributes like width, length, and grayscale distributions. **Figure 6**(c) shows thin-wall melt pool boundaries and coaxial melt pool isotherm images at different layers. OM images of the top and bottom layers show a direct association between columnar grain size and cooling and solidification rates. The melt pool at the bottom layer, subjected to higher cooling rates, increases columnar grain size, whereas the top layer, exposed to re-melting, shows lower cooling rates [161]. This correlation implies that strategic control of the melt pool size could potentially guide grain sizes and, subsequently, the material properties. **Figure 6**(d) further illustrates this point by showing a rising trend in melt pool size attributable to heat accumulation when the laser power remains constant [161].

Traditional image binarization approaches for detecting melt pool edges may be inaccurate, as melt pool boundaries frequently appear at different thresholds. To address this issue, novel image enhancing techniques have been proposed to measure melt pool width more accurately than existing emissivity-based edge detection algorithms [162]. Advanced computer vision algorithms can also be used to speed up and optimize the melt pool signature extraction process [138]. For instance, Liu et al. [134] proposed an image-enhancement generative adversarial network to improve the contrast ratio of thermal images for melt pool segmentation. Gravitational super-pixels algorithm [137] was used to reduce the dimensionality of infrared images and perform real-time melt pool segmentation with less uncertainty.

**Table 3**. List of physics-informed melt pool geometric and temperature features.

| Feature name | Description | Ref |
|---|---|---|
| Melt pool contour area ($m_{00}$) | Represents the area of the melt pool, calculated using pixel intensities. | [70] |
| Melt pool centroid position ($\overline{x}, \overline{y}$) | Indicates the centre of gravity of the melt pool, with deviations signifying process instabilities. | [70] |
| Central Moments ($\mu_{ji}$) | Describes the melt pool's probability distribution about its centre of gravity. | [70] |
| Convex hull area ($C$) | Measures the smallest convex polygon encompassing the melt pool, indicative of its geometry. | [70] |
| Bounding rectangle width $W$ and length $L$ | Dimensions derived from rotated bounding boxes, offering another perspective on the melt pool extent. | [70] |
| Melt pool width ($a$) and length ($b$) | Dimensions of the best-fit ellipse to the melt pool, representing its shape. | [163,164] |
| Spatter area ($A_{sp}$) | Total surface area of spatter particles, linked to material ejection and energy input. | [164] |
| Mean Spatter Temperature ($\overline{T_{sp}}$) | Average temperature of spatter particles, informative of the energy in the ejected material. | [164] |
| Distribution (variance) of melt pool temperature ($Var(T)$) | Highlights heat flow and cooling rates, with higher variance indicating significant temperature gradients. | [165] |
| Melt pool temperature skewness ($Skew(T)$) | Assesses asymmetry in temperature distribution, an indicator of melt pool stability. | [165] |
| Melt pool temperature kurtosis ($Kurt(T)$) | Measures the "tailedness" of the temperature distribution, with higher values indicating more variance due to extreme deviations. | [165] |
| Peak temperature ($T_{peak}$) | Maximal temperature in the melt pool, crucial for understanding thermal history and its effects on material properties. | [165,166] |
| Time over threshold of solidification temperature ($t_{solid}$) | Captures cooling behaviour above the solidification temperature, important for process analysis. | [166] |

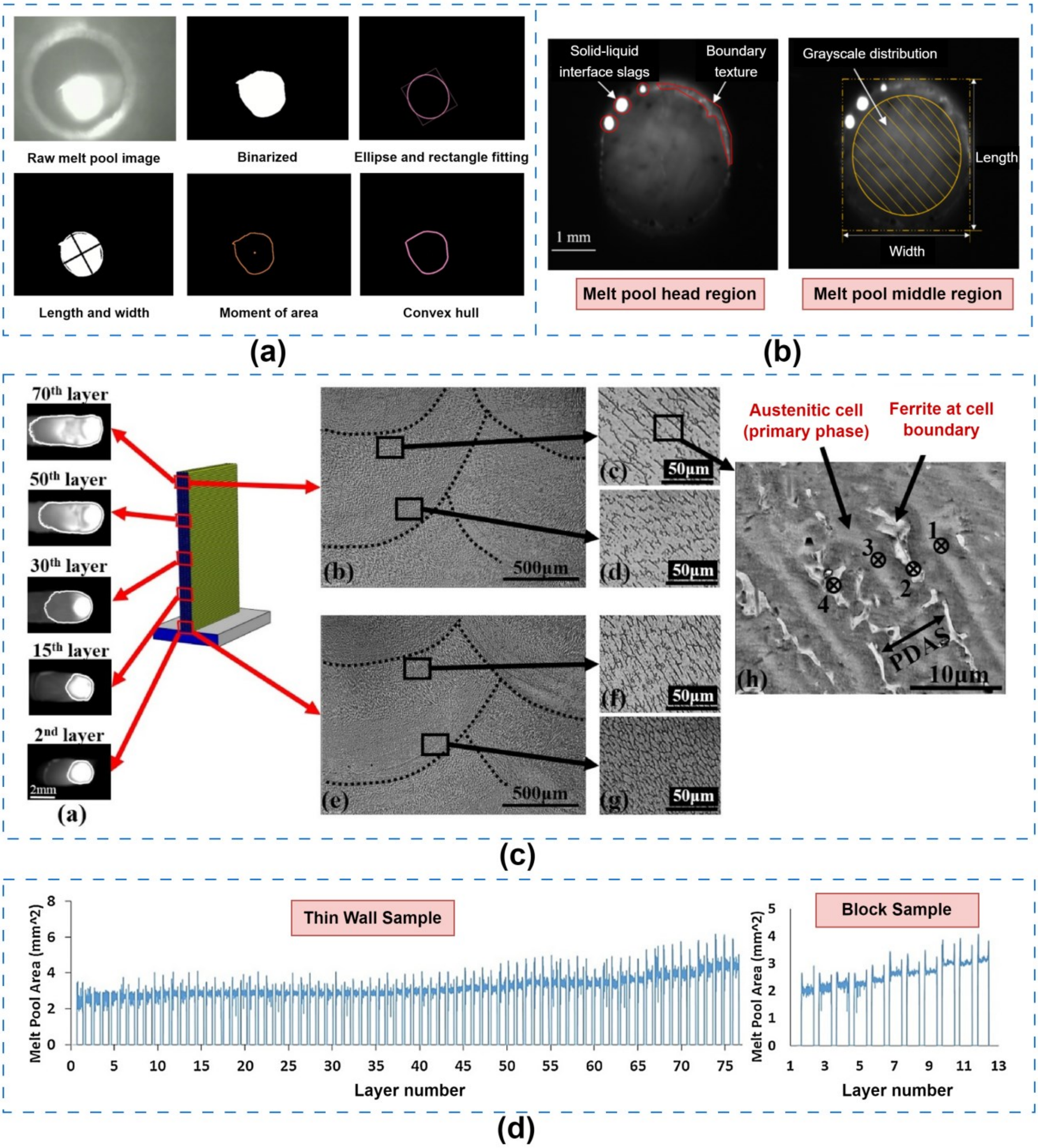

**Figure 6.** Coaxial Melt Pool Visualization in LDED Process. (a) Coaxial image processing and feature extraction, including binarization, ellipse fitting, and area calculations [70,160]. (b) Melt pool head and middle region visual characteristics, highlighting solid-liquid interface, boundary textures, and key metrics like width and grayscale distribution [129]. (c) Real-time coaxial melt pool isotherm images across layers in a thin-wall sample, correlating grain size variations with cooling rates [161]. (d) Melt pool size expansion in thin-wall (left) and block samples (right) under constant laser power [161].

Infrared (IR) thermal cameras offer additional advantages in capturing melt pool dynamics by providing temperature-related features, often more informative than those from monochrome CCD cameras in the visible spectrum. **Figure 7** shows several examples of these IR thermal images of the melt pool in the LAM. **Figure 7**(a) displays the melt pool's brightness temperature distribution under two contrasting conditions, captured using a coaxial IR camera [167]. Variations in laser power and gas flow rate are seen to influence the temperature distribution significantly. **Figure 7**(b) depicts the thermal image of a melt pool during the LDED process with titanium alloy. The separation of solidus and liquidus regions is clearly visible, offering real-time insight into the material's phase change during processing. These examples demonstrate the benefits of infrared thermal imaging in capturing melt pool dynamics and heat transfer, paving the path for more precise control and optimization of LAM processes.

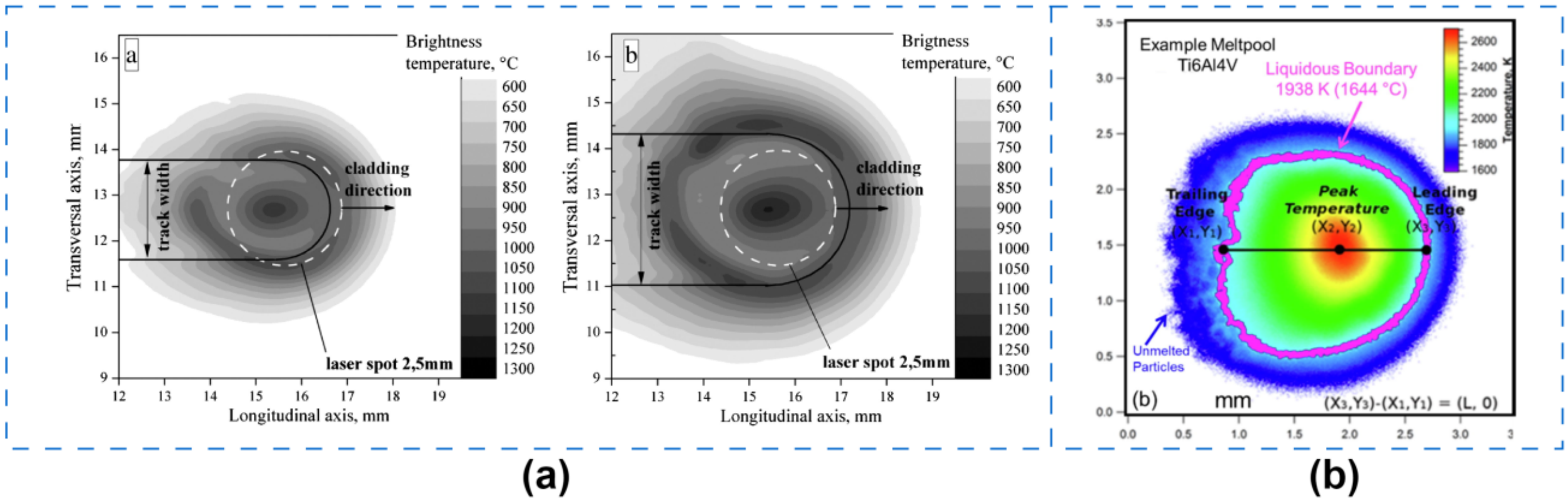

**Figure 7**. Infrared (IR) thermal images of melt pool in LAM: (a) Coaxial IR camera captures contrasting brightness temperature distributions under two distinct conditions (Left: P=500 W, f=6 l/min; Right: P=750 W, f=9 l/min) [167]. (b) Thermal image of a melt pool during LDED of titanium alloy material. A smooth boundary (purple line) separates the solidus and liquidus region of 1938 K. Un-melted particles are observed along the trailing edge.

### *3.1.2. In-Situ Defect Detection With Melt Pool Features*

Apart from extracting melt pool geometric features as quality indicators, melt pool images can be used for defect prediction through state-of-the-art ML and deep learning (DL) algorithms [168–174]. Several benchmarking papers, summarized in **Table A1**, focus on AI-assisted vision-based in-situ defect detection in LAM. Preliminary studies aimed to establish mappings between melt pool images and process conditions. For instance, Kwon et al. [175] used a high-speed camera to collect melt pool images during the SLM process, which were fed into a deep neural network (DNN) model to predict laser power levels. Spatters and temperature from the melt pool can also be correlated to process parameters such as scanning speed and laser power using deep convolutional neural network (DCNN) [176,177]. Since laser power and scanning speed are two of the most influential parameters, the suggested method demonstrates the feasibility of using AI techniques for quality prediction.

In recent years, AI-assisted vision-based porosity detection has been the main focus of ongoing research and development. Porosity can be induced by lack of fusion, gas entrapment, and particle movement within the melt pool; hence, high-level features retrieved from infrared or high-speed digital images of the melt pool can reveal the complex pore formation mechanism. Extensive research findings have been reported on vision-based porosity predictions [133,178–185]. **Figure 8** compares vision-based in-situ defect detection methods using traditional supervised ML, unsupervised learning, and deep learning methods. For example, Khanzadel et al. [180] used melt pool images to distinguish pores from normal printing states using traditional supervised ML learning methods such as Support Vector Machine (SVM), Decision Tree (DT), K-nearest neighbour (KNN), and Quadratic discriminant analysis (QDA). Melt pool morphological features such as circumference and coordinates were extracted. Functional principal component analysis (FPCA) was utilized to select essential features which were then used to train the ML models. Scime and Beuth [186] employed SVM to classify melt pool images to locate LoF pores, as shown in **Figure 8**(a). In contrast to Khanzadel et al. [180], they extracted melt pool signatures using an unsupervised learning technique, namely the Scale Invariant Feature Transform (SIFT) and histogram of oriented gradients (HOG). Other unsupervised learning methods, such as self-organizing maps (SOMs), can also be used to predict porosity locations in melt pool thermal images [181], as shown in **Figure 8**(b).

However, traditional supervised ML techniques have significant drawbacks: (1) the feature definition and extraction must be conducted manually, which cannot accurately and adequately depict the complex pore formation mechanism; (2) the feature selection procedure is subject to human judgment, potentially limiting model accuracy or leading to overfitting; (3) traditional ML techniques are ineffective for training large datasets. In contrast, DL techniques such as CNN do not require time-consuming and labour-intensive manual feature engineering but to automatically learn and extract features with greater efficiency and accuracy. For example, Zhang et al. [131] utilized a co-axial high-speed camera to classify porosity occurrences during the DED process using CNN, achieving 91.2% accuracy. Micropores below 100 μm can be predicted for samples fabricated by sponge Titanium powders. Similarly, Chen et al. [187,70] used CNN (VGG-16) for in-situ defect identification using coaxial melt pool image, as shown in **Figure 8**(c).

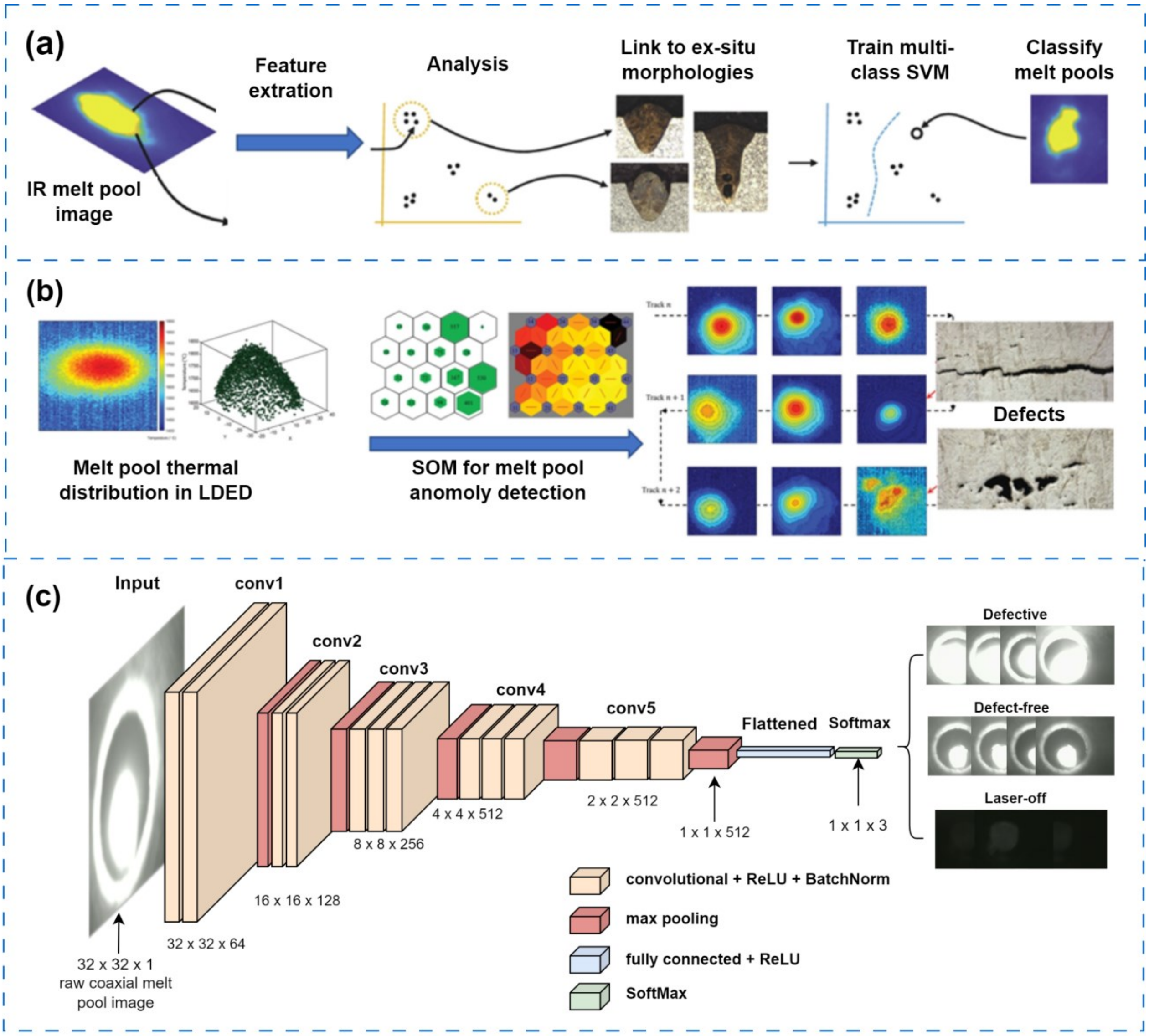

**Figure 8.** ML-assisted in-situ defect detection methods using coaxial melt pool images: (a) Flowchart outlining the application of a traditional supervised ML method for melt pool image classification [186,188]. (b) Flowchart depicting the use of an unsupervised Self-Organizing Map (SOM) method to distinguish porosities using melt pool images [181]. (c) The application of a deep learning model (e.g., CNN) for in-situ defect identification using coaxial melt pool image [187,70].

Optical-based monitoring approaches can also identify other types of defects. CNN models can classify melt pool images to predict humping, spatter, or normal fabrication states [189]. Gonzalez-Val et al. [111] proposed a CNN model to extract quality indicators from raw images, which were then used to quantify dilution and predict defective spots. The authors used a high-speed Medium Wavelength Infrared (MWIR) camera to monitor the melt pool during the DED process, demonstrating better performance than traditional CMOS cameras [190]. Although only a single track was deposited, the idea of online quality prediction was successfully demonstrated using state-of-the-art DL techniques.

However, the studies mentioned above do not fully achieve in-situ and real-time capabilities. Most studies employed offline ML model training (batch learning), while online model deployment and inference latency were not adequately addressed. Recently, several studies have attempted to bridge this gap. Knaak et al. [190–192] developed a real-time defect detection system with low latency (1.1ms) on low-power embedded computing boards for the laser welding process. The authors proposed an ensemble DL architecture that can extract spatio-temporal features from time-series melt pool infrared images. Various welding defects, such as sagging, lack of fusion, irregular width, and lack of penetration, can be identified on-the-fly. However, due to the complex layer-by-LAM process, real-time online defect detection capability is still underdeveloped.

#### *3.1.3. Anomaly Detection And Fault Diagnosis*

Optical-based sensing approaches can be used for various purposes beyond melt pool monitoring, including process anomaly detection [193–201], LDED powder flow monitoring and fault diagnosis [65,202,203], powder bed anomaly detection and classification [204–208], and identification of surface roughness and geometric distortions of additive manufactured components [209–211]. **Figure 9** demonstrate a recent example of anomaly detection in LPBF. Nguyen et al. [64] proposed a semi-supervised learning method to identify anomalies including LoF, overheated, and unfused powder with more than 95% accuracy. Similarly, Wang et al. [212] proposed a CenterNet-based anomaly detection method, where the exact locations of different anomalies can be identified and classified within a bounding box. Scime and Beuth [206] proposed a Multiscale CNN (MsCNN) that simultaneously detect and classify various powder bed anomaly as shown in **Figure 10**. An off-axis camera captures the powder bed image and MsCNN identifies anomalous area including recoater hopping, streaking, debris, super-elevation, part damage, and incomplete spreading. The identified anomalies for each layer can be stacked together and create a 3D quality visualization, which closely match the actual part quality. Moreover, Abdelrahman et al. [213] developed a layer-wise fault identification system for LPBF using high-resolution optical imaging. Optical data was employed to identify powder bed flaws by comparing images taken under different illumination conditions and at various layers. Another layer-wise imaging methods using DL for in-situ flaw detection in LPBF have also been demonstrated in [214]. Locations of powder bed anomalies can be detected and used to study critical defect formation [215]. Fischer et al. [216] developed a higher resolution imaging system that achieved a classification accuracy of 99.15% using a recoater-based line sensor to capture images of a powder bed. The author trained CNN to classify the following eight types of defects: balling, incomplete spreading, groove, ridge, spatters, protruding part, scattered powder, and homogenous. Lu et al. [217] linked powder bed in-situ images to the mechanical properties of as-built SLM component, suggesting that visual monitoring can predict component strength and density in addition to identifying defects. Additionally, image-based feature extraction can be employed for real-time measurement of deposition height [218,219], and combined with raw melt pool images and temperature profiles to predict microstructure characteristics such $\alpha$ lath thickness and $\beta$ grain size, as well as bead geometry in LW-DED [220,221].

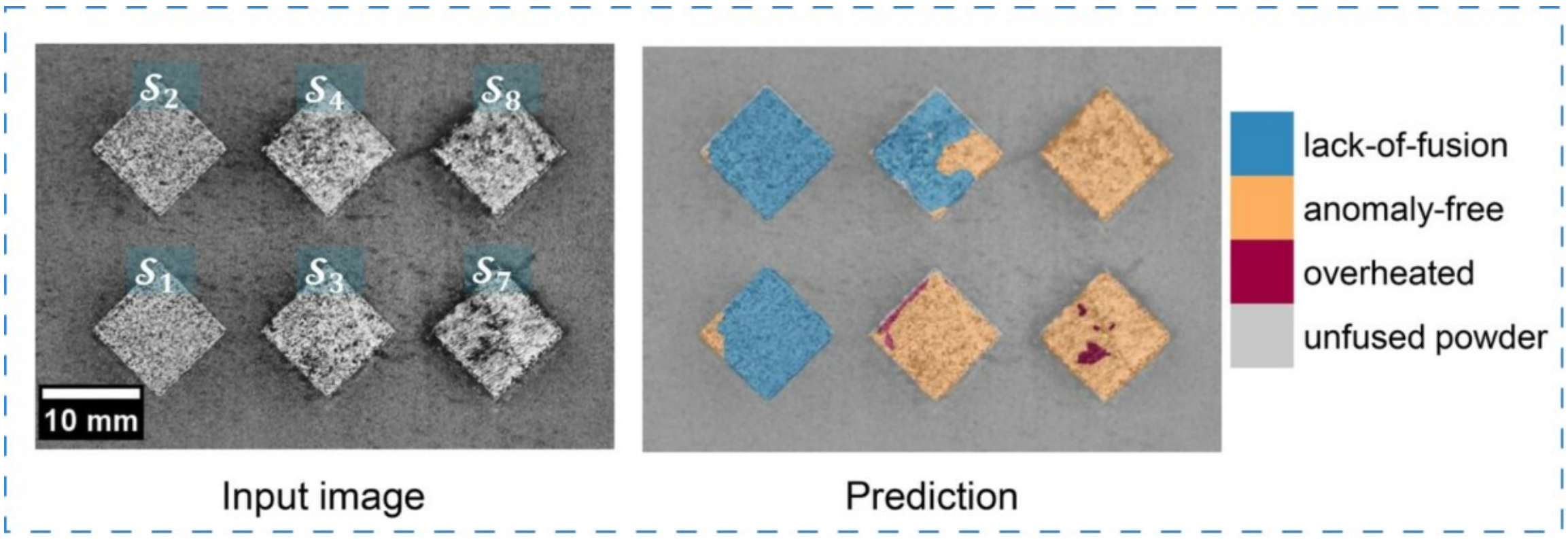

**Figure 9**. Vision-based anomaly (lack-of-fusion, overheated, unfused powder) detection in LPBF using semi-supervised ML on cubes printed with varying printing parameter settings [64].

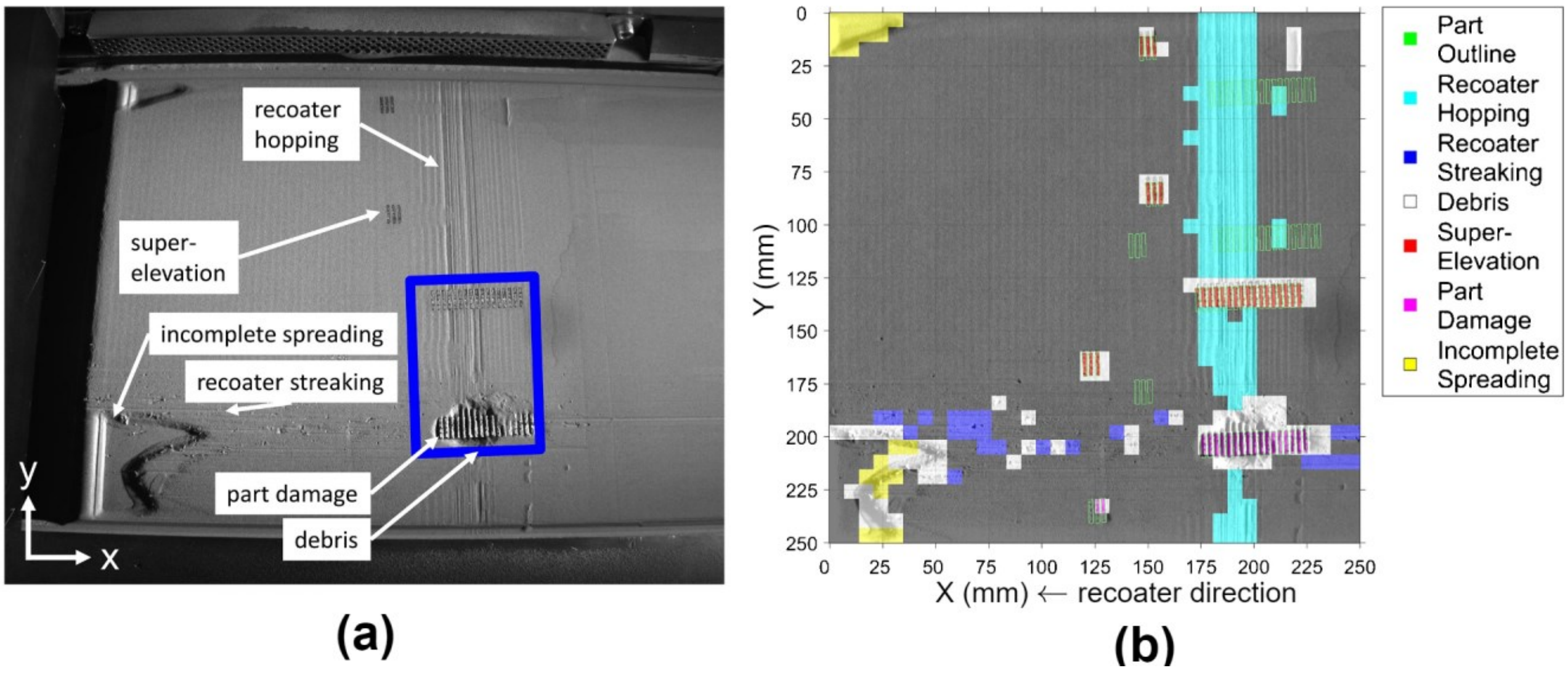

**Figure 10**. Powder bed anomaly detection and classification using multi-scale CNN (MsCNN) [206]: (a) Example of powder bed image during the process with manually annotated anomalies. (b) Powder bed anomalies detected and classified by the MsCNN.

### *3.1.4. Infrared Thermal Imaging For Localized Quality Predictions*

This section discusses recent advancements in localized quality prediction using in-situ infrared thermal imaging. Unlike vision-based melt pool monitoring, the IR camera is positioned off-axis to monitor the part surface temperature rather than the melt pool region. Thermal history is defined as time-series temperature data inside a specified ROI. Examples of localized quality prediction using IR thermal history is shown in **Figure 11**. Lough et al. [182] used SWIR temperature history data to conduct localized quality prediction in LPBF. By registering SWIR data with μCT data, thermal feature-based porosity probability maps can be constructed (as shown in **Figure 11**(a)), which effectively estimate the likelihood of porosity occurrences within each voxels of the part (as shown in **Figure 11**(b)). Similarly, Estalaki et al. [222] demonstrated the capability of using IR thermal history to predict voxelized micropore defects in LPBF with ML. The researchers identified two fundamental thermal history features: time above the apparent melting threshold ($\tau$) and the maximum IR radiance ($T_{max}$). The F1 scores of the ML models trained on these features achieved above 0.96, with feature importance analysis highlighting $T_{max}$ as the most significant feature for predicting the state of a voxel. Furthermore, Xie et al. [223] proposed a mechanistic data-driven framework, which integrated wavelet transforms and CNN using the IR thermal history data to predict localized quality, as shown in **Figure 11**(c). They focused on predicting spatial variations of as-built mechanical properties (e.g., ultimate tensile strength UTS), which is caused by localized heating and cooling during the LAM. The framework allowed for multi-resolution analysis and importance analysis, revealing crucial mechanistic features extracted by wavelet transform underpinning the LDED process. This methodology showcased strong predictive capabilities even with a small amount of noisy experimental data, laying a robust foundation for predicting the spatial and temporal evolution of mechanical properties in LAM.

These examples highlight the potential of in-situ infrared thermal imaging coupled with advanced data analysis techniques for predicting location-dependent part quality in LAM. However, the existing IR thermal monitoring methods still lacks a sufficient industrial readiness. One of the key limitations is the use of most existing thermal imaging is restricted to thin-walled parts, while the temperature profile for more complex geometries is difficult to be captured. This is particularly true for complex geometries where the thermal camera's field of view is blocked by the built proportion. One possible solution is to use multiple thermal cameras placed at different positions to capture the temperature distribution of the part from different viewpoints.

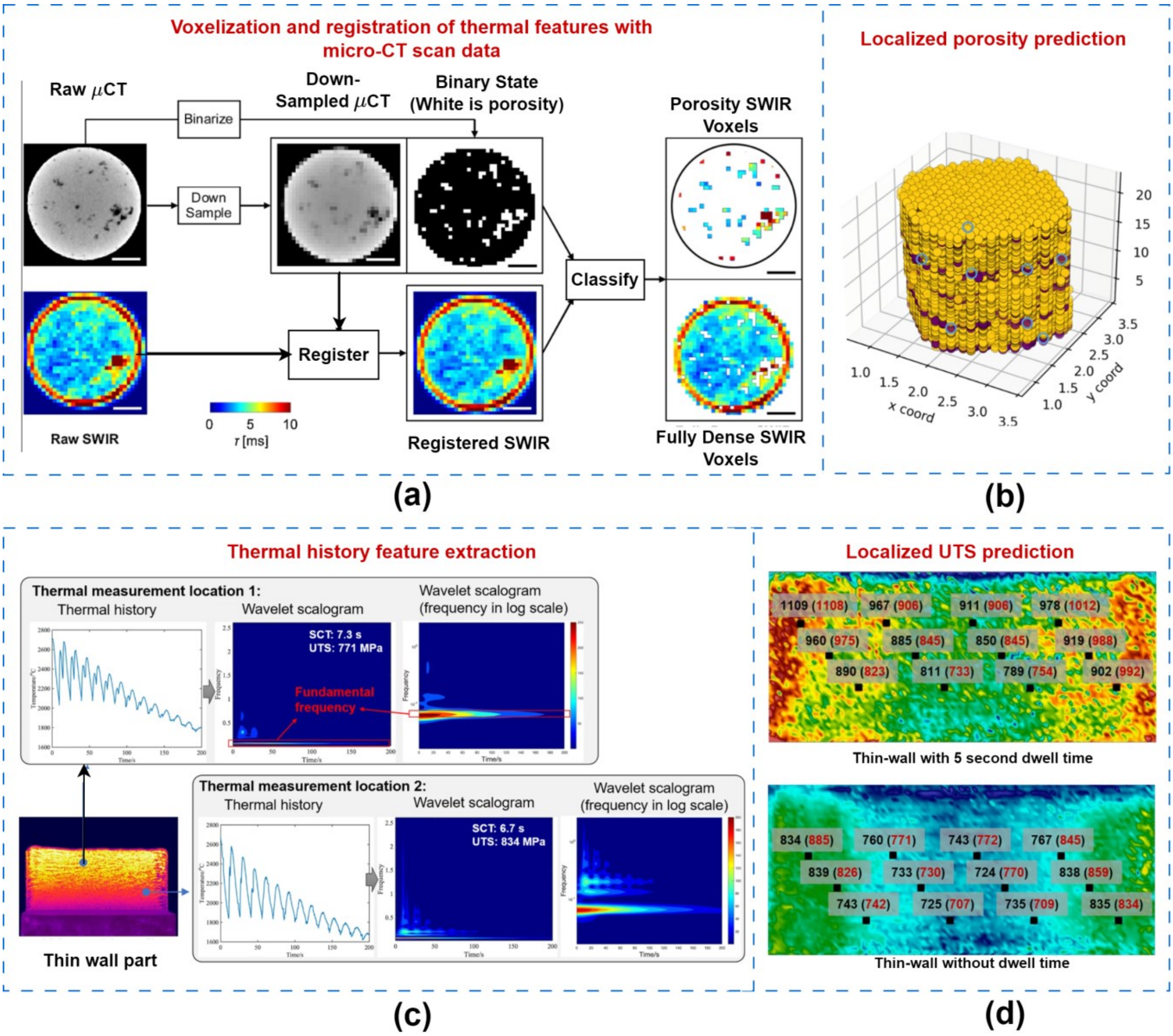

**Figure 11.** Thermal feature extractions for location-depended quality predictions: (a)-(b) Voxelization and registration of thermal features with micro-CT scanned data to predict locations of porosities in LPBF [182,222]. (c)-(d) Thermal history extraction and wavelet transform to predict location-specific UTS in LDED [223].

### *3.1.5. Challenges, Limitations, And Future Development*

This sub-section discusses the challenges and limitations of optical (vision)-based monitoring in LAM. Despite the significant advancements in vision-based monitoring techniques, there are still several significant challenges that need to be addressed to fully exploit their potential.

i. ***Real-Time Image Processing:*** Real-time image processing remains one of the most significant challenges in vision-based monitoring in LAM. The high-speed nature of LAM results in an immense volume of data generated by the monitoring cameras. This data demands not only robust computational resources but also highly efficient algorithms for real-time processing.

Approaches such as pruning neural networks, which streamline the predictive process by eliminating less significant neural connections, could be leveraged to enhance the speed of image processing.

ii. ***Sensing Capabilities and Cost***: The sensing abilities of different visual sensors—encompassing dynamic temperature sensitivity range, acquisition frequency, field of view, focal distances, and more—differ significantly. Balancing the accuracy of sensing, the cost of sensors, and the complexity of integration is often a major concern for industry end-users. Altenburg et al. [224] conducted a comprehensive evaluation of MWIR, SWIR, and high-speed NIR cameras for in-situ monitoring of the LDED process. While a high-speed camera in the visible (VIS) spectrum with a NIR bandpass filter offered the highest temporal and spatial resolution, its limited dynamic temperature range could fall short for process monitoring. Hence, identifying the optimal sensor that balances defect detection accuracy and cost remains a major challenge.

iii. ***Temperature Measurement and Emissivity Calibration***: SWIR and MWIR cameras provide a wide dynamic temperature range, which can be further expanded through proper selection of attenuation filters and integration times. However, to ensure accurate temperature readings, these IR thermal cameras necessitate emissivity calibrations [225,226]. This is particularly challenging due to variations in metal emissivity with temperature, wavelength, material phase, surface roughness and other factors [227]. As a result, precise temperature profiles around the melt pool remain elusive [141]. If calibration is not performed correctly, uncertainties and errors can arise, leading to inaccuracies in the monitored data [228].

iv. ***Sensor Integration Difficulties***: The integration of vision sensors into LAM systems poses its own set of challenges. The installation of coaxial vision sensors necessitates custom laser head design, and off-axis melt pool monitoring may require image transformation, potentially resulting in less reliable and accurate outcomes. Further, the problem of occlusion, where certain parts of the process are hidden from the camera's view, is a significant issue, especially in off-axis camera setups.

Despite these challenges, vision-based monitoring in LAM has made significant advancements in recent years and continues to be the most popular in-situ monitoring approach. Alternative monitoring techniques, such as laser profilometry scanning and acoustic-based monitoring, could mitigate some of the drawbacks of vision sensors, which will be discussed in the subsequent sections.

### 3.2. Acoustic-Based Monitoring

Acoustic monitoring in LAM provides flexible sensor setups, fast dynamic responses, and reduced hardware costs, making it an attractive alternative to other sensing techniques. This method leverages the acoustic signals produced during laser-material interactions, which encode valuable insights into physical phenomena such as melting, solidification, crack propagation, and pore formation [229]. Acoustic sensors are particularly adept at detecting structure-borne sounds, such as those generated by crack propagation within the workpiece, outperforming vision and temperature sensors in this regard [230]. The ease of integration with existing LDED and LPBF equipment further adds to its appeal within the AM community. This section elaborates on the advancements in acoustic-based monitoring for LAM, highlighting promising outcomes and innovative methodologies in the recent years.

#### 3.2.1. Origins Of Acoustic Emissions And Acoustic Feature Extraction

Despite the comparatively limited research on acoustic monitoring in LAM, it has long been employed to assess weld quality. The technique is inspired by experienced welders, who often use arc sound as an informal gauge of weld penetration quality. The basic premise lies in the fact that any mechanical interaction, such as the contact of a laser or arc with a substrate in welding and AM, generates acoustic waves. These waves can be captured and processed to yield valuable process insights. Accordingly, extensive efforts have been made to correlate acoustic signatures with weld quality [231–239]. For instance, Song et al. [231] and Lv et al. [232] studied arc welding sound in relation to distinct penetration states, employing statistical features to train ML models for penetration state recognition. As illustrated in **Figure 12**, arc sound varies significantly for different weld penetration state, demonstrating the feasibility of acoustic monitoring for penetration state recognition in practice. Hauser et al. [67] used Mel spectrum of AE signal to identify track deviation anomalies during WAAM, establishing that process anomalies and acoustic emissions were interrelated, primarily due to the size of the arc. It was also discovered that stable processes exhibit a consistent mean intensity in acoustic emissions, whereas anomalies show significant variations in acoustic intensities. For LAM, acoustic emissions predominantly arise from laser-material interactions, with the expansion of melted powder particles generating air pressure. However, it is challenging to distinguish the LAM sounds due to the ambient noises including the protective gas flow, machine sound, and powder stream noise.

Acoustic feature extraction, a pivotal step in acoustic monitoring, involves capturing, digitizing, and processing the raw acoustic signal to retrieve relevant information. These features span across the time-domain, frequency-domain, and time-frequency representations. The mathematical definitions and descriptions of these features are provided in **Table 4**. Time-frequency representations [240] are particularly effective, as they enable the calculation of relative energy densities across frequency bands, representing acoustic signatures in both frequency and time domains. Methods such as short-time Fourier transform (STFT) [241], wavelet transforms (WT) [242], and the Mel-frequency cepstrum coefficient (MFCC) [243] have been used with promising results. For instance, Lv et al. [232] extracted the cepstrum coefficient of arc sound and input it into a Back Propagating Neural Network (BPNN) model, achieving an accuracy of 80%-90% in predicting welding penetration. Shevchik et al. [235] demonstrated that spectral features extracted using WT can achieve exceptional performance for online welding quality monitoring. These promising results lay the foundation for further research into acoustic sensing technologies in LAM.

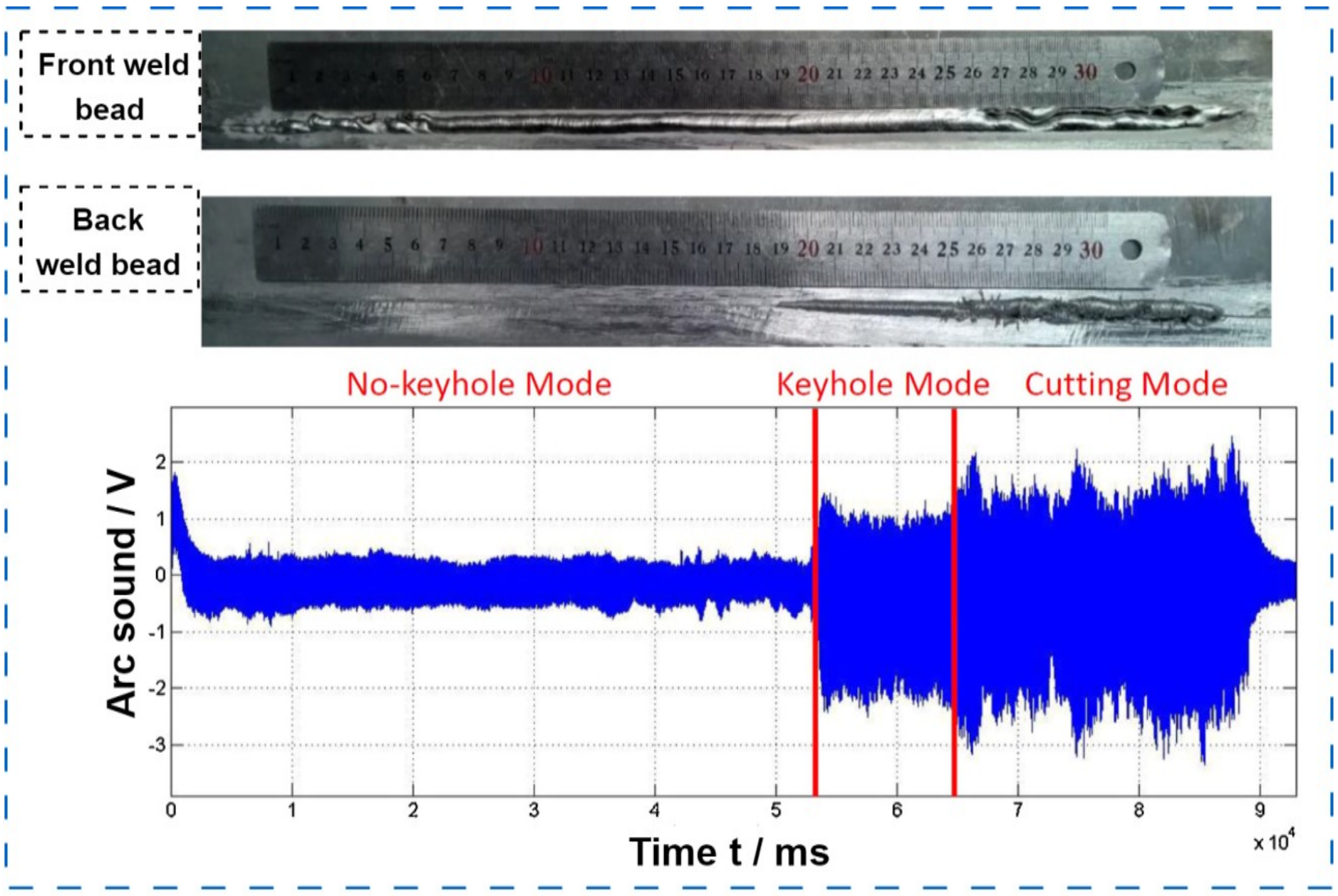

**Figure 12**. Acoustic Signal Characteristics in Arc Welding: Acoustic signatures corresponding to three distinct penetration states in plasma arc welding: no-keyhole mode, keyhole mode, and cutting mode [231].

**Table 4.** Summary of common acoustic features in time-domain, frequency-domain, and time-frequency representations.

| Feature type | Feature name | Description | Ref |
|---|---|---|---|
| Time-domain | Amplitude Envelope (AE) | Outlines signal amplitude changes over time. | [68] |
| | Root-Mean-Square Energy (RMS) | Indicates signal power, reflecting loudness. | [68] |
| | Zero Crossing Rate (ZCR) | Identifies signal sign changes, distinguishing sound types. | [68] |
| | Clearance Factor (CLF) | Measures peakiness relative to RMS, useful in fault diagnosis. | [102] |
| | Crest Factor (CF) | Indicates signal peakiness, useful in transient event identification. | [102] |
| | Impulse Factor (IF) | Measures signal impulsiveness, important for impulsive event detection. | [102] |
| Frequency-domain | Spectral Centroid (C) | Indicates the spectral mass centre, essential for frequency band analysis. | [244] |
| | Spectral Roll-off (SR) | Frequency below which a certain percentage of total energy lies, differentiating harmonic and non-harmonic parts. | [245] |
| | Spectral Bandwidth (SBW) | Average of frequency band distances from the centroid. | [244] |
| | Spectral Flatness (SF) | Geometric to arithmetic mean ratio of spectrum, assessing noise-like versus tone-like quality | [246] |
| | Band Energy Ratio (BER) | Ratio of low to high-frequency band power. | [247,102] |
| | Spectral Contrast (SC) | Difference between spectral peaks and valleys. | [68,247] |
| | Spectral Variance ($\mu_2$) | Standard deviation around the spectral centroid. | [68,248] |
| | Spectral Skewness ($\mu_3$) | Measures energy distribution symmetry around the centroid. | [68,248] |
| | Spectral Kurtosis ($\mu_4$) | Indicates the "tailedness" of the spectrum, useful for outlier detection. | [68,248] |
| | Spectral Crest (Crest) | Ratio of spectrum's maximum to its mean. | [68,248] |
| | Spectral Entropy (H) | Measures spectrum peakiness. | [68,249] |
| | Spectral Flux (Flux) | Assesses spectrum variability over time, useful in audio segmentation. | [250] |
| Time-frequency representations | Mel Frequency Cepstral Coefficients (MFCCs) | Represents sound phonemes in audio. | [68] |
| | short-time Fourier transform (STFT) | Time-frequency representation of signal frequencies. | [241] |
| | Continuous Wavelet transforms (CWT) | Provides a flexible time-frequency signal analysis. | [66,251] |
| | Chroma Feature | Projects spectrum onto 12 semitone bins of a musical octave. | [252] |

### *3.2.2. In-Situ Acoustic Monitoring In LPBF*

The application of in-situ acoustic monitoring in LPBF processes has gained significant attention due to its potential for defect detection in recent years. This section offers a holistic review of the acoustic sensor setups, acoustic signal processing techniques, and the correlation between acoustic signal features and defect occurrences in LPBF.

As depicted in **Figure 13**, the literature reports a variety of acoustic sensor setups for LPBF. **Figure 13**(a) and (b) display an airborne acoustic sensor (PAC AM4I microphone) connected to a Data Acquisition (DAQ) device, which is used to capture the acoustic waves emanating from laser-material interactions during LPBF [66,112]. The data acquisition was automatically triggered with a photodiode, and a post-process low-pass Butterworth filter with a cut-off frequency of 100 kHz is utilized to reject noise in the raw signal. The key advantage of the microphone sensor is the flexible sensor configurations and low cost. However, the positioning angle, distance to process zone, and types of microphone (e.g., polarization, frequency response range) are the critical factors influencing the quality and reliability of collected acoustic data.

Another prevalent approach involves the use of an acoustic emission (AE) sensor affixed to the substrate during LPBF, as demonstrated in **Figure 13**(c) and (d). The AE sensor picks up the elastic wave signal generated by laser interaction with the powder bed and substrate. This signal is then amplified by the preamplifier and stored by the data acquisition system [253]. The AE sensors excel in detecting structure-borne sounds, like those emitted during crack propagation within the workpiece [254], a merit unattainable by vision and thermal-based sensing methods. However, AE sensor-collected signals are highly susceptible to environmental noise, such as machine movement and gas flow, and the setup lacks the flexibility offered by the microphone sensor.

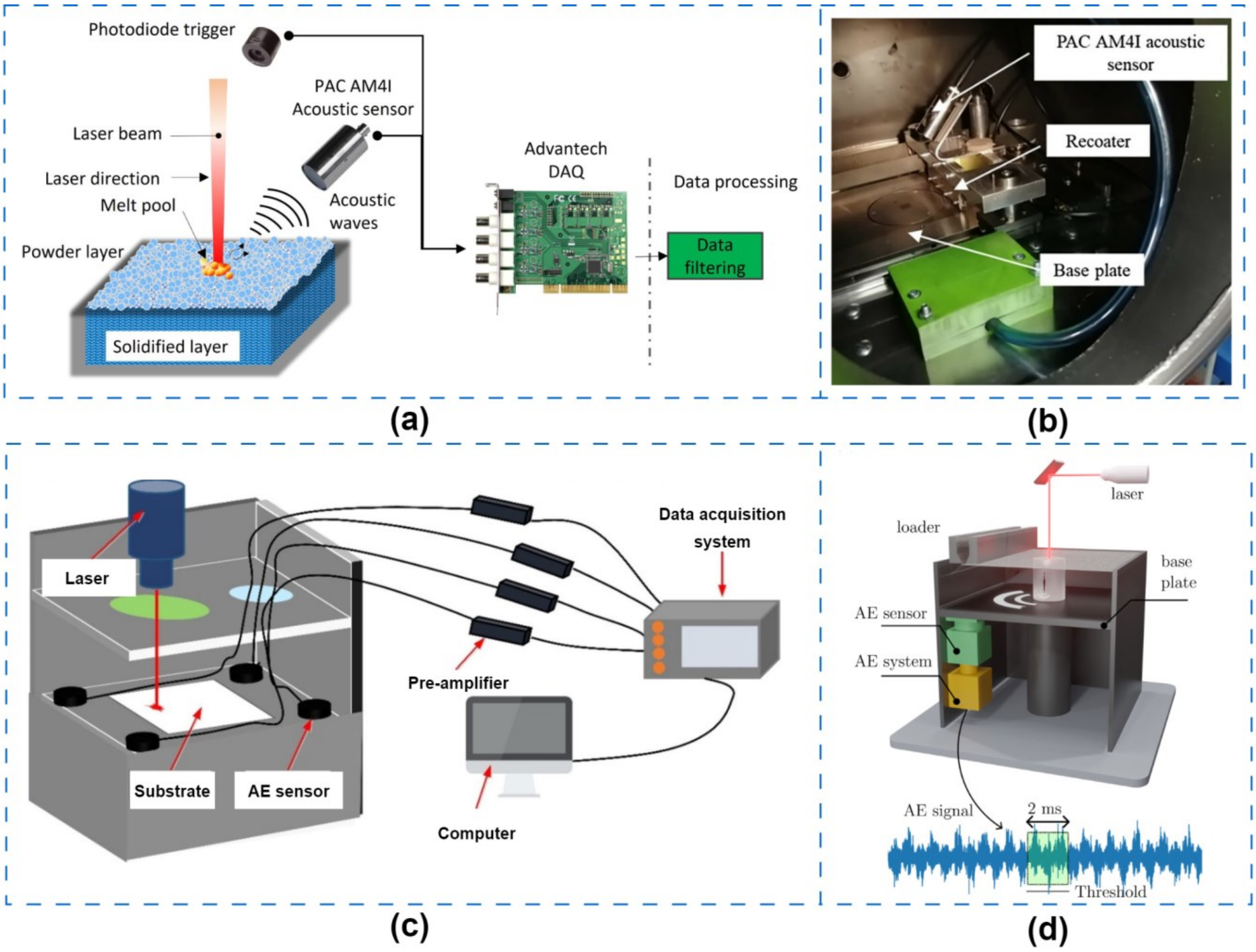

**Figure 13.** Various acoustic monitoring setups in LPBF. (a) Airborne acoustic sensor (PAC AM4I microphone) connected with a DAQ device for LPBF process monitoring [112]. (b) Image of LPBF process with installed PAC AM41 acoustic sensor [66]. (c) AE sensor placed near the substrate for SLM process monitoring [253]. (d) AE sensor setup implemented for LPBF process monitoring [254].

**Figure 14** provides a thorough examination of key factors that influence the acoustic signals captured by a microphone sensor during the LPBF process. It shows the sensitivity of the raw acoustic signal to multiple factors including the microphone installation angle, ambient noise, and the dynamics of laser-material interactions.

A crucial aspect is the frequency response and inclination angles of the microphone sensor, which influence the acoustic signal energy density during laser-material processing. **Figure 14**(a) elucidates the energy variances of the acoustic signal and static pressure under different microphone inclination angles. It highlights how the energy of static pressure fluctuations is highly susceptible to slight alterations in the microphone installation angle, resulting in significant changes in signal energy. The acoustic signal induced by laser-material processing, on the other hand, is unaffected by microphone inclination angles.

The raw signal, based on its frequency range, can be separated into two categories: the acoustic signal generated by laser-material processing and the signal reflecting static pressure fluctuations. As illustrated in **Figure 14**(b), the acoustic signal and static pressure variations can be discerned during part construction stages. Notably, static pressure fluctuations, typically occurred in a very low frequency band (<22.4 Hz) [255], can have significantly higher amplitude than the acoustic pressure variations.

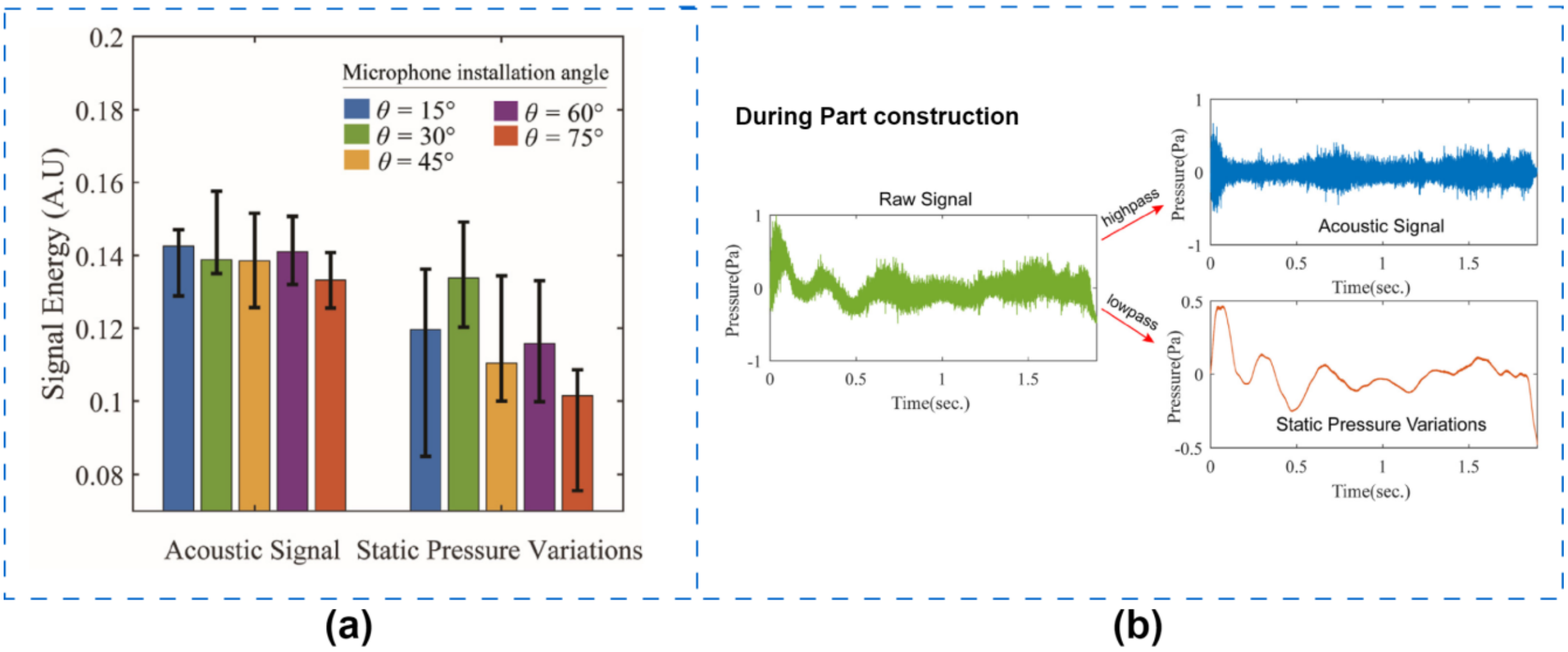

**Figure 14**. Acoustic signals in the LPBF process [255]: (a) Energy of acoustic signal and static pressure variations under different microphone inclination angles. (b) Acoustic signal and static pressure variations during part construction, with raw signal indicated in green.

The task of correlating acoustic signal characteristics with various defects in LPBF is critical in constructing ML models for defect identification. **Figure 15** illustrates this complex interaction. **Figure 15**(a) depicts the t-SNE representation of low-dimension feature space for three different regimes (LoF, conduction, keyhole pores) during LPBF processing of stainless steel, bronze, and Inconel [66]. The distinct clusters formed by different process regimes, regardless of alloy type, demonstrate the efficacy of using time-domain, frequency-domain, and time-frequency representations to distinguish between different defect regimes. Ito et al. [256] developed a wireless AE signals measurement method to detect microdefects (cracks and pores) at different locations in LPBF, as shown in **Figure 15**(b). The study successfully detected burst-type AE events during the process and confirmed the sources of these events as pores and microcracks within the specimen. Interestingly, a slight latency (0.4 to 0.8ms) was detected between the AE event and the occurrence of the defect, implying a delay in defect detection after the laser spot had moved away from the defect.

In **Figure 15**(c), the acoustic envelope and spectra of signals affiliated with keyhole pores and those not affiliated with keyhole pores in the LPBF process are compared [102]. The findings suggest that the acoustic signals generated from keyhole pores possess significantly higher energy than those from non-defective regions. **Figure 15**(d) presents a 3D wavelet representation of the AE signal for LoF, conduction, and keyhole pores in stainless steel, indicating absolute intensities in temporal frequency distribution [66]. This time-frequency representation clearly demonstrates the distinguishability of various defects and alloy materials, confirming the correlation of extracted acoustic signal features with defects in LPBF.

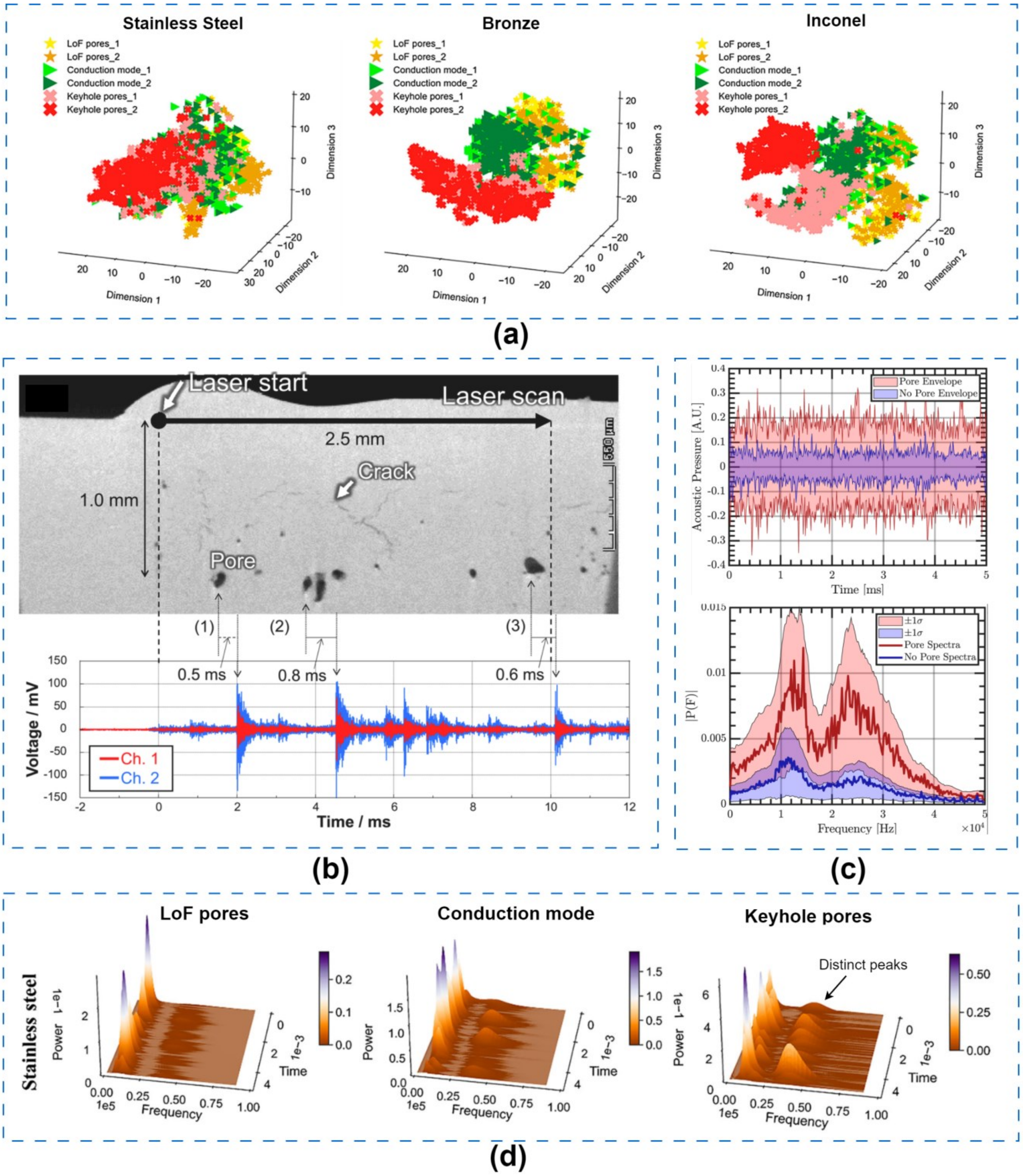

**Figure 15**. Acoustic signal signatures and their correlation to various defects in LPBF. (a) t-SNE representation

of low-dimension feature space for three different regimes (LoF, conduction, keyhole pores) in LPBF processing of stainless steel, bronze, and Inconel [66]. (b) Wireless AE signals associated with microdefects (cracks and pores) at different locations in LPBF [256]. (c) Acoustic envelope and spectra of pore-affiliated and non-pore affiliated signal in the LPBF process [102]. (d) 3D wavelet representation of the AE signal for three regimes (LoF, conduction, keyhole pores) in stainless steel, illustrating absolute intensities in temporal frequency distribution [66].

The acoustic signal processing and feature analysis described above demonstrate the feasibility of utilizing acoustic sensors to detect defects in the LPBF process. The features can be employed to effectively describe acoustic events, allowing appropriate ML models to be deployed for in-situ defect detection and classification tasks. Key researchers such as Wasmer and Shevchik et al. have made significant contributions to this field [257–259]. Using a highly sensitive optoacoustic sensor (fibre Bragg grating (FBG) sensor), they collected AE signals corresponding to various types and concentrations of pores in LPBF [259]. Different porosity levels and process regimes (LoF, tubular, large pores) were created by adjusting the input energy density. Wavelet transform (WT) was applied to extract relative energy in narrow frequency bands, with the wavelet spectrum characteristics serving as input features for a deep CNN model. The model achieved an accuracy of 89% in classifying the sounds of different porosities [257]. A similar approach using spectral clustering to obtain acoustic features was also explored [259]. A further study looked into reinforcement learning (RL) techniques for porosity differentiation [260].

Other notable studies include research by Ye et al. [261], who demonstrated that instead of extracting various acoustic features, raw acoustic data could be directly fed into a deep belief network (DBN) for classification tasks. The raw acoustic signal underwent band-pass filtering (500-90,000 Hz), and the DBN achieved a 93% accuracy in differentiating sounds associated with balling and overheating defects. A study by Tempelman et al. [102] revealed that keyhole pores could be predicted with 97% accuracy using a support vector machine (SVM) classifier. The pore locations were spatially and temporally registered with the recorded time-series of laser position and acoustic pressure to identify specific partitions of the acoustic signals which correspond to pore formation. There are several similar studies on AI-assisted acoustic-based defect detection in LPBF [262–264], proving the substantial potential and effectiveness of using acoustic sensing with ML for in-situ defect detection in the LPBF process.

Recent advancements in acoustic monitoring have expanded its applications in LPBF, as demonstrated in [66,112,251]. These innovative approaches include semi-supervised process monitoring [112], and transfer learning of AM mechanisms across different materials [251]. Drissi-Dauodi et al. [66] reduced the window size for extracting WT features from 160 ms to 5 ms, enabling faster and more precise defect identification. However, developing a CNN model for defect detection in a specific material can be expensive and time-consuming, requiring balanced datasets for each investigated regime and sufficient data to ensure high accuracy. To address this challenge, the authors proposed a semi-supervised learning approach [112] that differentiates defect-free regimes from defective ones by training ML algorithms only on the distribution of acoustic signatures corresponding to defect-free regimes. They developed a generative CNN model (GANomoly) with a generator and discriminator, achieving 97% accuracy. Furthermore, the authors demonstrated that CNN models can learn transferable features from one material to another with minimal training [251]. After acquiring acoustic knowledge from one material, the network can predict defects in another material by re-training only the final two fully connected layers. This significantly reduces the time and expense of DL model development for new materials. These novel applications highlight the potential of acoustic monitoring in addressing various challenges in LPBF.

In summary, in-situ acoustic monitoring has great promise for detecting anomalies and defects during LPBF processes. Acoustic signal processing combined with ML allows for the correlation of acoustic signals to defects, facilitating early detection and mitigation. The success of these techniques is dependent on the careful selection and positioning of sensors, the extraction and selection of acoustic signal features, and the deployment of appropriate ML models for defect prediction. Following that, the next sub-section will go through in-situ acoustic monitoring in LDED, which has different challenges than the LPBF process.

### *3.2.3. In-Situ Acoustic Monitoring In LDED*

The LPBF process experiences significant noise influence on acoustic signals from protective gas flow, recoating, and powder delivery systems. However, as laser-powder interactions occur on a small scale and systems such as recoating and powder delivery remain stationary during laser scanning, the primary noise source is the protective gas flow. This noise, mainly affecting static air pressure, is concentrated in the low frequency band and can be readily filtered out [255]. In contrast, LDED experiences a more complex noise composition due to protective gas flow and powder stream impacting the substrate, complicating the analysis of laser-material interaction sounds. Consequently, acoustic-based monitoring in LDED is particularly challenging.

Initial studies into in-situ acoustic monitoring for LDED by Hossain and Taheri et al. [265–267] developed a custom transducer-based sensing device attached to a part's substrate to collect AE signals. Using a K-Mean clustering algorithm, the authors demonstrated that different LDED build conditions generate distinct sounds [265]. They then evaluated the AE signal relationships with changing machine status and deposition parameters, affirming the AE signal's connection to LDED part quality through statistical methods [267]. However, the sensor setup lacked flexibility, and the investigation did not delve into in-situ defect detection.

Subsequent advancements have incorporated a low-cost microphone sensor for in-situ LDED monitoring [68,268,269]. **Figure 16** displays microphone sensors attached to the laser head across different LDED machines, demonstrating the sensor's flexibility compared to the transducer-based AE sensor. The microphone angle and distance to the heat source are consistently maintained across both cases.

To tackle the critical challenge of noise in LDED sound, Chen et al. [268] presented an end-to-end acoustic denoising method using deep learning. This approach, which included audio equalization, bandpass filtering, and Harmonic-Percussive Source Separation (HPSS), effectively minimized ambient noise and isolated the sound of laser-material interactions. It can also be generalized across various LDED machines and alloy materials. Bevans et al. [269,270] employed a novel wavelet-integrated graph theory approach, using wavelet transform filters [271] for acoustic signal denoising, which detected flaw onsets such as porosity, spatters, and line width variations in wire-based DED process (**Figure 16**(b)). A similar approach was also presented in [272].

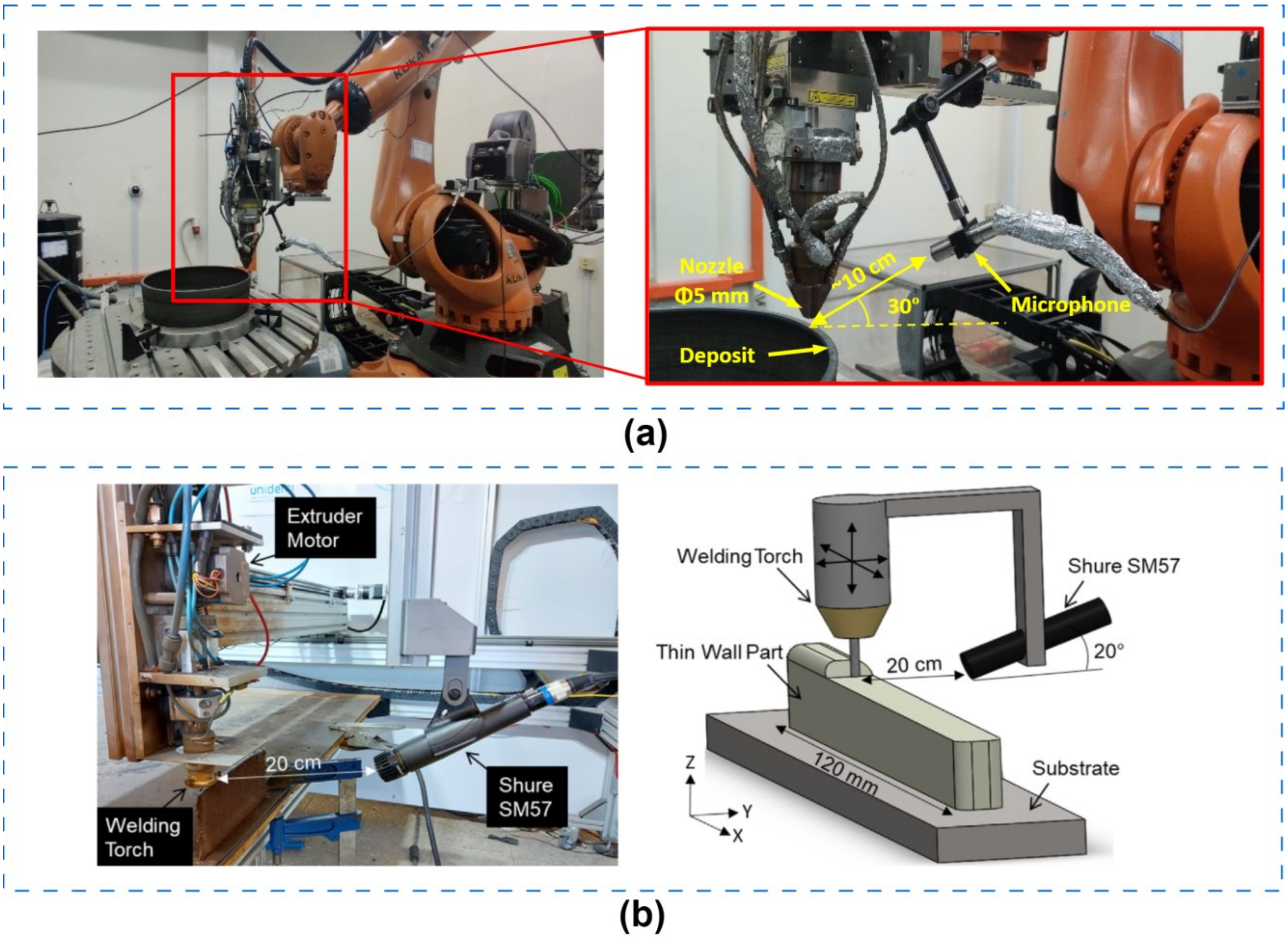

**Figure 16.** Various acoustic monitoring setups in LDED. (a) Microphone setup in a robotic LDED system [68]. (b) Image and schematic of an experimental setup featuring a Shure SM57 microphone attached to the welding torch in a wire-based DED process [269]. The microphone angle and distance to the heat source are consistently maintained across both cases.

Understanding acoustic signal characteristics and their correlation with various defects in LDED is crucial for defect detection. As shown in **Figure 17**, recent studies have analysed the acoustic signal signatures with different defect formation mechanisms. Hauser et al. [67] demonstrated a correlation between acoustic emissions, powder mass flow, and laser power for LDED processes deviating from optimal conditions. Higher laser power and mass flow rate resulted in increased AE intensity, indicating the transfer of laser energy to powder particles and stimulation of acoustic waves. The acoustic waves could be originated from the melting and volume contraction during solidification, which agitates the surrounding air (**Figure 17**(a)). Metal powder particles rapidly expand due to phase change from solid to liquid state, with volume expansion further intensified when the molten material is heated. The laser energy incites this expansion, stimulating surrounding air molecules. Consequently, the AE mean intensity rises with greater laser power and powder mass. Unstable processes often cause remelting of the deposited part, leading to geometrical fluctuations. These fluctuations alter the distance between the nozzle and the melt pool, and consequently, the interaction time between powder particles and the laser beam. Extended interaction time could result in fewer particles incorporating into the melt pool and a higher AE average intensity. Additional factors contributing to increased AE include heat accumulation, which causes increased spatters. These spatters are larger than powder particles, which could further increase AE when interacting with the laser beam.

Chen et al. [68] compared the acoustic signal signatures from defect-free, cracked, and keyhole pore areas using Fast Fourier Transforms (FFT) and MFCCs spectrums (**Figure 17**(b) and **Figure 17**(c)). They found that the magnitude of keyhole pore sound is considerably larger in the low frequency bands (0–5000 Hz), followed by crack sound and defect-free sound. The MFCCs value in low-frequency bands is lower in the defect-free deposition process. Cracks and keyhole pores, as indicated by a brighter colour in **Figure 17**(c), show a higher concentration of energy in the low-frequency bands. This is due to cracking being an energy-releasing process, allowing sound waves to readily distinguish such phenomena through unique patterns reflecting the abrupt increase in acoustic energy induced by crack propagation.

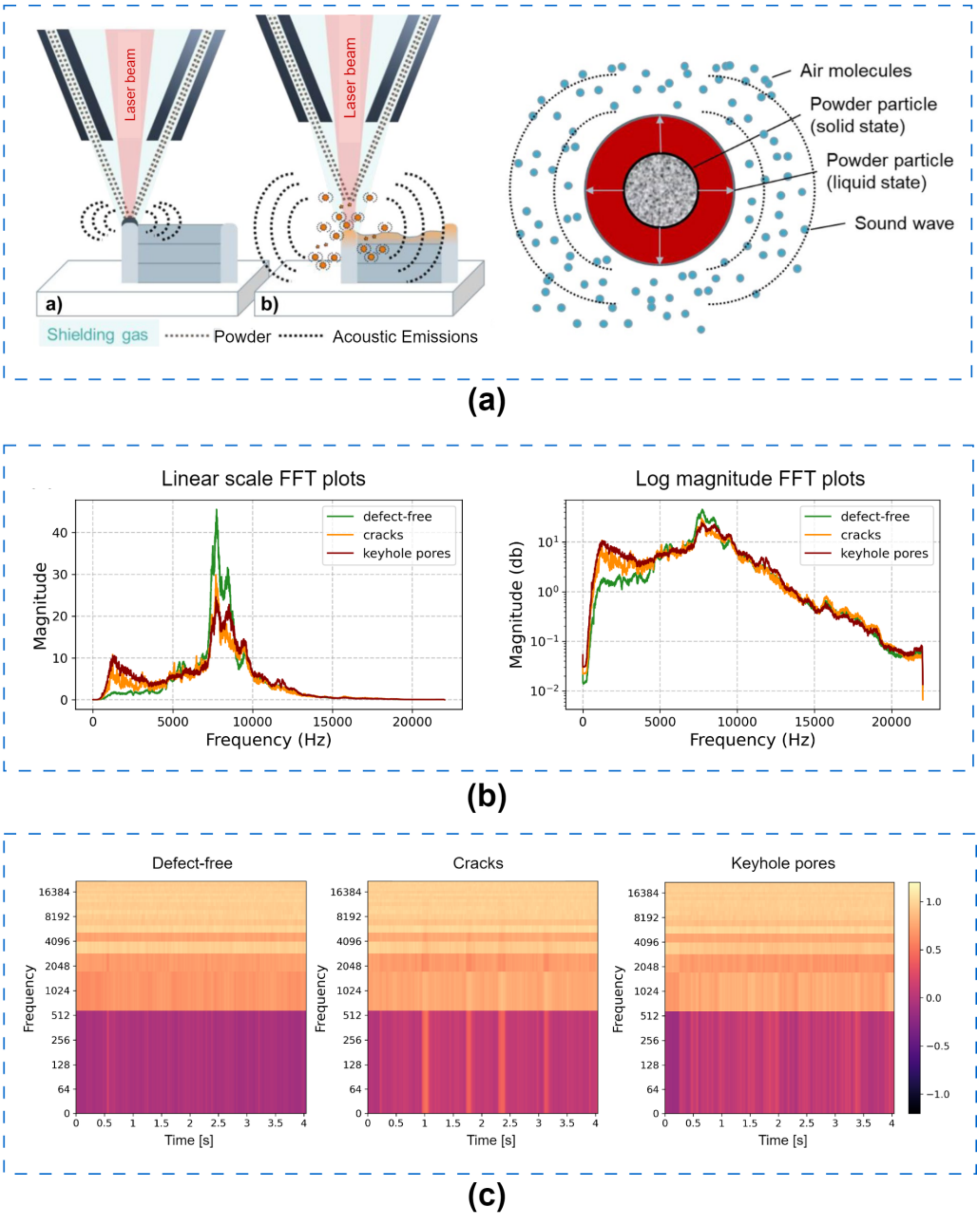

**Figure 17**. Acoustic signal analysis in LDED: origins, effects, and correlations to defects. (a) Source and influence of AE signal in stable and unstable LDED processes, highlighting increased spatters [67]. (b) FFT analysis of LDED sounds for different process conditions (i.e., defect-free, cracks, and keyhole pores) [68]. (c) MFCCs feature visualization across process regimes [68].

Existing studies have provided valuable insights into the acoustic signal signatures and their correlations to different process parameters and defects in the LDED process. Moreover, the research community has made substantial achievements in ML-assisted defect detection using these acoustic features. A survey on benchmarking ML models on acoustic-based defect detection are listed in **Table A2**. For instance, Gaja et al. [273] utilized unsupervised K-Means clustering to distinguish unique sound patterns in LDED. They found that, pores produce AE events characterized by higher energy, shorter decay time, and lower amplitude compared to cracks. Among various features, acoustic signal energy emerged as the most crucial in defect differentiation. Chen et al. [68] introduced a CNN model that leverages MFCCs acoustic features to classify LDED sounds for in-situ detection of cracks and keyhole pores, achieving an accuracy of 89%. As shown in **Figure 18**, the authors compared the performance of the MFCC-CNN model with traditional machine learning models (e.g., SVM, KNN, RF, GB, etc.) trained on both raw and denoised acoustic data. Their findings confirmed that denoising the acoustic signal can significantly enhance sound classification accuracy. The MFCC-CNN model excelled in overall accuracy (89%), keyhole pore prediction accuracy (93%), and AUC-ROC score (98%), outperforming the other models.

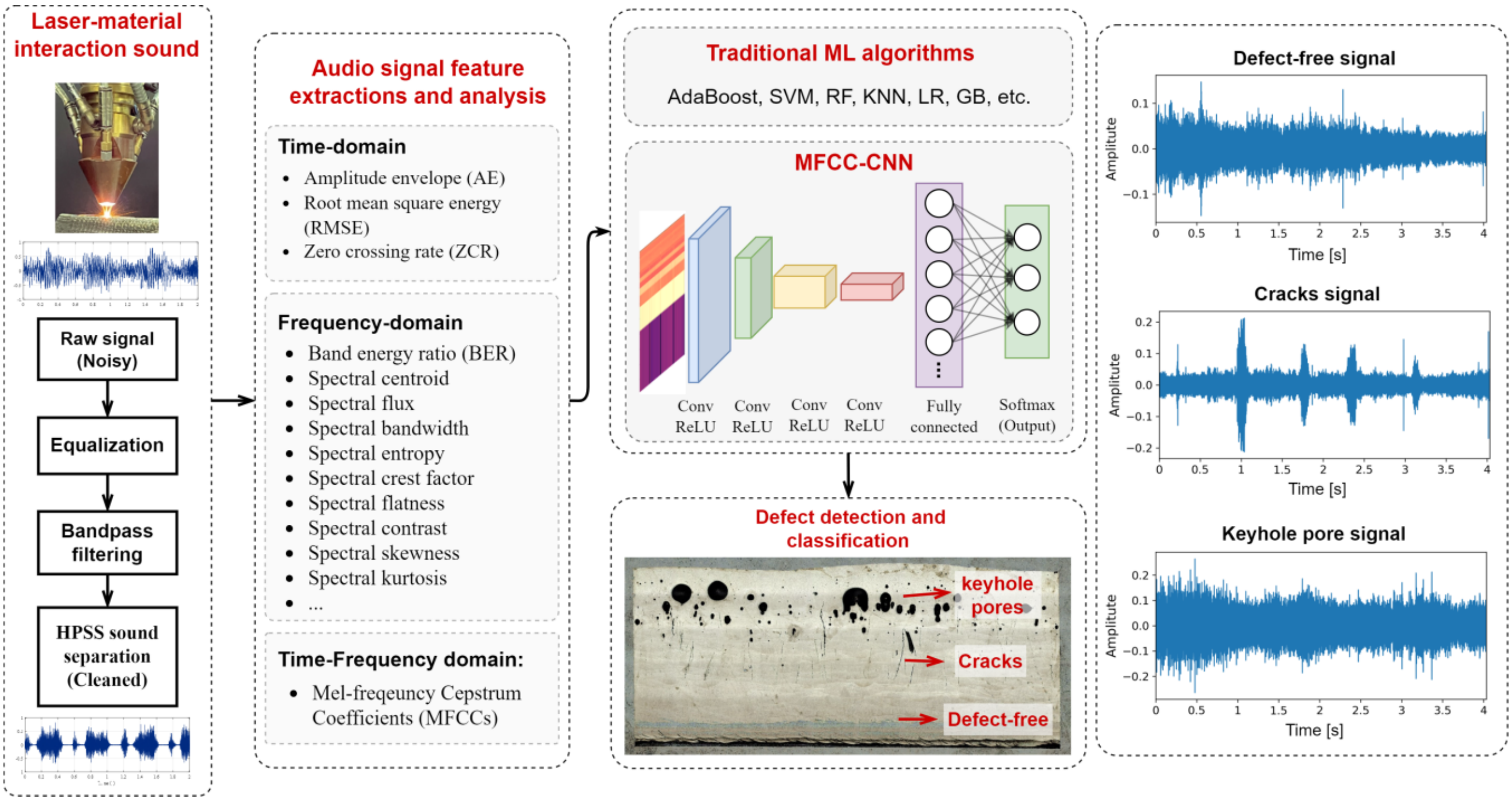

**Figure 18**. In-situ crack and keyhole pores detection framework through acoustic signal processing and deep learning [68].

### *3.2.4. Challenges, limitations, and future development*

Acoustic-based monitoring has shown significant promise in LAM for defect detection and process characterization. However, several challenges and limitations need to be addressed to make this technique more robust and adaptable for different LAM technologies. We provide an overview of the primary hurdles faced in the field, offering a basis for future research and development initiatives:

i. ***Noise Content Removal***: One of the most critical challenges in acoustic-based monitoring is the presence of extraneous noise. These noises can come from various sources such as the protective gas flow, machine movement, or even environmental factors. While several filtering techniques, including low-pass filters and advanced DL methods, have been used to mitigate this issue, there is still a need for more effective and reliable noise removal strategies that can enhance the accuracy of defect detection.

ii. ***Acoustic Monitoring for Multi-layer Multi-track Bulk Structures:*** Acoustic monitoring has been primarily used for single layer or single track analysis in LAM. However, the complexity increases significantly when it comes to multi-layer and multi-track structures due to the heat accumulation and thermal stress. The interaction between subsequent layers and tracks may result in unique acoustic signatures that are difficult to interpret. Further research is needed to understand these signals and develop effective strategies for monitoring multi-layer multi-track structures.

iii. ***Higher Temporal and Frequency Resolution***: The temporal resolution of current acoustic monitoring systems may not be sufficient to capture all relevant events during the LAM process. The laser-powder interaction happens across the frequencies, and the acoustic signal can change rapidly within microseconds. Therefore, higher temporal and frequency resolution of the ML model is needed to capture the full detail of these events, which could improve the sensitivity and accuracy of defect detection.

iv. ***Standardization and Reproducibility***: There is a lack of standardization in the setup of acoustic sensors in LAM, which can lead to inconsistencies and variations in the collected data. The positioning angle, distance to process zone, and types of microphone are critical factors influencing the quality and reliability of acoustic data. Therefore, more work is needed to establish standardized sensor setup guidelines and protocols to ensure the reproducibility and comparability of results across different studies and LAM systems.

v. ***Acoustic Monitoring for Fault Diagnosis***: Another promising but challenging area is the use of acoustic monitoring for fault diagnosis in the LAM process. Potential faults could include depletion of protective argon gas, powder depletion, nozzle clogging, oxidation in LDED, or anomalies in the powder bed in LPBF. These process disruptions could significantly affect the quality of the fabricated part and may even lead to system damage. Whiting et al. [274] presented a device for powder mass flow rate monitoring using acoustic emission signal in LDED. However, each fault may produce a unique acoustic signature, which can be drowned out by process noise or overlapped with the acoustic emissions from normal process. Therefore, the development of effective strategies for fault detection and diagnosis using acoustic monitoring remains a challenging task and requires further research.

vi. ***Integration with Other Sensing Techniques:*** Although acoustic-based monitoring provides valuable information about the LAM process, it is often insufficient on its own for comprehensive defect detection. The integration of acoustic monitoring with other sensing techniques, such as optical or thermal sensors, could provide a more complete picture of the process and improve the robustness of defect detection. However, this multi-modal sensing approach presents challenges in data fusion and interpretation, which will be discussed in subsequent sections.

### *3.3. Laser Line Scanning*

On-machine laser line scanning is a critical step for in-process surface defect identification and defects correction [275]. This can be achieved by a laser profilometer (also known as a laser displacement sensor), as shown in **Figure 19**. Laser line scanning is one of the most widely used approach for surface morphology monitoring [69,276–283]. In comparison to traditional vision-based techniques, laser displacement sensors provide higher accuracy and can produce precise height data without the need for a computationally costly 3D reconstruction process [283,284]. Laser line scanning relies fundamentally on the principle of laser triangulation [285–287], as illustrated in **Figure 19**(a). A laser line scanner projects a laser line onto an object's surface. An optical sensor, positioned at a known distance and angle, captures the laser light reflected from the surface. As the distance between the sensor and the object's surface changes due to surface irregularities, the position of the reflected laser line shifts accordingly. By calculating this shift, the system can generate a detailed topographical map of the surface, identifying potential defects.

The laser displacement sensor can be installed next to the laser head, as illustrated in **Figure 19**(a). Hand-eye calibration [288,289] is necessary to compute the coordinate transformation from workpiece coordination to the sensor's frame in order to determine the precise posture of the sensor relative to the robot. With the computed transformation matrix, the laser displacement sensor could obtain surface topography of the manufactured component represented by the 3D point cloud in the sensor coordinate system. The relationship between the local coordinates of the laser displacement sensor and the workpiece coordinate frame is shown in **Figure 19**(b).

Various point cloud data generated by on-machine laser line scanning is shown in **Figure 20**. Raw point cloud data obtained from the laser line scanning is typically noisy, containing outlier points and unwanted surfaces such as substrates. The shadow effect causes the sensor to be sensitive to the abrupt change of height, which causes inaccuracies at the edges and corners of the parts. These noise can affect the performance of the surface defect detection. The literature describes a variety of noise filtering and surface segmentation approach, with statistical outlier removal being a common technique [290]. This method calculates the distance of each point to its nearest neighbours, comparing it to the average of all such distances. Points that deviate significantly from the average are considered outliers and removed. K-Mean clustering [291] and Random sample consensus (RANSAC) algorithm [292] was used in [293] to clean the raw data and extract the point cloud of target surface. Moreover, the 3D point cloud data can be converted into a 2D depth images for further processing, which can speed up the computation and subsequent defect identification process. For example, Lyu and Manoochehri [194] used 2D depth images obtained from laser line scanning and applied a CNN to classify different surface defects, including over extrusion and under extrusion. Bulge and dent regions can be identified in a pixel-wise manner. Liu et al. [210] presented a window-based image processing approach to pair the depth information of a pixel and its corresponding local patch in 3D point cloud. Over- and under-built regions can be automatically extracted from the data.

Laser line scanning has shown substantial promise for in-process surface defect detection in LDED. However, it face several limitations and research gaps. The primary challenge lies in the resolution and reliability of surface defect detection. Although laser line scanning can detect surface defects, it may struggle to accurately identify minuscule or subtle defects, particularly those that do not significantly disrupt the projected laser line. Furthermore, scanning speeds may not keep up with the production rates in real-time manufacturing, leading to bottlenecks in the process. Ambient conditions such as dust, smoke, and varying light conditions can also affect the accuracy of the laser scanner, resulting in potential false readings or missed defects. From a research perspective, the development of more robust and advanced algorithms for data processing and defect identification from the point cloud data is a critical need. These algorithms need to be able to cope with highly noisy data and the variability inherent in LAM processes to provide reliable surface defect identification. Moreover, the integration of laser line scanning systems into existing LAM setups can be a challenging task due to alignment issues and space constraints. Current research and applications predominantly focus on the incorporation of laser line scanning into robotic LDED systems. The nature of these robotic systems allows for the easier implementation of the laser profilometer, given the additional degrees of freedom in the robotic arm to manipulate the sensor. However, similar integration is much more challenging for other types of systems, notably CNC-based LDED systems and LPBF systems. These types of systems usually don't provide readily available real-time coordinate data or TCP motion data, posing substantial hurdles for the implementation of in-process laser line scanning. The spatial limits of CNC-based LDED and LPBF systems complicate situations even more. The nature of LPBF systems, in particular, which manufacture components layer-by-layer using a high-energy laser beam to fuse metal powders, makes it difficult to identify the surface flaws by using laser line scanning. The powder bed environment could significantly disrupt the laser line, thereby affecting the measuring accuracy.

While laser line scanning has shown potential for surface defect identification in LDED, further research and technology development are required to address these challenges and fully exploit its capabilities. Future work should aim at developing more adaptable laser line scanning solutions and algorithms that can work in a broader range of LAM systems. This might involve improving the flexibility and miniaturization of the scanning systems to make them more compatible with CNC machine's spatial constraints. Research should also explore innovative methods for acquiring or estimating real-time coordinates and TCP motion data from these types of machines, potentially through the use of additional sensors or by capitalizing on advancements in machine learning methods. Another critical area of future research could be developing strategies to reduce the disruption caused by background noise, as well as more robust algorithms for surface defect identification.

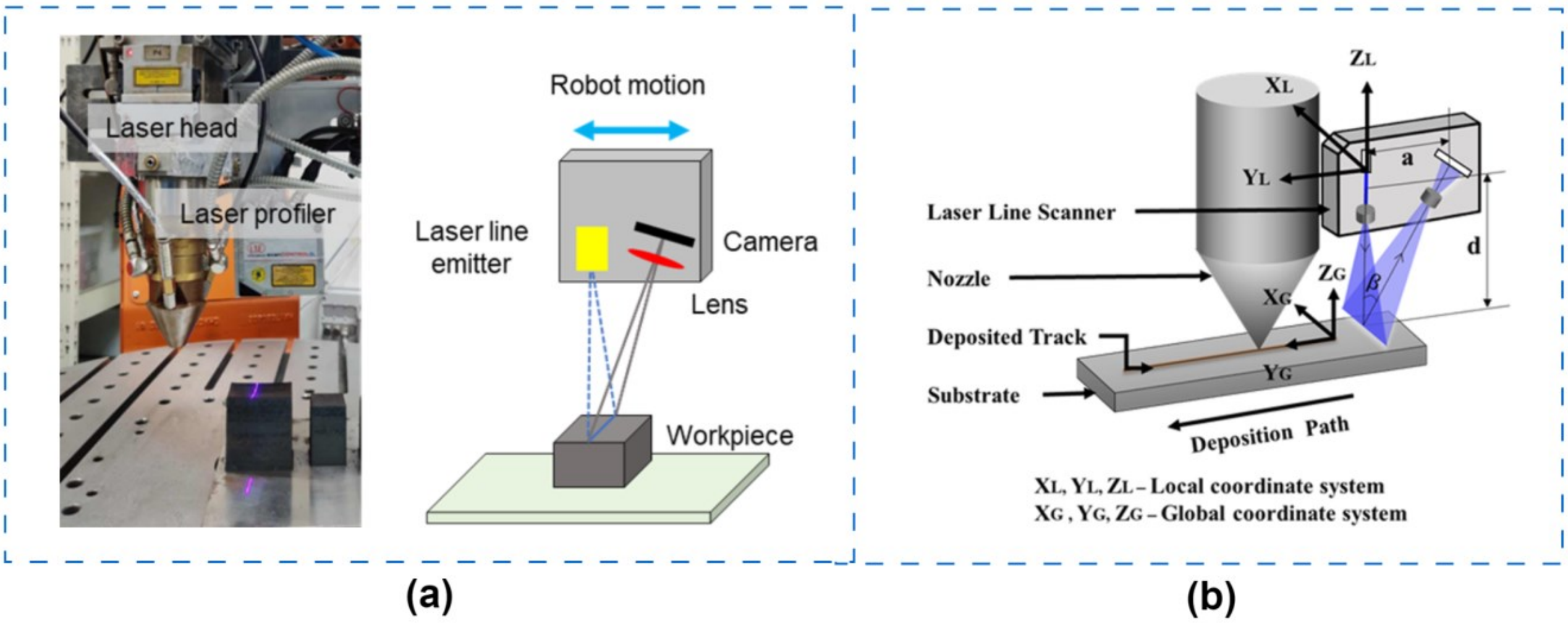

**Figure 19.** Sensor configurations for on-machine laser line scanning for surface morphology inspection: (a) on-machine laser line scanning for surface defect identification, adapted from [69], (b) laser displacement sensor

mounted alongside the DED laser head, adapted from [277].

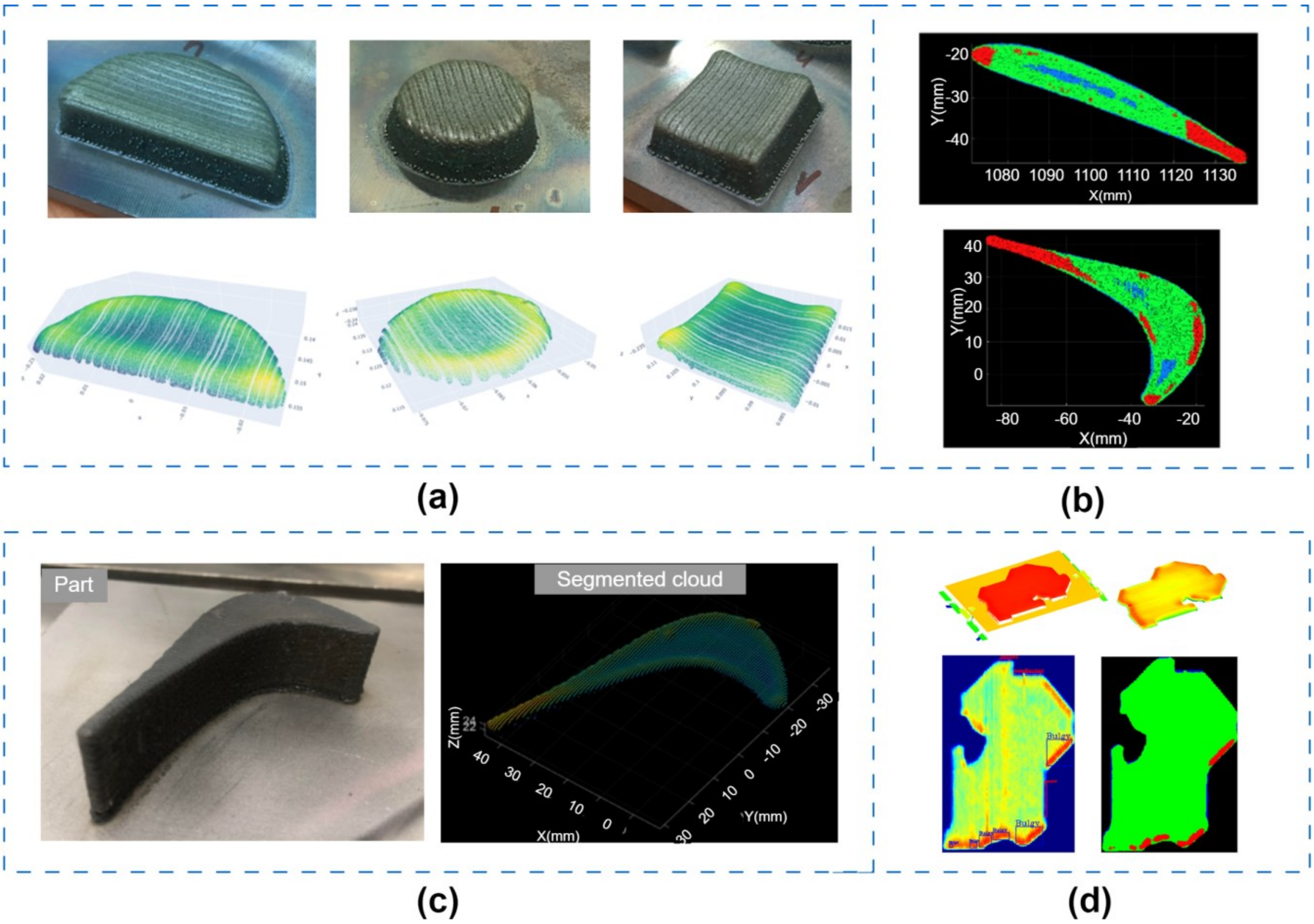

**Figure 20**. Point cloud data processing and surface defect identification: (a) 3D point cloud data of LDED parts with different geometries [69]. (b) Surface defect identification results using point cloud data and deep learning [278]. (Red: convex surface, blue: concave surface, green: normal surface). (c) As-printed LDED sample and point cloud data after noise filtering and substrate removal [278]. (d) Point cloud processing and defect detection results: the raw point cloud after substrate removal is used to identify bulge and dent area, as depicted in the pseudo-colour topography [276].

### *3.4. Other Emerging Monitoring Methods*

In addition to the aforementioned commonly adopted monitoring methods, some advanced emerging monitoring methods that utilized for in-situ defect detection in LAM are given in this section. Specifically, the focus will be on in-situ process monitoring using operando synchrotron X-ray imaging and inline coherent imaging (ICI).

Operando synchrotron X-ray imaging is a powerful non-destructive quality inspection tool that permits the real-time monitoring of materials during operation, thus providing insights into their structure, dynamics, and function [92]. The principles of operando X-ray techniques are rooted in the fundamentals of dynamic X-ray radiography (DXR) [294]. They involve the interaction of X-rays with a material, inducing scattering, absorption, or fluorescence phenomena, thereby revealing detailed information about the material's structure, composition, and changes during operation. The application of operando X-ray techniques in the field of LAM is a recent development, focused on real-time, in-situ defect detection, particularly the keyhole pore formation [295–299]. This kind of defect typically arises from the keyhole mode of laser melting, where high-intensity laser energy creates a deep, unstable vapor cavity (also known as keyhole) that can cause pore defects upon solidification. Operando X-ray provides accurate, high-resolution, and rapid detection of keyhole pores, which can be used to provide ground truth labels for annotating in-situ monitoring data [300]. For example, Ren et al. [301] utilized infrared thermal camera guided by high-speed synchrotron X-ray imaging to detect keyhole pore generation in LPBF of Ti-6Al-4V, as shown in **Figure 21**(a). The dynamics of the subsurface structure were revealed using high-energy X-rays passing through the single layer powder bed. Different forms of keyhole oscillations were discovered by combining the imaging technique with Multiphysics simulations, as shown in **Figure 21**(b). Operando X-ray imaging results were used to annotate these real-time captured data as "Pore" and "Non-pore", which served as ground truth labels to train supervised deep learning models with melt pool thermal images to detect keyhole pore generation events with sub-millisecond temporal resolution. Similar deep learning-assisted monitoring approach guided by operando X-ray were also demonstrated by Pandiyan et al. [302], where heterogenous sensors were used for defect identification and operando X-ray was used to provide observation on different process regimes: keyhole pores, conduction mode, LoF pores.

Operando X-ray imaging is similarly employed by Wolff et al. [303] in their piezo-driven LDED system, as shown in **Figure 21**(c). The authors captured the interaction of a laser beam and powder-blown deposition using in-situ high-speed X-ray imaging, unveiling how laser-matter interaction influences powder flow and porosity formation. The X-ray imaging setup allowed for the monitoring of individual powder particles flowing into a scanning melt pool. Similar to the LPBF process, it is evident that operando X-ray imaging can provide crucial insights for in-situ defect detection, particularly for tracking the powder flow and porosity formation.

Recently, Inline Coherent Imaging (ICI), an advanced sensing technique, has been proven effective for in-situ defect detection in LDED. Fleming et al. [304] demonstrated the applicability of ICI for in-situ monitoring of surface topology and cracks, as illustrated in **Figure 21**(d). For the first time, ICI was aligned off-axis (24° relative to laser), allowing for its integration into a LDED machine without alterations to the laser delivery optics. It showcased the ability to detect cracking events, including a sub-surface signal. A high correlation (>0.93) was reported between ICI surface topology and corresponding X-ray radiographs. Such in-situ correlative observation between ICI and synchrotron X-ray imaging has been instrumental in understanding the crack formation mechanism in LDED. The process monitored thin-wall builds of nickel super-alloy and revealed humping-induced cracking in surface valleys. Similar in-situ monitoring and process control method using optical coherent imaging (OCT) were also demonstrated in [305].

Operando X-ray imaging is a remarkable tool for in-situ defect detection in LAM. However, it faces a few key challenges. First, the availability and accessibility of X-ray imaging equipment can be a significant hurdle. The X-ray sources used for operando imaging, such as synchrotron sources, are not readily available in many facilities. This makes regular use or industrial scaling of this approach a formidable challenge. Furthermore, X-ray radiation poses safety risks, necessitating stringent precautions, protective gear, and specialized modifications to the environment where it is deployed. This can compound the complexity and expense of the process. Another substantial challenge comes in the form of data processing and interpretation. High-speed X-ray imaging yields large volumes of data that are computationally intensive and time-consuming to analyse and interpret. Most previous research is limited to single-track or thin-wall investigations, and it is difficult to extend operando X-ray to LAM of multi-layer multi-track complex structures.

Inline Coherent Imaging (ICI), on the other hand, also comes with various limitations. Integrating ICI systems into existing LAM setups can be difficult due to the systems' relative complexity. This is especially the case when off-axis alignment is needed. Despite ICI's impressive spatial and temporal resolution, it can still struggle to detect small defects or those situated deep within a material. Additionally, being an optical method, ICI is vulnerable to environmental conditions such as dust, smoke, or temperature fluctuations. The harsh operational conditions of LAM could negatively impact the system's accuracy and efficiency.

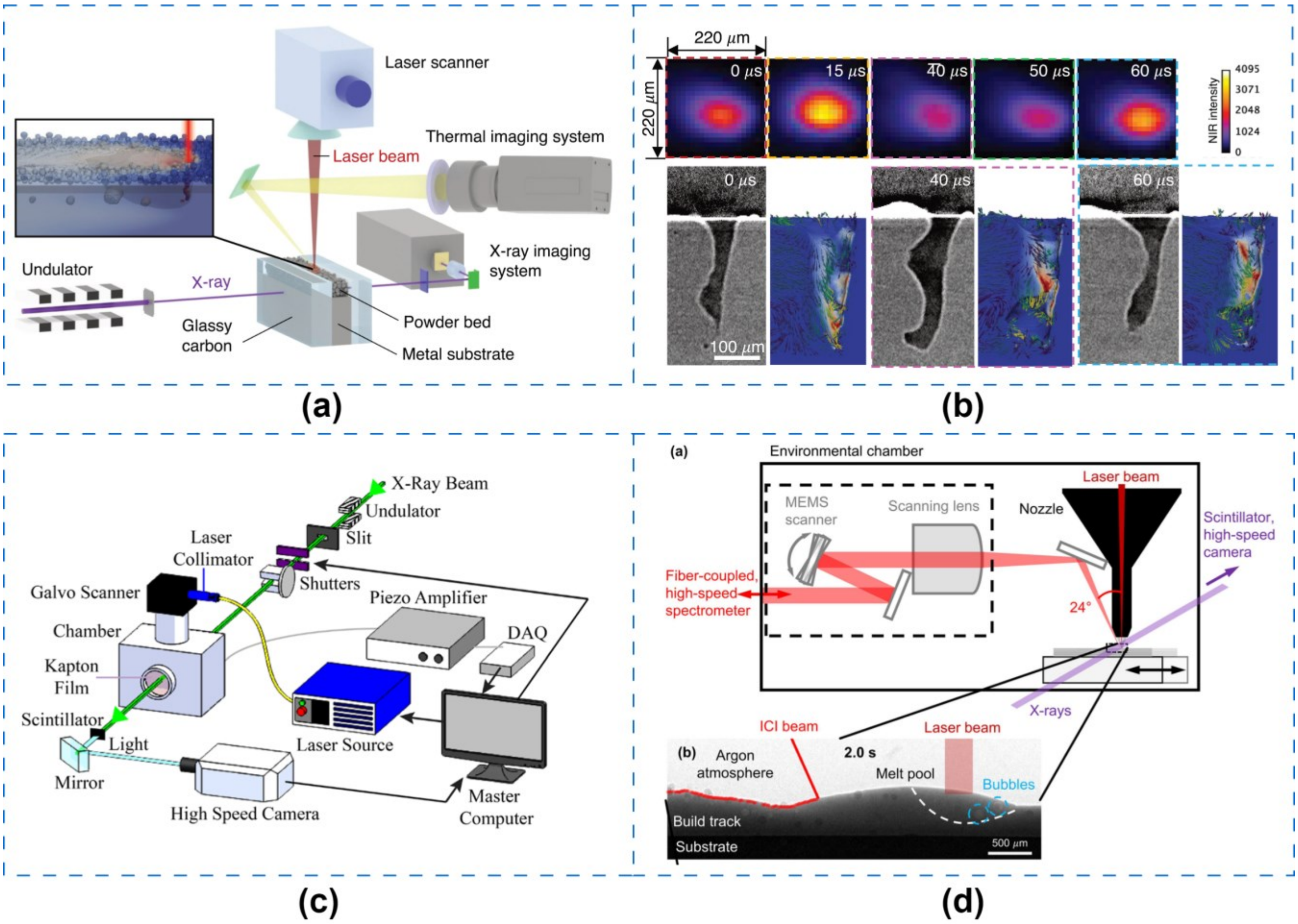

**Figure 21**. Other sensing techniques for in-situ defect detection in LAM. (a) Simultaneous synchrotron X-ray and thermal imaging for real-time keyhole pore detection in LPBF [301]. (b) Thermal images of intrinsic keyhole oscillation and high-speed X-ray images with corresponding Multiphysics simulation of intrinsic keyhole oscillation [301]. (c) In-situ high-speed X-ray imaging setup for a piezo-driven LDED system [303]. (d) Simultaneous operando X-ray and inline coherent imaging (ICI) for crack identification during the LDED process [304].

### *3.5. Multisensor Monitoring And Data Fusion*

Multisensor monitoring is preferable to traditional single-sensor monitoring because it leverages the unique strengths of different sensors to provide a more comprehensive and accurate understanding of the complex, multi-dimensional LAM process and defect formation mechanisms. Single sensors often excel in detecting specific types of defects (e.g., porosity, cracking, distortions, etc.) but struggle with others, making them less ideal for capturing the full picture of the process dynamics. By contrast, multisensor monitoring aggregates diverse sensor information, thereby counterbalancing individual sensor limitations, optimizing the signal-to-noise ratio, and enabling data fusion for enhanced predictability of process outcomes [306]. This section provides in-depth discussions on the advantages, techniques, and challenges of multisensor monitoring and data fusion in LAM.

### *3.5.1. Multisensor Monitoring Setups In LAM*

The latest advancements in sensor-based in-situ monitoring have led to the development of diverse multisensor monitoring setups aimed at detecting in-situ defects at multiple scales. As illustrated in **Figure 22**, these setups often incorporate multiple sensors that track different facets of the LAM process, thereby enhancing the reliability and accuracy in defect detection. A noteworthy example is the multisensor system employed in the robotic LDED shown in **Figure 22**(a) [70]. This system incorporates a coaxial visible spectrum melt pool camera, an off-axis infrared thermal camera, and a microphone sensor. Each of these sensors captures unique process signature, substantiating the superior performance of multisensor fusion-based defect detection over single sensor systems. In fact, coaxial melt pool cameras are often paired with off-axis infrared sensors to monitor both the melt pool dynamics and part thermal histories in many LDED systems [184,307]. Similarly, LPBF in-situ defect detection have been augmented with a multi-camera monitoring systems that includes a near-infrared thermal camera, a powder bed imaging camera, and a high-speed melt pool spatter dynamics imaging camera, as depicted in **Figure 22**(b) [308]. This heterogenous sensor data fusion enables detection of flaws in LPBF ranging from porosity at the micro-scale ($< 100$ µm), to layer-related inconsistencies at the meso-scale (100 µm to 1 mm) and geometry-related flaws at the macroscale ($> 1$ mm). There are also configurations that include four or more sensor types. For instance, **Figure 22**(c) [309] depicts a LPBF system that includes a microphone sensor, a multi-spectral emission sensor, a high-speed camera, and a laser scan trajectory recorder. The authors revealed that, while optical imaging provides substantial information for defect detection, adding additional sensing modalities significantly improves defect detection performance. **Figure 22**(d) depicts another LPBF setup that are equipped with back reflection (BR), visible, infra-red (IR), and structure borne AE sensor [302]. Interestingly, the study found that BR and AE sensors provide more important information to guide the decision-making process than IR and visible sensors.

Depending on the complexity of the LAM process being monitored and the type of defects encountered, each of these configurations offers distinct advantages. More comprehensive LAM process monitoring and multi-scale defect detection is feasible by using heterogenous sensor data, which enables in more effective defect identification and quality control.

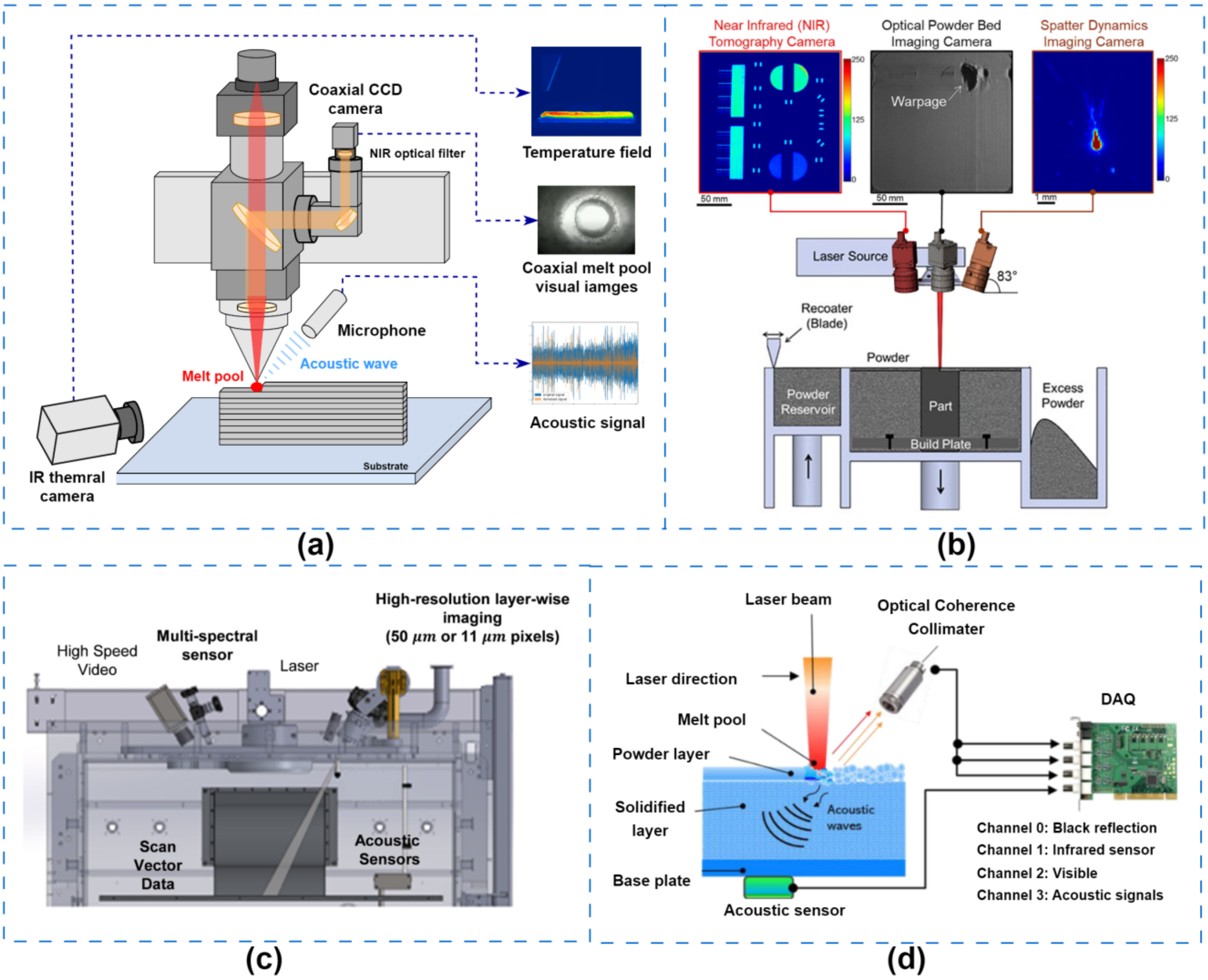

**Figure 22**. Various multisensor monitoring setups for in-situ defect detection in LAM process. (a) Multisensor setup in robotic LDED, featuring a coaxial melt pool camera, an off-axis infrared thermal camera, and a microphone sensor [70]. (b) Multi-camera monitoring system in LPBF, featuring three types of sensors installed on an optical table on top of the machine near the laser source: a near-infrared thermal imaging camera, a powder bed imaging camera and a high-speed melt pool spatter dynamics imaging camera [308]. (c) Multisensor monitoring system in LPBF, consisting of a microphone sensor, a multi-spectral emission sensor, a high speed camera, and laser scan trajectory recorder [309]. (d) Mini-LPBF setup equipped with four sensors, namely back reflection (BR), visible, IR, and structure borne AE sensor [302].

### *3.5.2. Multisensor feature visualization and correlation analysis*

In this section, we present the visualization and correlation analysis of multisensor features. Understanding these correlations and visual representations is paramount for making accurate predictions about potential defects and optimizing sensor utilization for maximum information capture. Gaikwad et al. [310] analysed the relationships between melt pool temperature features and geometric features retrieved from a thermal camera and a high-speed video camera. Based on both sensor features, three separate laser focus heights representing different process conditions could be recognized. This demonstrates that the coaxial temperature and geometric features are correlated. Furthermore, Liu et al. [311] developed an novel data-driven pipeline that correlates acoustic and thermal signals to infer and track dynamic melt pool visual characteristics, enhancing defect detection and quality control in LPBF processes.

**Figure 23** provides a detailed visualization and comparison of the multisensor features, highlighting unique characteristics from different sensing modalities. **Figure 23**(a) presents a spatial visualization of multisensor features, including coaxial melt pool visual geometric features, acoustic features, and temperature field features [70]. These features are juxtaposed against the physical quality of the LDED part as revealed through optical microscopy (OM). Anomalies in the LDED process, such as cracks and keyhole pores, can be linked to sudden shifts in multisensor feature values. Furthermore, features from different sensing modalities exhibit comparable trends, increasing or decreasing over time and layer height, in line with localized heat accumulation and quality deterioration. **Figure 23**(b) and (c) delineate the median saliency distribution per sensor for window lengths of 3.3 ms and 2.5 ms, respectively. This distribution highlights the relative importance of each sensor, where a rightward shift in the derivative distribution indicates higher sensor importance [302]. It was found that the BR and AE sensors carried the most informative content. Notably, with shorter window lengths, AE's importance over BR grows, indicating AE's ability to capture more time-resolved events that contribute to the decision-making process. However, the BR sensor, which requires a longer integration period, tends to provide more reliable results once granted a larger time window. This leads to higher accuracy, highlighting the correlation between BR and the stability of the LAM process.

In summary, multisensor feature visualization, correlation analysis, and selection improve sensor data utilization, ultimately leading to improved process monitoring and control. The information acquired through the correlation of these multisensor features lays the foundation for developing a digital twin of the process, which will be discussed further in the following section.

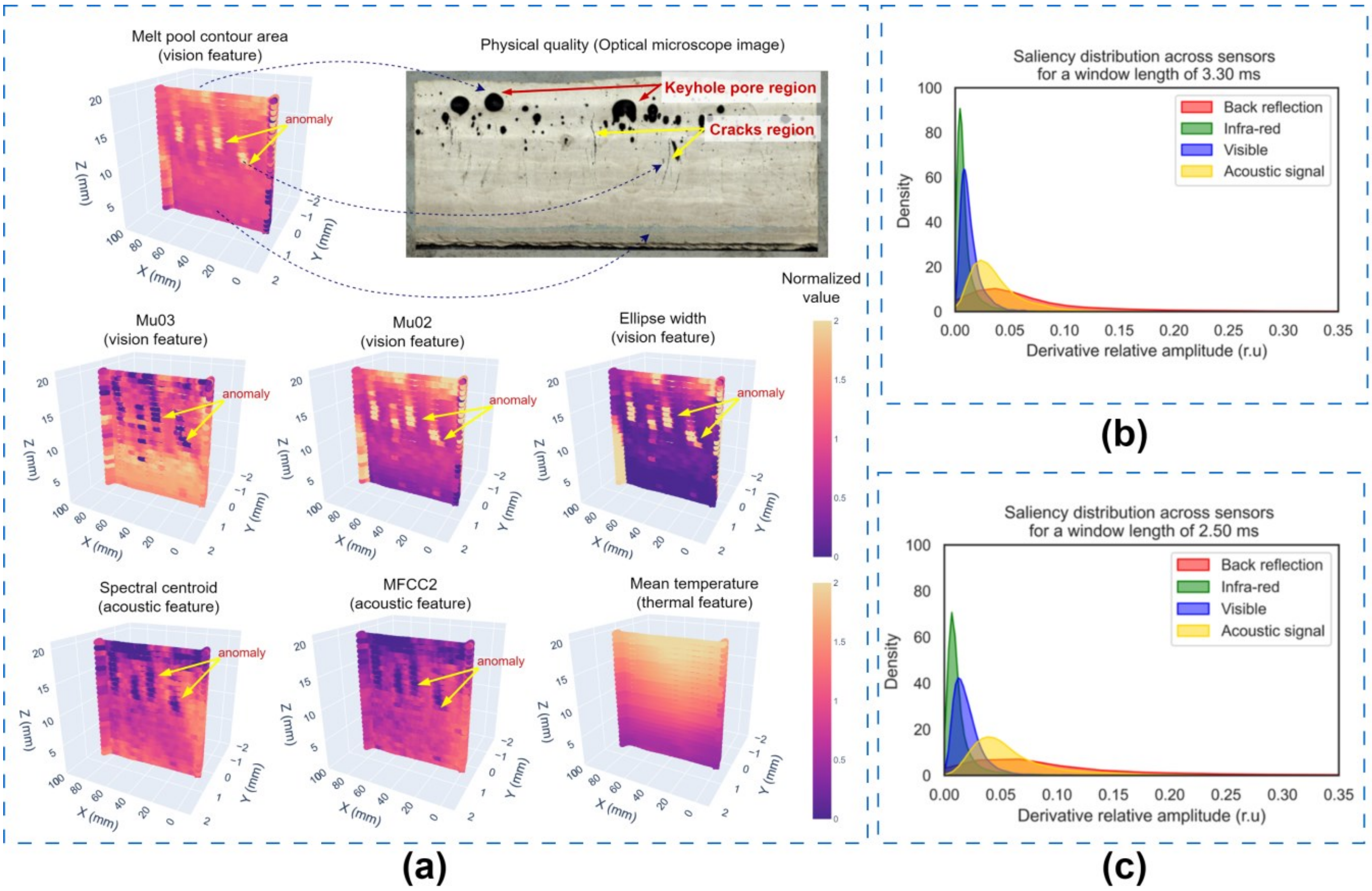

**Figure 23**. Visualization and correlation analysis of multisensor features. (a) Spatial multisensor feature visualization compared to the physical quality in an OM photo [70]. (b)-(c) Median saliency distribution per sensor for 3.3 ms and 2.5 ms window length, indicating sensor importance. The BR and AE sensors provide the most information [302].

### *3.5.3. Spatiotemporal data registration*

The synchronization and time-alignment of multisensor data is required for effective spatiotemporal multimodal data registration, which is a critical step in the development of an integrated dataset. This procedure establishes the foundation for detecting location-specific defects. However, this is particularly difficult due to two obstacles: (1) Since many commercial AM machines do not provide real-time TCP location information, spatial defect prediction is challenging; (2) different sensors may have different acquisition speeds, and any misalignment in multisensor data may jeopardize the reliability and robustness of multisensor-based defect detection and process control mechanisms. Despite its importance, much existing research falls short of effectively addressing these issues.

**Figure 24** elucidates two examples of spatiotemporal registration of multimodal data. Vandone et al. [312] proposed a data fusion strategy that combines multiple types of data collected both during and post-deposition process (**Figure 24**(a)). This dataset includes both online monitoring data, such as image features, statistics from visible light cameras, thermal measurements, and machine tracing data (e.g., laser power and laser spot positions), as well as offline inspection data, including 3D surface and volume reconstruction. Through the computation of the cross-correlation between the image-retrieved beam-on signal and the laser power signal, this comprehensive approach enables the time-alignment of melt pool images. The registered data, whether time- or space-referenced, is compiled in both temporal and spatial domains to form an integrated dataset. Similarly, Feng et al. [313] presented a method to register image-based in-situ monitoring data and ex-situ XCT 3D scan model. One notable limitation of this method is that spatial information is acquired offline, which makes real-time, location-dependent defect prediction challenging.

To address this limitation, Chen et al. [70] proposed a method for extracting features from multisensor inputs and registering them spatiotemporally with real-time robot TCP positions to facilitate location-specific quality mapping (**Figure 24**(b)). This system allows for the simultaneous synchronization and registration of multisensor features, ML prediction outputs, and robot TCP data, enabling localized quality prediction. The author used the ROS message filter module's Approximate Time Synchronizer algorithm [314] to align messages from different sensing modalities depending on their timestamps. This synchronization and registration process can estimate defective region boundaries despite minor variations due to computer program execution times. Furthermore, Kim et al [315] proposed a deep learning-based method that directly estimates melt pool coordinates in the machine coordinate system from melt-pool images. This methodology allows melt pool data registration using only machine-supplied process settings and melt pool images, with no need for system-to-system synchronization, providing a valuable alternative on the spatial registration of in-situ monitoring data, particularly in the context of an AM platform with limited access to real-time TCP position information.

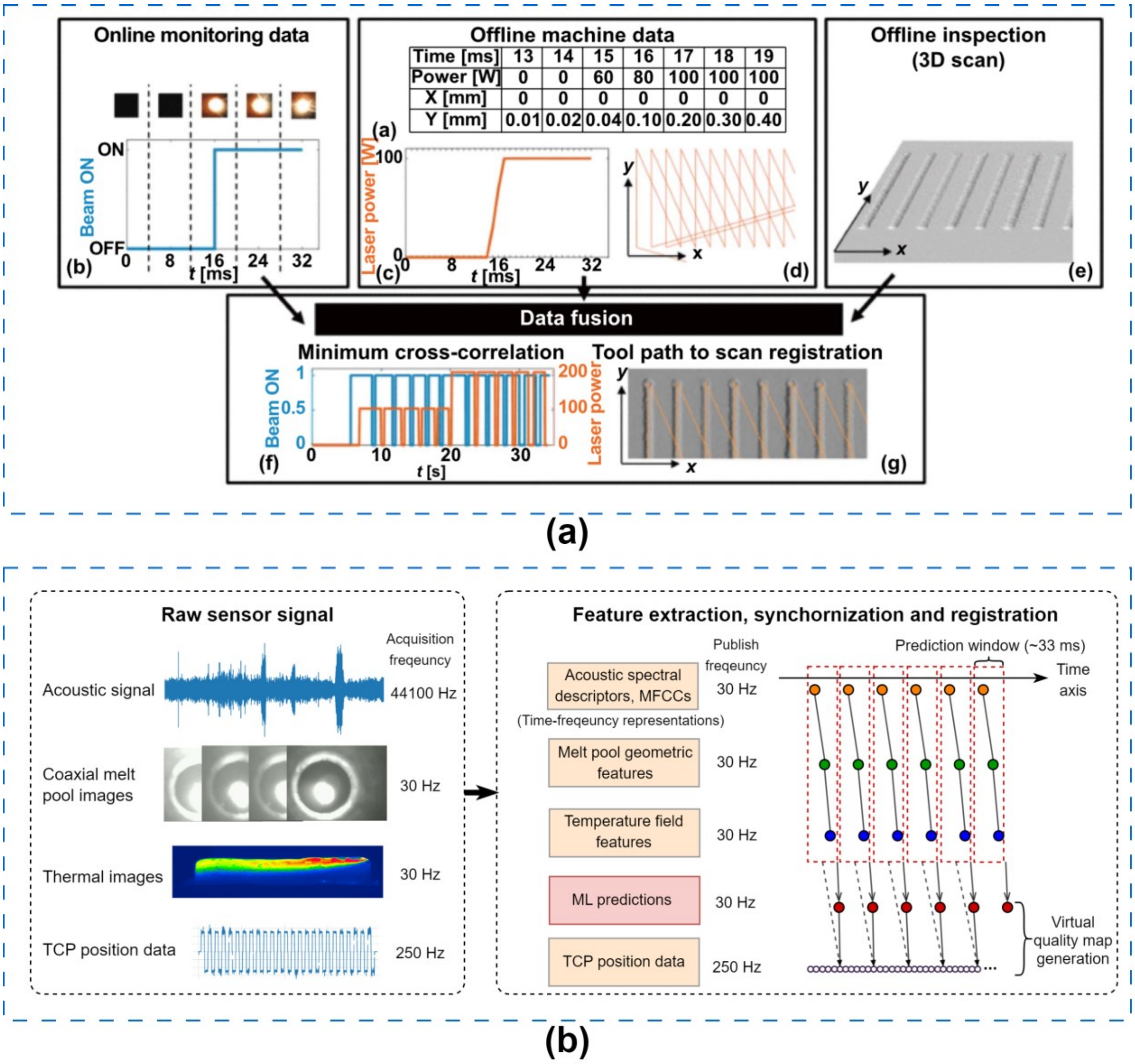

**Figure 24**. Multimodal data spatiotemporal registration. (a) Online and offline multimodal data registration: The machine log file records laser power and tool position at each timestamp, enabling the time-alignment of the melt pool images and the spatial registration of the tool trace to the 3D deposited geometry [312]. (b) Synchronization and registration of multisensor features, machine learning predictions, and real-time robot TCP positions: This process ensures that all features from different sensing modalities are extracted and collected at the same timestamp, enabling localized quality prediction [70].

### *3.5.4. Multimodal sensor fusion and quality prediction*

The multimodal sensor fusion for defect detection in LAM is a challenging task. Although there has been little research into this in the LAM field, it has been widely utilized in many other industrial applications, such as machine fault diagnosis [316] and autonomous vehicles [317]. Generally, multimodal sensor fusion can be classified into three levels:

i. ***Data-Level Fusion***: This fusion level directly combines raw data from multiple sensors, such as raw melt pool images and acoustic signal, into matrices or other representations. It offers high precision with minimal information loss. However, it lacks the ability to reject noise and disturbances and struggles with asynchronous and signal mismatching issues caused by different sensors, potentially leading to performance issues. It can also be computationally intensive due to the large volume of data.

ii. ***Feature-Level Fusion***: This level of fusion reduces the dimensionality of massive amounts of data, making it more suitable for real-time defect identification. This could involve, for example, combining the melt pool geometries with the acoustic signal spectrum features. Feature-level fusion can provide a more compact and meaningful representation of the data than data-level fusion. It is often used in conjunction with ML models to capture more abstract characteristics and achieve better defect detection outcomes. However, due to the varied units and dimensions of different types of sensors, it necessitates sophisticated data processing techniques.

iii. ***Decision-Level Fusion***: This is the highest level of fusion, where the outputs or "decisions" of multiple sensors or algorithms are combined to produce a final decision. For example, if one sensor determines that state is "defective" and another sensor determines that it is "defect-free", decision-level fusion might involve a voting scheme to resolve the discrepancy. Decision-level fusion can be less computationally intensive and provide precise fusion results, offering a strong robustness.

Several recent studies have demonstrated the benefits of multimodal sensor fusion in LAM defect detection, as summarized in **Table A3**. For example, Li et al. [318] developed the feature-level fusion technique, in which in-situ signal characteristics from a photodiode and a microphone were fused using a CNN-based model to estimate layer-wise LPBF quality. In addition, the author compared feature-level fusion with data-level and decision-level fusion [319]. The data-level fusion model is created through channel fusion. The feature-level fusion model is created by extracting and fusing signal features, and the decision-level fusion model is created by fusing the classification results from individual models. The results showed that the feature-level data fusion model outperformed the other two fusion models in terms of classification accuracy. **Figure 25** depicts an example of multimodal sensor fusion based on deep learning for in-situ quality monitoring in LAM. A hybrid CNN method was proposed which incorporates visual and audio feature extraction streams to achieve feature-level fusion for localized quality prediction [187]. A virtual quality map reflecting the physical quality values observed in the OM image can be constructed by registering the quality prediction outputs with the location information [70]. Recently, Perani et al. [320] proposed an online track geometry prediction utilizing a visual CNN model (VGG-based) with real-time power, velocity, and one-hot-encoded states (accelerating, stable, decelerating) data [320]. This methodology allowed the prediction of track geometry during transient conditions, which has the ability to reduce the dimension deviations of the deposited parts.

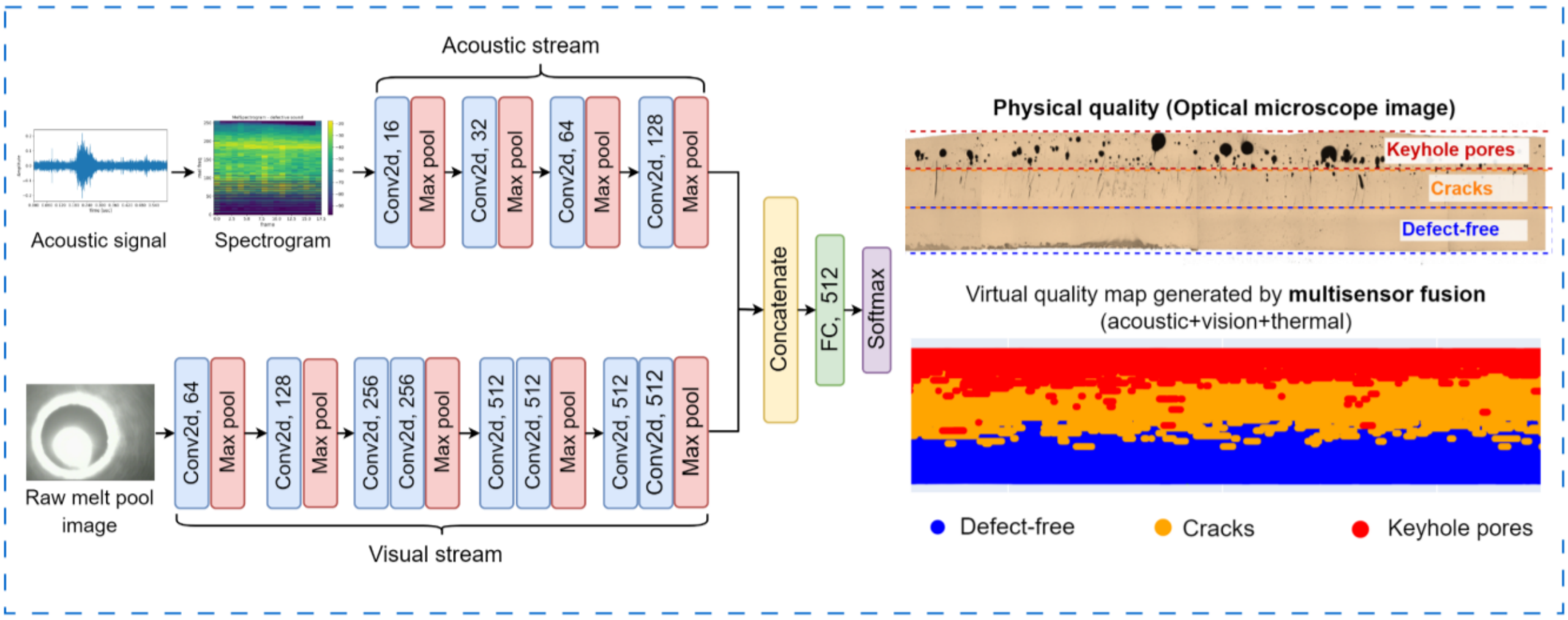

**Figure 25**. Examples of deep learning-based multimodal data fusion for in-situ quality monitoring in LAM. Hybrid CNN model concatenates the coaxial visual feature extraction stream and acoustic feature extraction stream to achieve feature-level fusion for localized quality prediction [187,70,160].

## 4. Adaptive Quality Enhancement

This section investigates the crucial aspect of adaptive quality enhancement in LAM, specifically focusing on addressing the limitations posed by open-loop process settings (i.e., using fixed process parameters) commonly found in most LAM systems. These settings often lead to defects due to factors like dynamic heat accumulation and variability in printing feature sizes. For instance, changes in the length of the raster line, a typical toolpath scanning strategy, can result in inconsistent heat distribution and over-melting in smaller features [321]. In addition, inaccuracies in robotic and laser positions, along with fluctuations in acceleration and deceleration, can deviate from set printing speeds [39,322]. Such variabilities are key contributors to quality issues such as keyhole pores, geometric deviations, and microstructure inhomogeneity. To tackle these challenges, the section is divided into two key areas: (1) closed-loop feedback control (**Section 4.1**), focusing on pre-empting defect generation for enhanced geometric accuracy and microstructure uniformity, and (2) Iin-process defect correction (**Section 4.2**) detailing methods to rectify defects using additive or subtractive approaches, thereby enhancing the overall quality of LAM products.

### *4.1. Closed-Loop Feedback Control*

The complexities of the LAM process necessitate a real-time closed-loop control system that can adjust process parameters to compensate for any deviations or errors occurring during the process. This system leverages raw sensor signals (e.g., melt pool temperature) or features derived from in-situ monitoring data (e.g., melt pool width) to adjust process parameters such as laser power, scanning speed, and powder feeding rate. As a result, the closed-loop control system can significantly enhance the quality of the manufactured part, improving its geometric accuracy, surface finish, and microstructure homogeneity.

**Table 5** presents a literature survey of the closed-loop control systems in LAM, including controlled variables, feedback signals, and algorithms employed in the control systems. It also provides insights into their strengths and weaknesses. **Figure 26** illustrates some examples of closed-loop control in LAM. In **Figure 26**(a), a classic closed-loop control structure is demonstrated, wherein the melt pool peak temperature serves as a feedback signal, and the laser power as the controlled variable. The aim of the controller is to maintain a consistent melt pool temperature, thereby reducing heat accumulation and suppressing the occurrence of porosity defects.

However, many of the controllers present in the literature require experimental system identifications to obtain plant models or layer-dependent adaptive control rules, a process which is both laborious and time-consuming. To address this, Chen et al. [163] introduced an adaptive PID controller with automatic parameter tuning unit, as depicted in **Figure 26**(b). This advanced controller seamlessly adapts to LDED processes involving varying part shapes, materials, toolpaths, and process parameters, thereby eliminating the need for manual parameter adjustment. The superior geometric accuracy achieved through this controller is demonstrated in **Figure 26**(c). Further examples on geometric accuracy enhancement by closed-loop control are shown in **Figure 26**(d)-(e). For instance, unstable ripple dynamics can be effectively removed through layer-by-layer height control [323]. Similarly, the geometric precision of spiral part can be enhanced with a feedforward PI controller [324]. Moreover, geometric stability can be significantly improved for variable-height parts with large overhanging angle [71]. These case studies highlight the potential for significant advancements in LAM quality through the application of closed-loop control systems.

Despite significant advancements in the field, several challenges remain in closed-loop control of LAM: (1) Existing methods are primarily validated within the confines of single-track or thin-wall studies, limiting their efficacy in practical LAM production where parts like rocket engine nozzles or propellers have complex and bulky geometries. (2) Most control methods employ approximations of a linear transfer function of the plant model, ignoring the inherently non-linear nature of the LAM process. (3) The existing literature predominantly explores the impact of closed-loop control on accuracy for specific materials and geometries, while neglecting the intricacies of part geometry and a comprehensive analysis of microstructural characteristics and mechanical properties.

Several recent advancements in closed-loop control are attempting to address these challenges [325]. For instance, Wang et al. [321] proposed a real-time laser power control in LPBF, utilizing a customized self-control L-PBF platform to tackle challenges such as long signal processing time, high device communication lag, and lack of signal-to-quality correlations. Meanwhile, Liao et al., [326] proposed a simulation-guided process design method for melt pool depth control in LDED. This technique enables the offline determination of the time-series laser power profile to achieve a desired melt pool depth, leading to improved surface finish and geometric accuracy. However, this method is not immune to inevitable systematic errors originating from the simulation. Another noteworthy contribution is from Ogoke and Farimani [327], who proposed a method using deep reinforcement learning for thermal control. This novel approach can process a vast amount of data and provide valuable physical insights. However, this method simplifies the heat source to consider only the effects of heat conduction on the resultant temperature field, neglecting the effects of convection and radiation heat transfer.

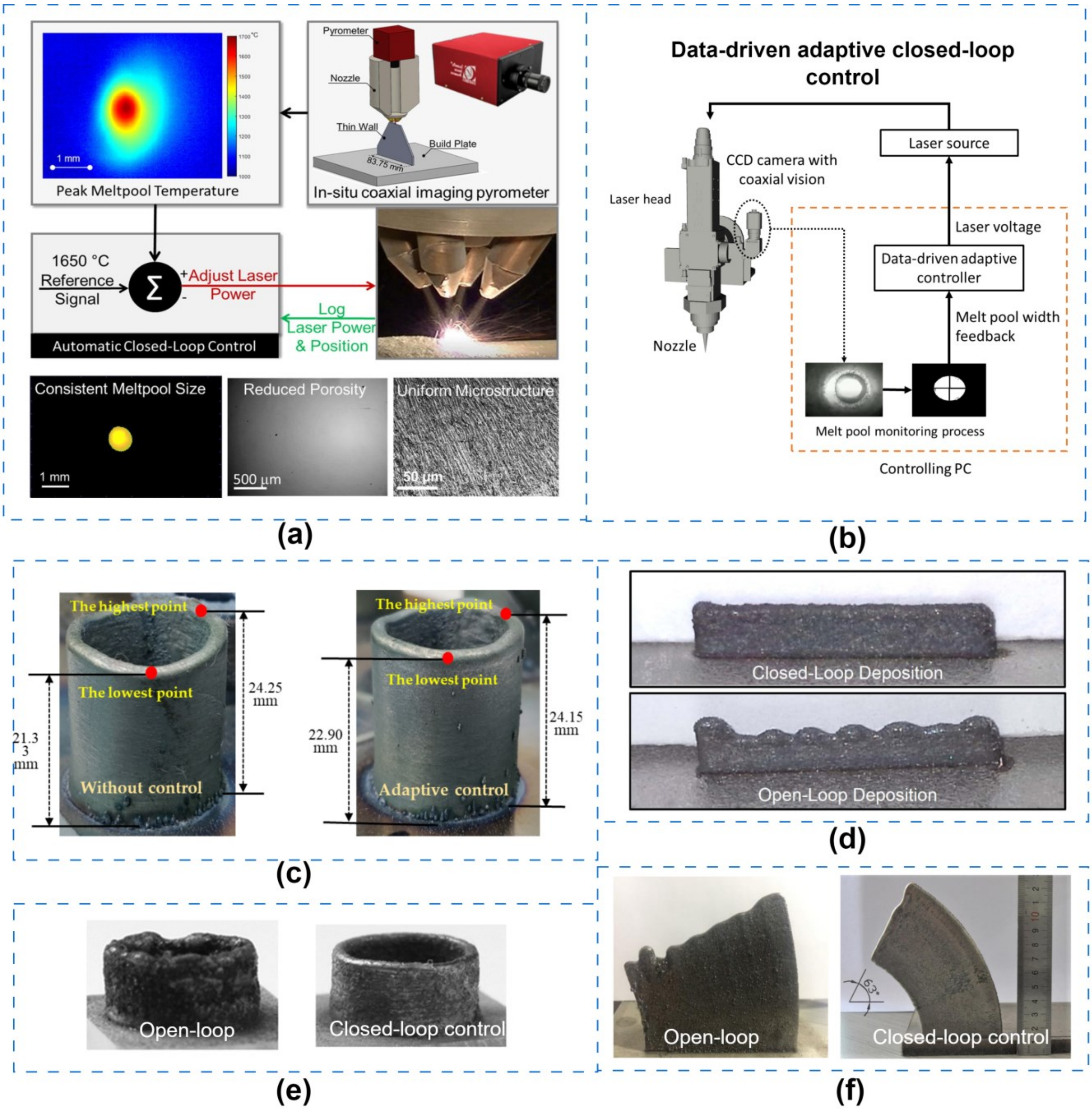

**Figure 26.** Examples of closed-loop control in LAM for improving geometric accuracy and microstructure homogeneity. (a) Closed-loop control of melt pool temperature in LDED to mitigate porosity defects and promote uniform microstructure [63]. (b) Data-driven adaptive closed-loop control strategy to adjust laser voltage signal based on melt pool size feedback [163]. (c) Adaptive closed-loop controlled samples shows improvement on geometric accuracy [163]. (d) Repetitive control of LDED process to track the reference height and remove the unstable ripple dynamic that naturally occurred in the open-loop deposition [323]. (e) Real-time control of melt pool geometry to using a feedforward PI controller in LDED [324]. (f) Precise control of variable-height DED parts to improve geometrical stability [71].

**Table 5.** A survey on closed-loop control in LAM, extended from [63].

| Ref | LAM process | Sensor types | Feedback signal | Controlled variable | Control algorithm | Outcomes | Limitations | Year |
|---|---|---|---|---|---|---|---|---|
| Wang et al., [321] | LPBF | A high-speed thermal sensor | Melt pool thermal emissions | Laser power | Custom PID | Prevents over-melt, balling, surface roughness | Extensive PID tuning needed | 2023 |
| Freeman et al., [328] | LDED | Coaxial camera | Melt pool width | Laser power | Rule-based | Stable melt pool size, better mechanical uniformity | Low frame rate/response speed | 2023 |
| Su et al., [329] | LDED | Coaxial IR camera | Melt pool width | Laser power | PID | Enhances tensile strength by ~59% | Specific to Fe-Ni-Cr alloy | 2022 |
| Smoqi et al., [63] | LDED | Coaxial two-wavelength imaging pyrometer | Peak melt pool temperature | Laser power | Rule-based | Reduced porosity, uniform microstructure | Thin-wall study | 2022 |
| Chen et al., [163] | LDED | Coaxial melt pool camera | Melt pool width | Laser power | Adaptive PID | Improved geometric accuracy | Microstructure not investigated | 2020 |
| Gibson et al., [330] | LDED | Coaxial IR melt pool camera | Melt pool size | Laser power, speed, deposition rate | Rule-based multi-modality control | Consistent track geometry | Single track study | 2020 |
| Liu et al., [331] | LDED | Coaxial NIR monochrome camera | Melt pool size | Laser power | Model predictive control | Consistent height | Thin-wall study | 2019 |
| Akbari and Kovacevic, [73] | LDED | Coaxial melt pool camera | Melt pool width | Laser power | Adaptive PI | Uniform and finer microstructure | Thin-wall study | 2019 |
| Yeung et al., [332] | LPBF | Coaxial high-speed camera | Melt pool intensity | Laser power | Rule-based | Improved surface roughness | Internal quality not addressed | 2019 |
| Shi et al., [333] | LDED | Off-axis CCD camera | Deposition height | Laser power, scanning speed | PI | Consistent height, uniform microstructure | Thin-wall study | 2018 |
| Hofman et al., [334] | LDED | Coaxial CMOS camera | Melt pool width | Laser power | PI | Constant dilution, hardness in clad layer | Single track study | 2012 |
| Song et al., [335,336] | LDED | Off-axis pyrometer | Melt pool temperature | Laser power | Generalized predictive controller | Improved geometric accuracy | Single point temperature measurement | 2012 |
| Tang and Landers, [337,91] | LDED | Coaxial single-wavelength pyrometer | Melt pool temperature | Laser power | Rule-based | Stable track morphology | Multi-layer, single track study | 2010 |

Closed-loop control has a significant impact on the microstructure of LAM-fabricated parts. **Figure 27** presents comparisons of the microstructures of samples fabricated with and without closed-loop laser power control. In **Figure 27**(a), the microstructure of an SS 316L sample produced with closed-loop control exhibits more homogenous microstructure than the one fabricated without control in the LDED process [73]. Notably, the microstructure of the sample with open-loop processing starts with a fine cellular structure and transitions gradually to a combination of columnar and cellular structures in the middle. The uppermost layers are dominated by coarse columnar grains and even dendrites with secondary arm spacing despite the randomness of the dendritic grain growth in certain areas, a directional solidification pattern is evident in particular regions. In contrast, the sample produced with closed-loop controlled laser power has a consistent cellular grain structure throughout all layers, with varying cell sizes at various cross-section positions. This is due to the controller's ability to supress heat accumulation, which produces a relatively finer, more homogeneous, and uniform microstructure. Therefore, the closed-loop controlled sample would have experienced less variation in temperature gradient and solidification rate during solidification. Meanwhile, closed-loop control can also affect the grain morphology and grain size. **Figure 27**(b) provides OM and SEM observations of the cross-section profile and microstructure of molten pools with and without closed-loop control [329]. In the open-loop processing, the middle grains form dendrites aligned with the heat dissipation direction due to rapid heat accumulation. With the closed-loop controller implemented, the cooling and solidification rates of the melt pool are accelerated by reducing laser power input, resulting in a microstructure dominated by uniform fine equiaxed grains.

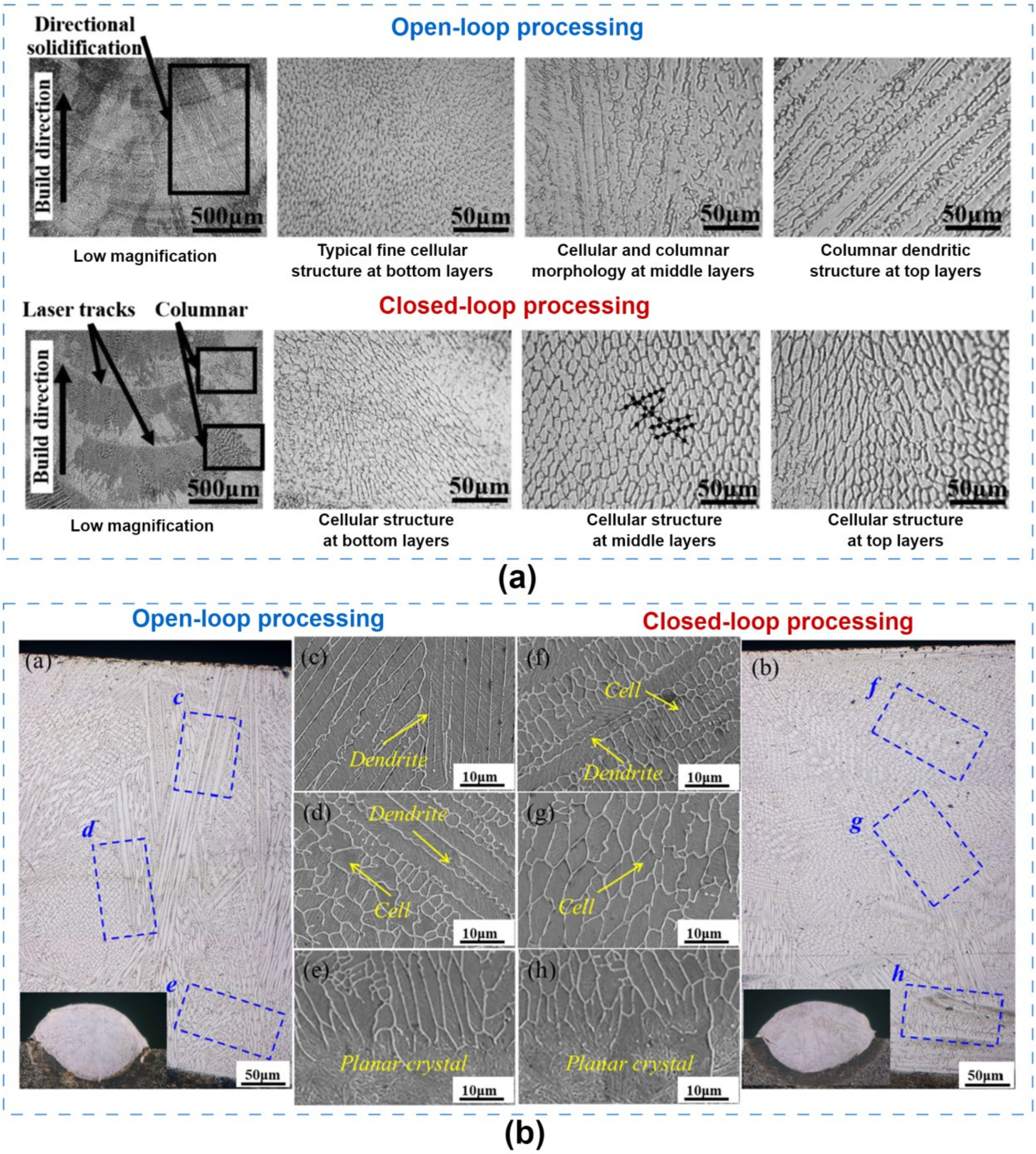

**Figure 27**. Microstructure Comparison in SS 316L Samples: Uncontrolled vs. Closed-Loop Control. (a) The uncontrolled sample shows a mix of cellular and columnar structures in middle layers, transitioning to coarser grains at the top. The closed-loop control sample maintains a consistent cellular grain structure [73]. (b) OM and SEM analyses reveal dendritic formations aligned with heat flow in the uncontrolled sample, contrasting with the uniform, small equiaxed crystals in the closed-loop controlled sample [329].

## *4.2. In-Process Defect Correction*

While closed-loop feedback control strategies effectively pre-empt defect generation and promote microstructure homogeneity, the complex nature of the LAM process still leaves room for defects to arise. Two key challenges inherent to these strategies include: (1) the dimensional inaccuracies and sub-surface defects that result from the interaction of multiple error sources or instabilities, which reduce the effectiveness of most single-input and single-output (SISO) closed-loop controllers for fabricating complex geometrical parts, and (2) the robustness of sensing signals, which is critical for a successful closed-loop control system. Defects can still arise if the in-situ monitoring signal is insufficient for providing adequate information on the process dynamics, or if the closed-loop control algorithms fall short in identifying optimal process parameter adjustments.

Given these challenges, in-process defect correction is a necessary complement to real-time closed-loop feedback control. This strategy, which is primarily accomplished by additive or subtractive methods, rectifies defects during the LAM process itself, improving dimensional accuracy and eradicating sub-surface defects before they potentially cause catastrophic failure. As a result, in-process defect repair acts as a backup mechanism in LAM to ensure the quality of as-fabricated parts.

Research in-process defect correction mainly addresses surface defects, such as dimensional deviations, and sub-surface defects, including pores and cracks. The use of cyclical additive manufacturing and subtractive machining stages in hybrid additive-subtractive manufacturing is a promising method. **Figure 28** provides a comprehensive depiction of this method, focusing particularly on surface defect correction. As shown in **Figure 28**(a), the process begins with surface monitoring using on-machine laser line scanning combined with in-situ point cloud processing [293,69]. Upon completing a deposition cycle, the LDED process halts, and the laser line scanning technique measures the intermediate layer to produce a high-resolution 3D point cloud of the as-built part, thereby capturing any geometric deviations or surface anomalies. Subsequently, as **Figure 28**(b) illustrates, the point cloud data feeds into a ML model to identify regions that deviate from the intended design [69]. The model accurately recognizes bulge (over-built) or dent (under-built) areas. Following this, as depicted in **Figure 28**(c), a corrective toolpath is generated based on the extracted defect boundaries [160,72,338,339]. A decision on whether using AM to repair the surface or using SM to remove the defective regions is made. LDED process deposits material in both concave and convex deviations until the highest point is reached, thus smoothing the surface texture. On the other hand, subtractive machining removes excessive material from bulging areas. This method can also be extended to eliminate subsurface flaws such as pores and cracks using robotic machining [70,160]. Similarly, Bernauer et al. [283] recently developed a layer height compensation technique for wire-based LDED to improve dimensional accuracy. The authors utilized a laser line scanner to map out the height profile and adjusted the wire feed rate accordingly with a segment-specific controller. Their method demonstrated that height variances could be effectively compensated within a few layers.

Integrating defect detection and correction can significantly enhances LAM quality. However, several challenges persist, such as unifying these processes within a single software platform, accurately detecting and correcting subtle defects at multiple scales, and improving adaptive toolpath generation for more precise corrections. Addressing these issues through further research will pave the way for in-process defect elimination to become a standard practice in LAM quality assurance.

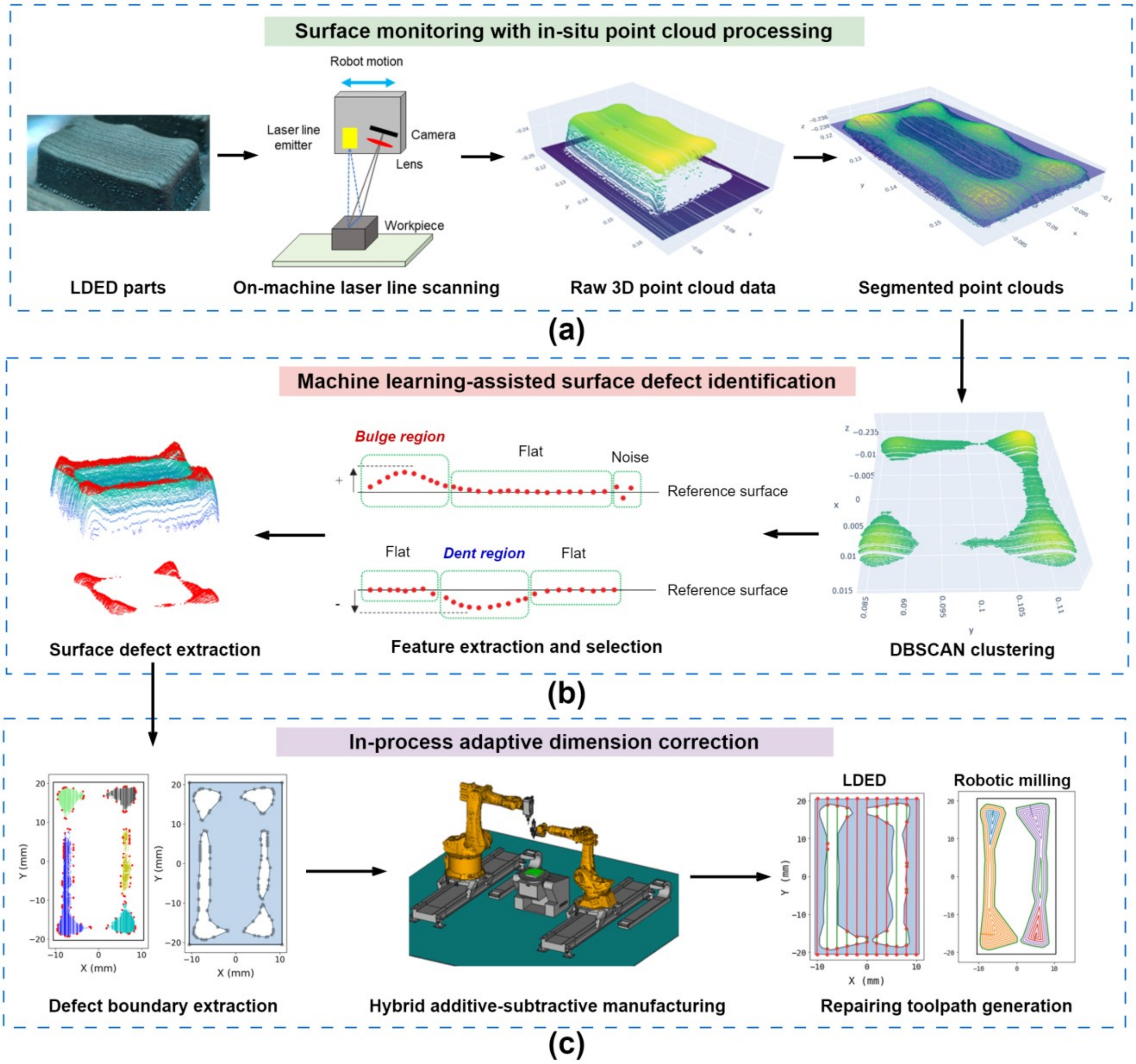

**Figure 28.** In-process adaptive dimension correction strategy in hybrid additive-subtractive manufacturing: (a) Surface monitoring using on-machine laser line scanning with in-situ point cloud processing [293,69]. (b) Machine learning-assisted surface defect identification and deviation boundary extraction [69]. (c) In-process dimension correction and repairing toolpath generation by LDED or robotic milling [160,72,338,339].

# 5. Summary And Future Perspectives

This section provides a summary and future outlook for in-situ process monitoring and control in LAM. In **Section 5.1**, key findings and developments from the preceding sections are concisely reviewed. The subsequent **Section 5.2** identifies and discusses major challenges, setting a direction for future advancements in the field.

## *5.1. Summary*

This work presents a comprehensive review of various aspects of in-situ process monitoring and adaptive quality enhancement in LAM. A range of mainstream monitoring methodologies are explored, including vision-based monitoring, acoustic-based monitoring, laser line scanning, and multisensor fusion. The incorporation of adaptive quality enhancement methods such as closed-loop feedback control and in-process defect correction further discussed this review.

In **Table 6**, a comparative analysis of different sensors for in-situ monitoring and closed-loop control in LAM is provided, evaluating them based on factors such as cost, sensitivity to process dynamics, noise rejection capability, sensor installation flexibility, and common functionalities. Vision-based monitoring, encompassing melt pool monitoring, thermal history tracking, powder bed anomaly detection, and powder stream fault diagnostics, offers a visually rich insight into the LAM processes. SWIR, MWIR, and visible spectrum cameras and pyrometers can capture visual data coaxially or off-axis. The use of deep learning algorithms, such as CNNs, enhances the precision of real-time quality predictions based on image data captured from LAM processes.

On the other hand, acoustic-based monitoring leverages time-domain, frequency-domain, and time-frequency representations of LAM sounds, providing valuable insights into potential defects associated with specific acoustic patterns. Although acoustic monitoring offers increased flexibility in sensor installations and a quicker response at a lower hardware cost compared to vision-based monitoring, its noise rejection capabilities require further enhancements.

Furthermore, the paper delves into laser line scanning techniques which generate point cloud data to represent part surface morphology and enable in-situ point cloud processing and machine learning-assisted surface defect identification. Other sensing approaches, such as operando X-ray and inline coherent imaging (ICI), have shown promising results with high accuracy in defect identification, despite substantial obstacles for industrial adoption due to high costs and safety concerns. The paper then discusses multi-sensor fusion, a technique that fuses heterogenous data from various sensors to predict defects with superior accuracy than traditional single-sensor-based monitoring approaches.

In general, in-situ defect detection in LAM can be categorized into three hierarchical levels:

*(i)* **Region-based defect identification**: This involves identifying the process regimes (e.g., conduction mode, lack of fusion mode, keyhole mode, etc.) that the current process falls under. A region can be defined as a segment of the signal, encompassing several layers or even whole parts. Region-based defect identification is relatively straightforward and typically yields higher detection accuracy. However, its limited spatial and temporal resolution makes it less suitable in a real production context.

*(ii)* **Layer-wise defect identification**: This level detects defect occurrences within each layer using historical sensor data gathered from the entire layer. It can effectively identify various defects such as dimensional deviations, overheating, or powder bed anomalies. Layer-wise identification can also be achieved through layer-wise imaging or on-machine laser scanning. However, it may result in lower efficiency as the system must pause during the inspection process.

*(iii)* **Real-time defect detection**: This level entails the immediate identification of defect onset, such as keyhole pores and cracks. Real-time detection provides the highest spatial and temporal resolution, enabling rapid defect detection as soon as it occurs. This level presents the most challenges due to the necessity for swift system response, high computational efficiency, and fast sensor acquisition frequency.

While real-time defect detection may have the highest temporal resolution and fastest system response, it may not be suitable for all cases due to the higher software and hardware costs. Depending on the specific application, a hierarchical multisensor monitoring system spanning all three levels enabling multi-scale defect detection might be more desirable. The concept of this multi-scale defect detection system will be elaborated further in the following section.

As the review progresses towards adaptive quality enhancement, it highlights the advantages of closed-loop feedback control. With its capability to reduce porosity, improve geometric accuracy, and enhance microstructure homogeneity, this approach plays a critical role for improving consistency and overall product quality in LAM. Furthermore, the novel field of in-process defect correction, which employs subtractive machining methods or additive methods to eliminate detected defects, further ensures the high quality of the final product.

Implementing in-situ process monitoring and adaptive quality enhancement in LAM offers multiple benefits: (i) It facilitates non-destructive early defect identification and correction, thereby assures as-built product quality and process repeatability. (ii) It allows real-time control and adaptation of the LAM process, thus enhancing the overall product quality. (iii) It enables superior control over the microstructure, resulting in materials with improved mechanical properties. (iv) Lastly, these advanced technologies help to minimize waste, increase efficiency, and ultimately, offer an economical advantage in the field of LAM.

**Table 6.** Comparisons of various sensors for in-situ process monitoring and closed-loop control in LAM.

| Sensor | Cost | Sampling frequency | Sensitivity to LAM process dynamics | Noise/disturbance rejection capability | Sensor installation flexibility | Functionality |
|---|---|---|---|---|---|---|
| Coaxial melt pool camera (visible spectrum) with NIR filter | Low | Low (30 – 50 Hz) | Limited sensitivity, might not detect subtle changes in process drift | Low | Medium (require customized mounting and installation) | Coaxial melt pool geometry monitoring |
| Off-axis high speed camera with NIR filter | High | Very high (1000 – 2000 Hz) | High sensitivity, but limited FoV | High | High | Off-axis melt pool and monitoring and spatters detection |
| SWIR camera (Off-axis) | High | Medium to High (60 – 1000 Hz, depending on FoV) | High sensitivity to changes in temperature and material properties, but limited FoV | High | Medium (require specific setup) | Melt pool temperature or part surface temperature monitoring |
| SWIR camera (Coaxial) | High | Medium to High (60 – 1000 Hz, depending on FoV) | High sensitivity to changes in temperature and material properties | High | Low (require customized and complex setup) | Coaxial melt pool temperature monitoring |
| MWIR camera (Off-axis) | High | Medium to High (60 – 1000 Hz, depends on FoV) | Very high sensitivity to changes in temperature and material properties | High | Medium (require specific setup) | Melt pool and part surface temperature monitoring |
| Dual-colour pyrometer (Coaxial) | Medium to High | Medium (100 - 200 Hz) | High sensitivity to temperature changes in the melt pool | Low to Medium | Low to Medium | Melt pool temperature monitoring |
| Airborne acoustic microphone sensor | Low | High (20 k – 100 kHz) | High sensitivity, but susceptible due to noisy environment | Low | High | Monitoring of acoustic emissions from the process, can indicate anomalies such as spatter events |
| Photodiode | Low | High (up to 10 kHz depending on the specific photodiode and its setup) | Medium, mainly used for light intensity measurements | Low to medium (depends on the type and setup) | High (small and relatively easy to mount) | Provides high temporal bandwidth relative brightness stemming from the melt pool region |
| Laser profilometry | Low | Medium (100 - 200 Hz) | High (20 – 100 µm resolution) | Medium (Sensitive to lighting and surface conditions) | High | Surface morphology monitoring; stand-off distance monitoring |
| Operando X-ray diffraction | Very high | Low (typically in the range of seconds to minutes depending on the X-ray source and detector) | High (capable of providing in-depth insight into the crystalline structure of the part being manufactured) | High (X-ray diffraction signals are typically distinct and not prone to noise) | Low (require complex setup and safety considerations) | Provides information on the phase and crystalline structure of the material in real-time during manufacturing. |

### 5.2. Research Gaps And Future Perspectives

The integration of sensor-based in-situ monitoring and adaptive control technologies is crucial for LAM's reliability and efficiency, yet their full deployment remains incomplete. We've pinpointed six major research gaps hindering the widespread use of these technologies in LAM. This section also discusses future R&D directions for in-situ monitoring and adaptive quality enhancement, which is visually represented in **Figure 29**.

#### 5.2.1. Standardization And Reproducibility

Standardization and reproducibility are critical for the adoption of in-situ monitoring and quality enhancement in LAM, yet they often receive inadequate attention, leading to inconsistent research outcomes and impeding effective industrial translation.

One major issue is sensor setup standardization, where variations in sensor positioning and type affect data quality. This is true for both acoustic and vision-based sensors, affecting data consistency. Standardizing sensor setup protocols is essential for reproducible results across studies and LAM systems.

Another concern is the lack of a standardized methodology for validating ML model accuracy in defect detection. Variations in validation practices, like using single-track parts or complex structures, and inconsistent process parameters and materials for ML model training, result in models with limited general applicability. Therefore, it is necessary to develop a standardized validation approach that can accurately assess the robustness and adaptability of these models across a range of conditions and materials. Recently, Snow et al. [340] introduced a new standard for evaluating ML model accuracy in defect detection, emphasizing probability-of-detection (POD) and probability-of-false-alarm (PFA) curves aligned with NDE standards. This method systematically compares detected subsurface flaws with post-build XCT data, introducing the $a_{90/95}$ metric, which represents the flaw size detectable with 90% confidence at the lower 95% interval of the POD curve. This approach directly evaluates detectability as a function of flaw size, offering a more relevant assessment compared to traditional ML metrics.

The design of test components for evaluating closed-loop control algorithms also requires standardization. Different studies use varying structures such as thin-wall parts or spiral structures. However, it is crucial to validate closed-loop control performance on complex geometries to fully assess their effectiveness, including geometric accuracy, microstructure integrity, and mechanical performance.

Finally, the establishment of standardized reporting practices in the research community can significantly enhance the reproducibility of studies in LAM. This includes standardized methods of presenting sensor data, ML model training, and validation protocols, and details of the LAM process parameters. Addressing these standardization and reproducibility aspects will drive cohesive advancement in LAM towards robust and industrially viable in-situ monitoring and adaptive quality enhancement solutions.

### *5.2.2. Location-Specific Quality Prediction*

LAM processes inherently exhibit spatial dependency. Nevertheless, methodologies for addressing location-specific quality prediction remain noticeably absent. The primary root of this research gap is the challenge involved in constructing models that can accurately correlate in-situ process monitoring data with localized quality characteristics. The majority of existing research tilts towards general quality prediction, frequently overlooking the vital aspect of spatial information. This oversight is particularly true in CNC-based LDED and LPBF processes. In these cases, the real-time position data is often not readily accessible. The complexity of understanding heat distribution, melt pool dynamics, and spatially-dependent residual stresses, presents significant challenges. As a result, existing ML models may offer less reliable predictions in a spatial context. Although recent studies have suggested that robot-based LDED systems may be better positioned by leveraging Tool Centre Point (TCP) position data to obtain real-time location information, more research is still needed. Addressing this gap necessitates the advancement of analytical methods and the development of ML algorithms capable of synthesizing both spatial and temporal data. This would enable more accurate, location-specific quality predictions, thereby enhancing the overall efficiency and reliability of LAM processes.

### *5.2.3. Reliable Defect Detection Models For Real Production*

Currently, supervised ML models intended for in-situ defect detection primarily utilize data generated from components intentionally produced with less than ideal process settings. These settings induce defects such as LoF pores, keyhole pores, and cracking. The prevalent approach for dataset creation in this field of research has focused on the fabrication of single-track or thin-wall components. While such an approach may be pragmatic for the initial development of models, it fails to reflect adequately the complexities inherent to real-world production scenarios.

In real production, LAM processes often use predefined parameters like laser power and scanning speed for complex geometries. Defects can arise unpredictably in different locations, posing prediction challenges. ML models, when trained on thin-wall structures, might underperform for multi-layer, multi-track components due to different defect mechanisms. Thin walls are more defect-prone from rapid temperature changes, while complex structures have sparser, unpredictable defects, complicating in-situ monitoring.

Further complexity arises when considering multi-material LAM. Different materials have distinct thermal properties and behaviours under LAM processing conditions, leading to different defect formation mechanisms. An ML model trained for one material may fail to accurately predict defects in another material, emphasizing the need for more versatile and adaptable models. Transfer learning of defect formation mechanism can be applied to enhance the ML model in multi-material LAM [251].

Therefore, the next wave of research should concentrate on developing and validating ML models using data from real production scenarios. Such data should encapsulate the stochastic nature of defect occurrence, the different modes of defect generation, and multi-material manufacturing. These aspects are paramount to significantly enhancing the robustness and reliability of defect prediction models in real-world LAM production scenarios.

#### *5.2.4. Multisensor Data Fusion*

While multisensor monitoring and data fusion present promising benefits for enhancing the accuracy and robustness of in-situ defect detection in LAM, several complexities hinder their industrial adoptions. The key research gap is the determination of optimal sensor combinations and the development of effective data fusion strategies. An effective implementation of multisensor monitoring systems is collectively challenged by data heterogeneity, sensor noise characteristics, and the variable relevance of different sensor data to different quality attributes. Moreover, current studies often overlook the investigation of noise characteristics in multisensor data, which significantly affects the sensitivity, reliability, and resultant robustness of the developed models.

It is essential to understand sensor noise characteristics for optimizing signal processing techniques and designing robust multisensor monitoring systems. It is also fundamental to quantitatively assess the uncertainty induced by sensor data noise, crucial for determining the reliability of in-situ monitoring and control systems for industrial deployment.

The issue of sensor redundancy and cost is another consideration that needs further attention. Determining the minimum set of sensors that can effectively monitor the LAM process can not only reduce the cost of monitoring but also reduce potential noise and disturbances. This necessitates further investigation into the interplay between different sensors and their respective contributions to the accuracy, reliability, and robustness of defect detection.

Therefore, future research should concentrate on detailed noise analysis and the development of noise-resistant signal processing and model development techniques. Additionally, the focus should be on the creation of tailored sensor integration and data fusion techniques that account for these complexities, thus enabling more efficient, reliable, and cost-effective process monitoring.

### *5.2.5. Decision-Making Strategies Beyond Early Stopping*

Most existing process control strategies in LAM are primarily focused on the early detection of anomalies and subsequent process termination to prevent further deterioration of part quality. However, the question of how to mitigate defects effectively remains a challenge. This limitation is mainly due to the lack of a comprehensive understanding of the physics behind defect formation and propagation. In addition, current decision-making procedures requires optimization in striking a balance between maximizing productivity (i.e., reducing process interruptions for in-process inspections) and minimizing material waste (i.e., early identification of potential anomalies) [341]. Therefore, the challenge lies in developing strategies that can promptly detect defects, while also dynamically adjust process parameters to mitigate and rectify these defects in real-time.

Future research should investigate physics-informed and data-driven methodologies [342–344], or employ rule-based approaches [39] for real-time decision-making. These strategies should aim to enhance not just early defect detection, but also facilitate dynamic process adjustments to alleviate and rectify defects. Such an approach could substantially improve overall part quality, production efficiency, and consequently, the cost-effectiveness of LAM operations.

A further noteworthy area of exploration is the shift from reactive measures to a more proactive defect prediction and elimination strategy. Unlike conventional methods that focus on detecting and correcting defects after their occurrence, future methodologies should proactively anticipate defects before they occur. This would entail predicting potential defects and making preventive adjustments to process parameters to prevent the emergence of these defects. Such a proactive approach offers significant advantages, such as reduced waste from the removal of defective material, increased productivity due to less time spent on defect correction, and overall enhanced efficiency of the LAM process. The technology enabling this proactive approach is still in its infancy, and substantial research is needed to fully understand and exploit its potential for improving LAM operations.

### *5.2.6. Hierarchical Multi-Scale Defect Identification And Prediction*

The necessity for hierarchical multi-scale defect identification and prediction underscores a critical research gap. Although the concept of a multi-level defect identification has been mentioned in past research [206,308,345], there is a clear need to explore this further. A comprehensive system that integrates in-situ monitoring across all levels including region-based, layer-wise, and real-time quality monitoring is still largely unexplored and thus presents a significant research opportunity.

At the region-based level, cyclic in-process laser line scanning could be utilized for geometric distortion and surface defect detection. This technique would involve scanning the part surface after every few layers of deposition to maintain a continuous check on quality. Layer-wise defect detection and quality control could benefit from the use of historical data, such as past layer melt pool image data or acoustic signals, to predict potential flaws. Lastly, real-time quality monitoring is needed for the immediate identification and correction of defects like keyhole pores, LoF pores, or cracks. This type of detection is challenging due to the dynamic nature of the LAM process and the smaller scales of the defects.

The majority of existing research has tended to focus on only one of these levels, while what is lacking is a comprehensive, integrated monitoring and control system that concurrently operates on all levels. This would involve a multisensor approach where each sensor, or a combination of sensors, would specialize in identifying different types of defects at varying scales, contributing to defect detection at their respective operational levels. The adoption of such a system would not only create a holistic picture of the entire LAM process but also greatly enhance the robustness, reliability, and overall quality control in LAM. The challenge lies in the fusion of multisensor, multimodal data in a way that minimizes redundancy and optimizes cost-effectiveness, while ensuring the reliability and robustness of the defect detection system. It is an ambitious objective that pushes the boundaries of current LAM research.

These research gaps collectively delineate a roadmap for future investigations in LAM. Addressing these areas could lead to significant enhancements in process predictability, quality, efficiency, and reliability, driving the broader application of LAM in various industries. In the next sub-section, we propose a research roadmap towards fully autonomous, self-adaptation LAM system.

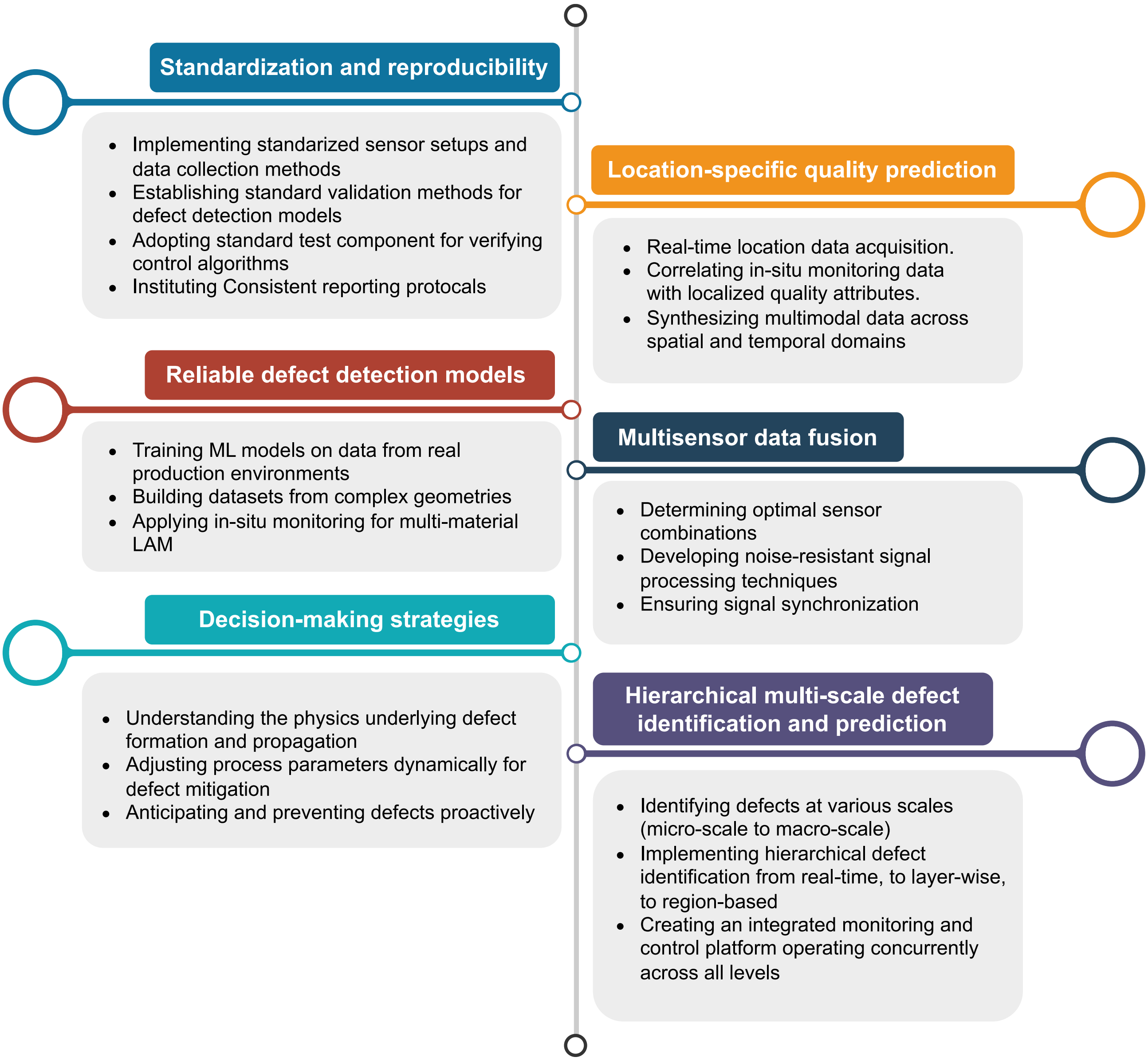

**Figure 29**. The perspectives in future R&D of in-situ process monitoring and adaptive quality enhancement in LAM.

### *5.3. Roadmap Towards Fully Autonomous And Self-Adaptation LAM Processes*

The future of LAM necessitates a paradigm shift: moving from defect detection to proactive defect prediction. This evolution relies on harnessing the richness of multimodal sensor data and understanding the underlying spatiotemporal dependencies. Such an ambitious undertaking outlines a trajectory towards a fully automated, defect-free, and self-adaptation LAM system that can operate continuously without human supervision. In line with this, we propose a research roadmap that delineates the steps towards achieving a fully autonomous and self-adaptative LAM process.

*i.* **Multisensor Monitoring and Data Fusion Framework**: Harnessing the collective intelligence of multiple sensors through multimodal data fusion is a pivotal step. This union of the complementary strengths of various sensors not only improves the accuracy and reliability of the predictive models but also furnishes a more comprehensive understanding of the LAM process. Future research should prioritize the development of noise-resistant signal processing techniques. Additional emphasis should be placed on creating tailored sensor integration and data fusion techniques that utilize a minimal set of sensors to effectively monitor the LAM process, reducing sensor redundancies. This not only cuts down on monitoring costs but also reduce potential noise and disturbances, thus enabling more efficient, reliable, and cost-effective process monitoring.

*ii.* **Proactive Defect Prediction and Mitigation Models**: The path forward involves the creation of robust machine learning and deep learning models that can accurately anticipate potential defects using multisensor data before their occurrences. Rather than simply detecting the defects, these advanced models would analyse patterns and trends in the historical and real-time sensor data to predict future process anomalies. The models' actionable insights, synthesizing physics-informed, spatio-temporal knowledge like thermal histories, would trigger automatic modifications to process parameters, such as laser power and speed, thereby pre-empting defect occurrence. This self-regulating system, informed by spatio-temporal physics knowledge, can significantly enhance the LAM process's quality consistency and reliability. Such a proactive approach offers significant advantages, such as waste reduction from the elimination of defective material, improved productivity due to less time spent on defect correction, and overall efficiency enhancement of the LAM process.

*iii.* **Multi-scale Hierarchical Defect Identification and Rectification Models**: Despite proactive defect mitigation measures, when a defect does occur, the system should engage an automated solution rather than merely halting. For instance, integrating robotic machining into the process chain could remove detected defects, facilitating seamless production. Defect identification should be approached from a multi-scale perspective, spanning from micro-scale defects (e.g., porosities, cracks, microstructure flaws) to macro scales (e.g., surface unevenness, distortions). This approach would require the multisensor monitoring system to conduct defect detection hierarchically: from real-time detection to layer-wise detection, and finally to region-based detection. Each sensor or a combination thereof would specialize in identifying different types of defects at various scales, contributing to defect detection at their respective operational levels. The adoption of such a system would not only provide a comprehensive view of the entire LAM process but also significantly enhance robustness, reliability, and overall quality control in LAM.

This roadmap envisions an era of autonomous and self-adaptation LAM systems, with each step contributing to a holistic approach for ensure product quality and process efficiency. The implementation of a multisensor monitoring and data fusion framework forms the foundation, allowing for comprehensive process understanding and more accurate and reliable predictive models. Coupled with this, proactive defect prediction and mitigation models based on machine learning and deep learning techniques promise preventive measures and process adjustments. Lastly, the multi-scale hierarchical defect identification and rectification models pave the way for continuous, unsupervised operation, ensuring product quality despite potential anomalies.

The fusion of AI, sensor technology, and robotics has the potential to revolutionize LAM, particularly when applied in a concerted manner as outlined in this roadmap. While the path forward may present challenges, the endeavour is both necessary and promising for the sustainable and broad-scale adoption of LAM technologies across various industries. This roadmap thus offers a structured approach towards this future, laying the groundwork for research and innovation to bring about the next generation of LAM systems.

## Appendix A. Surveys on Machine Learning-Assisted Defect Detection in Laser Additive Manufacturing

**Table A1.** A survey on ML-assisted vision-based defect and fault detection in LAM.

| Ref | Sensor types | Process | Model Input | Objective | Algorithms | Performance metrics | Limitations | Year |
|---|---|---|---|---|---|---|---|---|
| Zhang et al. [346] | High-speed camera (coaxial) | LPBF | Melt pool videos | Quality level prediction (thinness, regularity, etc.) | Super frame feature pyramid transformer | Accuracy: 0.97 | - Not location-specific<br>- Single track study | 2023 |
| Yin et al. [347] | Coaxial CCD camera | LDED | Melt pool images | Local defects (porosity) detection | Multibranch Fusion CNN | Accuracy: 90.18% | - Limited temporal and spatial resolution<br>- Thin-wall study | 2023 |
| Kim et al. [348] | IR thermal camera (coaxial) | LDED | Melt pool temperature, geometry | Layer height, surface unevenness estimation | ANN | RMSE: 25.44 µm, R^2: 12.62% | - Single track study | 2023 |
| Nguyen et al [64] | Off-axis monitoring camera | LPBF | Powder bed image | Surface appearance classification | Semi-supervised CNN | Accuracy > 95% | - Layer-wise inspection<br>- Limited materials | 2023 |
| Oster et al. [166] | SWIR thermography | LPBF | Thermal history | Keyhole porosity prediction | CNN | Accuracy: 0.96<br>F1-Score: 0.86 | - Lacks spatial information | 2023 |
| Estalaki et al. [222] | SWIR camera (off-axis) | LPBF | Thermal features extracted from SWIR imaging | Localized (voxelized) porosity prediction | Classic ML models | F1: 0.966 | - Layer-wise inspection<br>- Limited spatial and temporal resolution | 2022 |
| Lough et al. [182] | SWIR camera (off-axis) | LPBF | Thermal features | Localized (voxelized) porosity prediction | Statistical method | TPR – 0.9<br>FPR – 0.15 | - Simple geometries<br>- Dependence on specific thermal features | 2022 |
| Smoqi et al. [164] | Dual-wavelength pyrometer (off-axis) | LPBF | Physics-informed melt pool signatures | Porosity type and severity classification | KNN,CNN | F1 score (KNN): 95%<br>F1 score (CNN): 89–97% | - Limited material types and test geometries | 2022 |
| Larsen and Hooper [193] | High-speed imaging sensor (coaxial) | LPBF | Melt pool images | Anomaly detection (quality degradation, porosity) | Variational RNN | ROC AUC: 0.944 | - Limited process conditions | 2022 |
| | | | | | | | - Lacks generalizability | |
| Xie et al. [223] | IR thermal camera (off - axis) | LDED | Thermal history data features | Location-dependent Mechanical properties prediction | CNN (ResNet) | R^2 score: 0.7 | - Thin wall study<br>- Limited sensor flexibility | 2021 |
| Cai et al. [349] | High-speed camera (off - axis) | Laser welding | Melt pool images | Keyhole penetration state monitoring | CNN | Accuracy: 98.37%<br>Latency: 2.9 ms | - Limited to single track | 2021 |
| Snow et al. [214] | High resolution digital camera | LPBF | Layer-wise images | Anomaly classification | CNN | F1: 86.6%<br>Accuracy: 87.3% | - Layer-wise inspection<br>- Limited generalizability | 2021 |
| Knaak et al. [191] | MWIR camera (coaxial) | LDED | Melt pool images | Online defect prediction (e.g., sagging, LoF, sound weld, etc.) | Ensemble CNN-GRU | Latency: 1.1 ms<br>Accuracy: 95.1%<br>F1 score: 95.2% | - Single track study<br>- Lacks generalizability | 2021 |
| Kwon et al. [175] | High speed camera (coaxial) | LPBF | Melt pool images | Laser power level prediction | DNN | Failure rate < 1.1% | - Not location-specific<br>- Lacks generalizability | 2020 |
| Gonzalez-Val et al. [111] | MWIR camera (coaxial) | LDED | Melt pool images | Dilution estimation; LoF pore identification | CNN (ResNet) | F1 Score: 0.974<br>Accuracy: 0.967 | - Single track study<br>- Not location-specific | 2020 |
| Zhang et al. [350] | High speed camera (off-axis) | LPBF | Melt pool features | Classification of melting states | CNN | Accuracy: 0.997 | - Single track study | 2020 |
| Zhang et al. [131] | High-speed digital camera (coaxial) | LDED | Melt pool images | Porosity detection | CNN | Accuracy: 91.2%<br>RMSE: 1.32% | - Single track study<br>- Not location-specific<br>- Large pore size | 2019 |
| Scime and Beuth [186] | High-speed camera (off-axis) | LPBF | Melt pool morphologies | Keyhole pores, balling instability detection | SVM | Accuracy: 85.1% | - Limited accuracy<br>- Single track study<br>- Lacks spatial information | 2019 |
| Zhang et al. [132] | CMOS camera | LPBF | Melt pool, plume, spatter features | Quality level prediction | SVM, CNN | Accuracy: 92.7%<br>Precision : 92.8% | - Not location specific<br>- Single track study | 2018 |
| Scime and Beuth [206] | Digital Camera (off-axis) | LPBF | Layer-wise grayscale images | Powder bed anomaly classification | Multi-scale CNN | Overall accuracy: 97% | - Layer-wise inspection | 2018 |
| Khanzadeh et al. [180] | Infrared camera (off -axis) | LDED | Melt pool thermal images | Porosity prediction | KNN, SVM, DT, SOM, etc | Recall: 98.44%<br>Accuracy: 96% | - Thin wall study<br>- Non-location-specific | 2018 |

**Table A2.** A survey on ML-assisted acoustic-based defect detection in LAM.

| Ref | Sensor types | Process | Model input | Objectives | Algorithms | Performance metrics | Limitations | Year |
|---|---|---|---|---|---|---|---|---|
| Chen et al. [68] | Microphone | LDED | MFCCs | Process regime classification (keyhole pores, cracks, defect-free) | CNN | Accuracy: 89%, Keyhole pore accuracy: 93%, AUC-ROC: 98% | - Limited temporal and spatial resolutions<br>- Thin-wall study | 2023 |
| Kononenko et al. [254] | AE sensor | LPBF | Spectra principal components | In-situ crack detection | Classic ML (SVM, GPR, LR, etc.) | Accuracy: 99% | - Lacks sensor installation flexibility<br>- Lacks location-specific information | 2023 |
| Drissi-Daoudi et al. [351] | Microphone | LPBF | Spectrograms | Processing regime classification, processing maps | CNN | Accuracy > 96% | - Not location-specific | 2023 |
| Bevans et al. [269] | Microphone | LW-DED | Graph Laplacian Fiedler number | Flaw onset detection (porosity, line width, spatter) | Wavelet integrated graph theory | False alarm rate < 2% | - Limited to specific settings<br>- Lacks material/process generalizability | 2023 |
| Pandiyan et al. [251] | Microphone (PAC AM4I) | LPBF | Wavelet spectrogram | Balling, LoF, conduction, keyhole pore detection | ResNet18 and VGG16 | Accuracy ≈ 96% | - Lacks spatial information | 2022 |
| Drissi-Daoudi et al. [66] | Microphone (PAC AM4I) | LPBF | Raw signal | Material/process regime differentiation | CNN (VGG-16) | Accuracy: 93% | - Not location-specific<br>- Limited focus on inference time | 2022 |
| Tempelman et al. [102] | Microphone | LPBF | Acoustic signals | Keyhole pore detection | SVM | Accuracy: 97% | - Single track study<br>- Limited spatial and temporal resolution | 2022 |
| Pandiyan et al. [112] | Microphone (PAC AM4I) | LPBF | Raw acoustic signal | Anomaly detection (balling, LoF, keyhole pores, etc.) | VAE and GAN | Accuracy: 96% - 97% | - Limited focus on model inference time<br>- Not location-specific | 2021 |
| Hossain and Taheri [267] | Transducers | LDED | Acoustic emission signal | Build condition classification | CNN | Accuracy: 95% | - Process condition classification<br>- Not defect-correlated | 2021 |
| Shevchik et al. [259] | Fibre Bragg grating sensor | LPBF | Wavelet spectrograms | Quality classification | CNN | Accuracy: 78% - 91% | - General quality classification<br>- Limited types/locations of defects | 2019 |
| Shevchik et al. [257] | Fibre Bragg grating sensor | LPBF | Spectrogram | Pore concentration level prediction | CNN | Accuracy: 83% - 89% | - General quality classification<br>- Not location-specific<br>- Limited temporal resolution | 2018 |
| Ye et al. [261] | Microphone | LPBF | Raw acoustic signal | Melting state classification | DBN | Accuracy: 93% | - Single track study<br>- General classification<br>- Not location-specific | 2018 |

**Table A3.** A survey on ML-assisted multisensor fusion-based defect detection in LAM.

| Ref | Sensors | Process | Model Input | Objectives | Algorithms | Performance metrics | Limitations | Year |
|---|---|---|---|---|---|---|---|---|
| Bevans et al. [308] | Infrared camera, spatter imaging camera, optical powder bed imaging camera | LPBF | Spectral graph-based process signatures | Flaw detection across micro, meso, and macroscales | Classic ML | F-score: 93% | Limited to specific flaws and materials | 2023 |
| Chen et al. [70] | Microphone, SWIR camera (off-axis), coaxial CCD camera | LDED | Acoustic, coaxial melt pool, SWIR image features | Crack and keyhole pore detection | Classic ML | - Accuracy: 96%<br>- ROC-AUC: 99%<br>- False alarm rate: 4.4% | - Limited spatial and temporal resolution<br>- Thin-wall study | 2023 |
| Gaikwad et al. [310] | Two coaxial high-speed video cameras, temperature field imaging system | LPBF | Melt pool temperature, shape, size, spatter intensity | Laser defocusing detection due to thermal lensing | SVM | - False positive rate: 0.1-0.001<br>- True positive rate: 90% | - Not location specific<br>- Limited to certain materials/settings | 2022 |
| Pandiyan et al. [302] | Back reflection (BR), Visible, Infra-Red (IR) sensor, Acoustic Emission (AE) | LPBF | Acoustic, IR, BR signals | Keyhole, conduction, LoF regime detection | CNN+LSTM | Accuracy: 98.2%-99.9% (0.5-4.0 ms timescales) | - Not location specific<br>- High sensor installation complexity | 2022 |
| Li et al. [318] | AE sensor, photodiode | LPBF | AE, photodiode signals | Quality classification (poor, medium, high) | CNN | Overall accuracy: 99.08% | - General quality focus<br>- Not location/type-specific | 2022 |
| Jamnikar et al. [220,221] | Pyrometer, off-axis melt pool camera | LW-DED | Melt pool images, temperature features | Microstructure and bead geometry prediction | CNN | NRMSE: 10%-20% | - Thin wall study<br>- Limited to certain materials/geometries | 2022 |
| Petrich et al. [309] | High-speed camera, High-resolution off-axis camera, acoustic sensor (microphone) | LPBF | Pre/post laser scan imagery, acoustic, multi-spectral emissions | Localized flaw detection | NN | Accuracy: 98.5% | - Layer-wise inspections | 2021 |
| Guo et al. [183–185] | Coaxial pyrometer, off-axis infrared camera | LDED | Pyrometer, infrared thermal images | In-situ porosity and pore size prediction | Physics-informed CNN | Accuracy: 100%<br>MAPE: 6.91% | - Lacks spatial information<br>- General anomaly detection (not type-specific) | 2020 |

## Declaration of competing interest

The authors declare that they have no known competing financial interests or personal relationships that could have appeared to influence the work reported in this paper.

## Acknowledgments

This research is funded by the Agency for Science, Technology and Research (A*STAR) of Singapore through the Career Development Fund (Grant No. C210812030), and RIE2025 MTC IAF-PP grant (Grant No. M22K5a0045). It is also supported by Singapore Centre for 3D Printing (SC3DP), the National Research Foundation, Prime Minister's Office, Singapore under its Medium-Sized Centre funding scheme. It is also supported by "The Belt and Road" Innovative Talent Exchange Foreign Experts Project (Grant No. DL2022030010L).

## References

[1] F42 Committee. Terminology for Additive Manufacturing Technologies. ASTM International; n.d. https://doi.org/10.1520/F2792-12A.

[2] Yap CY, Chua CK, Dong ZL, Liu ZH, Zhang DQ, Loh LE, et al. Review of selective laser melting: Materials and applications. Applied Physics Reviews 2015;2:041101. https://doi.org/10.1063/1.4935926.

[3] Li Z, Sui S, Ma X, Tan H, Zhong C, Bi G, et al. High deposition rate powder- and wire-based laser directed energy deposition of metallic materials: A review. International Journal of Machine Tools and Manufacture 2022;181:103942. https://doi.org/10.1016/j.ijmachtools.2022.103942.

[4] Gong G, Ye J, Chi Y, Zhao Z, Wang Z, Xia G, et al. Research status of laser additive manufacturing for metal: a review. Journal of Materials Research and Technology 2021;15:855–84. https://doi.org/10.1016/j.jmrt.2021.08.050.

[5] Su J, Jiang F, Teng J, Chen L, Yan M, Requena G, et al. Recent Innovations in Laser Additive Manufacturing of Titanium Alloys. Int J Extrem Manuf 2024. https://doi.org/10.1088/2631-7990/ad2545.

[6] Liu T, Chen P, Qiu F, Yang H-Y, Tan NYJ, Chew Y, et al. Review on Laser Directed Energy Deposited Aluminum Alloys. Int J Extrem Manuf 2023. https://doi.org/10.1088/2631-7990/ad16bb.

[7] Schmidt M, Merklein M, Bourell D, Dimitrov D, Hausotte T, Wegener K, et al. Laser based additive manufacturing in industry and academia. CIRP Annals 2017;66:561–83. https://doi.org/10.1016/j.cirp.2017.05.011.

[8] Leal R, Barreiros FM, Alves L, Romeiro F, Vasco JC, Santos M, et al. Additive manufacturing tooling for the automotive industry. Int J Adv Manuf Technol 2017;92:1671–6. https://doi.org/10.1007/s00170-017-0239-8.

[9] Gisario A, Kazarian M, Martina F, Mehrpouya M. Metal additive manufacturing in the commercial aviation industry: A review. Journal of Manufacturing Systems 2019;53:124–49. https://doi.org/10.1016/j.jmsy.2019.08.005.

[10] Tan C, Weng F, Sui S, Chew Y, Bi G. Progress and perspectives in laser additive manufacturing of key aeroengine materials. International Journal of Machine Tools and Manufacture 2021;170:103804. https://doi.org/10.1016/j.ijmachtools.2021.103804.

[11] Blakey-Milner B, Gradl P, Snedden G, Brooks M, Pitot J, Lopez E, et al. Metal additive manufacturing in aerospace: A review. Materials & Design 2021;209:110008. https://doi.org/10.1016/j.matdes.2021.110008.

[12] Mohanavel V, Ashraff Ali KS, Ranganathan K, Allen Jeffrey J, Ravikumar MM, Rajkumar S. The roles and applications of additive manufacturing in the aerospace and automobile sector. Materials Today: Proceedings 2021;47:405–9. https://doi.org/10.1016/j.matpr.2021.04.596.

[13] Lee J, Chua PC, Chen L, Ng PHN, Kim Y, Wu Q, et al. Key Enabling Technologies for Smart Factory in

Automotive Industry: Status and Applications. International Journal of Precision Engineering and Manufacturing-Smart Technology 2023;1:93–105.
[14] Vasco JC. Chapter 16 - Additive manufacturing for the automotive industry. In: Pou J, Riveiro A, Davim JP, editors. Additive Manufacturing, Elsevier; 2021, p. 505–30. https://doi.org/10.1016/B978-0-12-818411-0.00010-0.
[15] Zhao N, Parthasarathy M, Patil S, Coates D, Myers K, Zhu H, et al. Direct additive manufacturing of metal parts for automotive applications. Journal of Manufacturing Systems 2023;68:368–75. https://doi.org/10.1016/j.jmsy.2023.04.008.
[16] Chen L, Ng NPH, Jung J, Moon SK. Additive Manufacturing for Automotive Industry: Status, Challenges and Future Perspectives. 2023 IEEE International Conference on Industrial Engineering and Engineering Management (IEEM), 2023, p. 1431–6. https://doi.org/10.1109/IEEM58616.2023.10406820.
[17] Vaneker T, Bernard A, Moroni G, Gibson I, Zhang Y. Design for additive manufacturing: Framework and methodology. CIRP Annals 2020;69:578–99. https://doi.org/10.1016/j.cirp.2020.05.006.
[18] Jiang J, Xiong Y, Zhang Z, Rosen DW. Machine learning integrated design for additive manufacturing. J Intell Manuf 2022;33:1073–86. https://doi.org/10.1007/s10845-020-01715-6.
[19] Xiong Y, Tang Y, Kim S, Rosen DW. Human-machine collaborative additive manufacturing. Journal of Manufacturing Systems 2023;66:82–91. https://doi.org/10.1016/j.jmsy.2022.12.004.
[20] Weng F, Chew Y, Zhu Z, Yao X, Wang L, Ng FL, et al. Excellent combination of strength and ductility of CoCrNi medium entropy alloy fabricated by laser aided additive manufacturing. Additive Manufacturing 2020;34:101202. https://doi.org/10.1016/j.addma.2020.101202.
[21] Zhu Z, Ng FL, Seet HL, Lu W, Liebscher CH, Rao Z, et al. Superior mechanical properties of a selective-laser-melted AlZnMgCuScZr alloy enabled by a tunable hierarchical microstructure and dual-nanoprecipitation. Materials Today 2021. https://doi.org/10.1016/j.mattod.2021.11.019.
[22] Tan C, Liu Y, Weng F, Ng FL, Su J, Xu Z, et al. Additive manufacturing of voxelized heterostructured materials with hierarchical phases. Additive Manufacturing 2022;54:102775. https://doi.org/10.1016/j.addma.2022.102775.
[23] Ahn D-G. Direct metal additive manufacturing processes and their sustainable applications for green technology: A review. Int J of Precis Eng and Manuf-Green Tech 2016;3:381–95. https://doi.org/10.1007/s40684-016-0048-9.
[24] Niaki MK, Torabi SA, Nonino F. Why manufacturers adopt additive manufacturing technologies: The role of sustainability. Journal of Cleaner Production 2019;222:381–92. https://doi.org/10.1016/j.jclepro.2019.03.019.
[25] Tan C, Li Q, Yao X, Chen L, Su J, Ng FL, et al. Machine Learning Customized Novel Material for Energy-Efficient 4D Printing. Advanced Science 2023:2206607. https://doi.org/10.1002/advs.202206607.
[26] Moon SK, Tan YE, Hwang J, Yoon Y-J. Application of 3D printing technology for designing light-weight unmanned aerial vehicle wing structures. Int J of Precis Eng and Manuf-Green Tech 2014;1:223–8. https://doi.org/10.1007/s40684-014-0028-x.
[27] Huang R, Riddle M, Graziano D, Warren J, Das S, Nimbalkar S, et al. Energy and emissions saving potential of additive manufacturing: the case of lightweight aircraft components. Journal of Cleaner Production 2016;135:1559–70. https://doi.org/10.1016/j.jclepro.2015.04.109.
[28] Zhu J, Zhou H, Wang C, Zhou L, Yuan S, Zhang W. A review of topology optimization for additive manufacturing: Status and challenges. Chinese Journal of Aeronautics 2021;34:91–110. https://doi.org/10.1016/j.cja.2020.09.020.
[29] Nguyen PCH, Kim Y, Choi Y. Lightweight design with metallic additively manufactured cellular structures. Journal of Computational Design and Engineering 2022;9:155–67. https://doi.org/10.1093/jcde/qwab078.
[30] Kim J-E, Cho N-K, Park K. Computational homogenization of additively manufactured lightweight structures with multiscale topology optimization. Journal of Computational Design and Engineering 2022;9:1602–15. https://doi.org/10.1093/jcde/qwac078.
[31] Zou J, Xia X. Topology optimization for additive manufacturing with strength constraints considering anisotropy. Journal of Computational Design and Engineering 2023;10:892–904. https://doi.org/10.1093/jcde/qwad028.
[32] Wang D, Liu L, Deng G, Deng C, Bai Y, Yang Y, et al. Recent progress on additive manufacturing of multi-material structures with laser powder bed fusion. Virtual and Physical Prototyping 2022;17:329–65. https://doi.org/10.1080/17452759.2022.2028343.
[33] Nazir A, Gokcekaya O, Md Masum Billah K, Ertugrul O, Jiang J, Sun J, et al. Multi-material additive manufacturing: A systematic review of design, properties, applications, challenges, and 3D printing of materials and cellular metamaterials. Materials & Design 2023;226:111661. https://doi.org/10.1016/j.matdes.2023.111661.

[34] Su J, Tan C, Ng FL, Weng F, Chen L, Jiang F, et al. Additive manufacturing of novel heterostructured martensite-austenite dual-phase steel through in-situ alloying. Materials Today Communications 2022;33:104724. https://doi.org/10.1016/j.mtcomm.2022.104724.

[35] Putra NE, Mirzaali MJ, Apachitei I, Zhou J, Zadpoor AA. Multi-material additive manufacturing technologies for Ti-, Mg-, and Fe-based biomaterials for bone substitution. Acta Biomaterialia 2020;109:1–20. https://doi.org/10.1016/j.actbio.2020.03.037.

[36] Delic M, Eyers DR. The effect of additive manufacturing adoption on supply chain flexibility and performance: An empirical analysis from the automotive industry. International Journal of Production Economics 2020;228:107689. https://doi.org/10.1016/j.ijpe.2020.107689.

[37] Choong YYC, Tan HW, Patel DC, Choong WTN, Chen C-H, Low HY, et al. The global rise of 3D printing during the COVID-19 pandemic. Nature Reviews Materials 2020;5:637–9. https://doi.org/10.1038/s41578-020-00234-3.

[38] Tan HW, Choong YYC. Additive manufacturing in COVID-19: recognising the challenges and driving for assurance. Virtual and Physical Prototyping 2021;0:1–6. https://doi.org/10.1080/17452759.2021.1975882.

[39] Yeung H, Kim FH, Donmez MA, Neira J. Keyhole pores reduction in laser powder bed fusion additive manufacturing of nickel alloy 625. International Journal of Machine Tools and Manufacture 2022;183:103957. https://doi.org/10.1016/j.ijmachtools.2022.103957.

[40] Moon SK, Ng NPH, Chen L, Ahn D-G. A novel quality inspection method for aerosol jet printed sensors through infrared imaging and machine learning. CIRP Annals 2023. https://doi.org/10.1016/j.cirp.2023.03.029.

[41] Tan C, Li R, Su J, Du D, Du Y, Attard B, et al. Review on field assisted metal additive manufacturing. International Journal of Machine Tools and Manufacture 2023;189:104032. https://doi.org/10.1016/j.ijmachtools.2023.104032.

[42] Wang M, Kashaev N. On the maintenance of processing stability and consistency in laser-directed energy deposition via machine learning. Journal of Manufacturing Systems 2024;73:126–42. https://doi.org/10.1016/j.jmsy.2024.01.005.

[43] Tapia G, Elwany A. A Review on Process Monitoring and Control in Metal-Based Additive Manufacturing. Journal of Manufacturing Science and Engineering 2014;136:060801. https://doi.org/10.1115/1.4028540.

[44] Reutzel EW, Nassar AR. A survey of sensing and control systems for machine and process monitoring of directed-energy, metal-based additive manufacturing. Rapid Prototyping Journal 2015;21:159–67. https://doi.org/10.1108/RPJ-12-2014-0177.

[45] Everton SK, Hirsch M, Stravroulakis P, Leach RK, Clare AT. Review of in-situ process monitoring and in-situ metrology for metal additive manufacturing. Materials & Design 2016;95:431–45. https://doi.org/10.1016/j.matdes.2016.01.099.

[46] Chua ZY, Ahn IH, Moon SK. Process monitoring and inspection systems in metal additive manufacturing: Status and applications. Int J of Precis Eng and Manuf-Green Tech 2017;4:235–45. https://doi.org/10.1007/s40684-017-0029-7.

[47] Lu QY, Wong CH. Additive manufacturing process monitoring and control by non-destructive testing techniques: challenges and in-process monitoring. Virtual and Physical Prototyping 2018;13:39–48. https://doi.org/10.1080/17452759.2017.1351201.

[48] Tang Z, Liu W, Wang Y, Saleheen KM, Liu Z, Peng S, et al. A review on in situ monitoring technology for directed energy deposition of metals. Int J Adv Manuf Technol 2020;108:3437–63. https://doi.org/10.1007/s00170-020-05569-3.

[49] Xia C, Pan Z, Polden J, Li H, Xu Y, Chen S, et al. A review on wire arc additive manufacturing: Monitoring, control and a framework of automated system. Journal of Manufacturing Systems 2020;57:31–45. https://doi.org/10.1016/j.jmsy.2020.08.008.

[50] Grasso M, Remani A, Dickins A, Colosimo BM, Leach RK. In-situ measurement and monitoring methods for metal powder bed fusion: an updated review. Meas Sci Technol 2021;32:112001. https://doi.org/10.1088/1361-6501/ac0b6b.

[51] AbouelNour Y, Gupta N. In-situ monitoring of sub-surface and internal defects in additive manufacturing: A review. Materials & Design 2022;222:111063. https://doi.org/10.1016/j.matdes.2022.111063.

[52] Herzog T, Brandt M, Trinchi A, Sola A, Molotnikov A. Process monitoring and machine learning for defect detection in laser-based metal additive manufacturing. J Intell Manuf 2023. https://doi.org/10.1007/s10845-023-02119-y.

[53] Wang P, Yang Y, Moghaddam NS. Process modeling in laser powder bed fusion towards defect detection and quality control via machine learning: The state-of-the-art and research challenges. Journal of Manufacturing Processes 2022;73:961–84. https://doi.org/10.1016/j.jmapro.2021.11.037.

[54] Qin J, Hu F, Liu Y, Witherell P, Wang CCL, Rosen DW, et al. Research and Application of Machine Learning for Additive Manufacturing. Additive Manufacturing 2022:102691. https://doi.org/10.1016/j.addma.2022.102691.
[55] Xiong Y, Tang Y, Zhou Q, Ma Y, Rosen DW. Intelligent additive manufacturing and design: state of the art and future perspectives. Additive Manufacturing 2022;59:103139. https://doi.org/10.1016/j.addma.2022.103139.
[56] Segovia Ramírez I, García Márquez FP, Papaelias M. Review on additive manufacturing and non-destructive testing. Journal of Manufacturing Systems 2023;66:260–86. https://doi.org/10.1016/j.jmsy.2022.12.005.
[57] Guo S, Agarwal M, Cooper C, Tian Q, Gao RX, Guo W, et al. Machine learning for metal additive manufacturing: Towards a physics-informed data-driven paradigm. Journal of Manufacturing Systems 2022;62:145–63. https://doi.org/10.1016/j.jmsy.2021.11.003.
[58] Cai Y, Xiong J, Chen H, Zhang G. A review of in-situ monitoring and process control system in metal-based laser additive manufacturing. Journal of Manufacturing Systems 2023;70:309–26. https://doi.org/10.1016/j.jmsy.2023.07.018.
[59] Azamfirei V, Psarommatis F, Lagrosen Y. Application of automation for in-line quality inspection, a zero-defect manufacturing approach. Journal of Manufacturing Systems 2023;67:1–22. https://doi.org/10.1016/j.jmsy.2022.12.010.
[60] Mu H, He F, Yuan L, Commins P, Wang H, Pan Z. Toward a smart wire arc additive manufacturing system: A review on current developments and a framework of digital twin. Journal of Manufacturing Systems 2023;67:174–89. https://doi.org/10.1016/j.jmsy.2023.01.012.
[61] Chua ZY, Ahn IH, Moon SK. Process monitoring and inspection systems in metal additive manufacturing: Status and applications. Int J of Precis Eng and Manuf-Green Tech 2017;4:235–45. https://doi.org/10.1007/s40684-017-0029-7.
[62] Gunasegaram DR, Barnard AS, Matthews MJ, Jared BH, Andreaco AM, Bartsch K, et al. Machine learning-assisted in-situ adaptive strategies for the control of defects and anomalies in metal additive manufacturing. Additive Manufacturing 2024;81:104013. https://doi.org/10.1016/j.addma.2024.104013.
[63] Smoqi Z, Bevans BD, Gaikwad A, Craig J, Abul-Haj A, Roeder B, et al. Closed-loop control of meltpool temperature in directed energy deposition. Materials & Design 2022;215:110508. https://doi.org/10.1016/j.matdes.2022.110508.
[64] Nguyen NV, Hum AJW, Do T, Tran T. Semi-supervised machine learning of optical in-situ monitoring data for anomaly detection in laser powder bed fusion. Virtual and Physical Prototyping 2023;18:e2129396. https://doi.org/10.1080/17452759.2022.2129396.
[65] Lee H, Heogh W, Yang J, Yoon J, Park J, Ji S, et al. Deep learning for in-situ powder stream fault detection in directed energy deposition process. Journal of Manufacturing Systems 2022;62:575–87. https://doi.org/10.1016/j.jmsy.2022.01.013.
[66] Drissi-Daoudi R, Pandiyan V, Logé R, Shevchik S, Masinelli G, Ghasemi-Tabasi H, et al. Differentiation of materials and laser powder bed fusion processing regimes from airborne acoustic emission combined with machine learning. Virtual and Physical Prototyping 2022;0:1–24. https://doi.org/10.1080/17452759.2022.2028380.
[67] Hauser T, Reisch RT, Kamps T, Kaplan AFH, Volpp J. Acoustic emissions in directed energy deposition processes. Int J Adv Manuf Technol 2022;119:3517–32. https://doi.org/10.1007/s00170-021-08598-8.
[68] Chen L, Yao X, Tan C, He W, Su J, Weng F, et al. In-situ crack and keyhole pore detection in laser directed energy deposition through acoustic signal and deep learning. Additive Manufacturing 2023;69:103547. https://doi.org/10.1016/j.addma.2023.103547.
[69] Chen L, Yao X, Xu P, Moon SK, Bi G. Rapid surface defect identification for additive manufacturing with in-situ point cloud processing and machine learning. Virtual and Physical Prototyping 2020;16:50–67. https://doi.org/10.1080/17452759.2020.1832695.
[70] Chen L, Bi G, Yao X, Tan C, Su J, Ng NPH, et al. Multisensor fusion-based digital twin for localized quality prediction in robotic laser-directed energy deposition. Robotics and Computer-Integrated Manufacturing 2023;84:102581. https://doi.org/10.1016/j.rcim.2023.102581.
[71] Shi T, Shi J, Xia Z, Lu B, Shi S, Fu G. Precise control of variable-height laser metal deposition using a height memory strategy. Journal of Manufacturing Processes 2020;57:222–32. https://doi.org/10.1016/j.jmapro.2020.05.026.
[72] Xu P, Yao X, Chen L, Zhao C, Liu K, Moon SK, et al. In-process adaptive dimension correction strategy for laser aided additive manufacturing using laser line scanning. Journal of Materials Processing Technology 2022;303:117544. https://doi.org/10.1016/j.jmatprotec.2022.117544.
[73] Akbari M, Kovacevic R. Closed loop control of melt pool width in robotized laser powder–directed energy

deposition process. Int J Adv Manuf Technol 2019;104:2887–98. https://doi.org/10.1007/s00170-019-04195-y.
[74] Kies F, Wilms MB, Pirch N, Pradeep KG, Schleifenbaum JH, Haase C. Defect formation and prevention in directed energy deposition of high-manganese steels and the effect on mechanical properties. Materials Science and Engineering: A 2020;772:138688. https://doi.org/10.1016/j.msea.2019.138688.
[75] King WE, Anderson AT, Ferencz RM, Hodge NE, Kamath C, Khairallah SA, et al. Laser powder bed fusion additive manufacturing of metals; physics, computational, and materials challenges. Applied Physics Reviews 2015;2:041304. https://doi.org/10.1063/1.4937809.
[76] Wilson JM, Piya C, Shin YC, Zhao F, Ramani K. Remanufacturing of turbine blades by laser direct deposition with its energy and environmental impact analysis. Journal of Cleaner Production 2014;80:170–8. https://doi.org/10.1016/j.jclepro.2014.05.084.
[77] Rao J, Sing SL, Lim JCW, Yeong WY, Yang J, Fan Z, et al. Detection and characterisation of defects in directed energy deposited multi-material components using full waveform inversion and reverse time migration. Virtual and Physical Prototyping 2022;17:1047–57. https://doi.org/10.1080/17452759.2022.2086142.
[78] Xie Y, Chen L, Yao X, Feng W, Moon SK. Adaptive Voxelization and Material-dependent Process Parameter Assignment for Multi-material Additive Manufacturing. 2023 IEEE International Conference on Industrial Engineering and Engineering Management (IEEM), 2023, p. 1462–6. https://doi.org/10.1109/IEEM58616.2023.10406952.
[79] Heer B, Bandyopadhyay A. Compositionally graded magnetic-nonmagnetic bimetallic structure using laser engineered net shaping. Materials Letters 2018;216:16–9. https://doi.org/10.1016/j.matlet.2017.12.129.
[80] Liu Y, Weng F, Bi G, Chew Y, Liu S, Ma G, et al. Characterization of wear properties of the functionally graded material deposited on cast iron by laser-aided additive manufacturing. Int J Adv Manuf Technol 2019;105:4097–105. https://doi.org/10.1007/s00170-019-03414-w.
[81] NASA looks to large-scale DED Additive Manufacturing for future rocket engines. Metal AM Magazine 2020. https://www.metal-am.com/nasa-looks-to-large-scale-ded-additive-manufacturing-for-future-rocket-engines/ (accessed May 11, 2023).
[82] Gu D, Shi X, Poprawe R, Bourell DL, Setchi R, Zhu J. Material-structure-performance integrated laser-metal additive manufacturing. Science 2021;372:eabg1487. https://doi.org/10.1126/science.abg1487.
[83] Sun G, Zhou R, Lu J, Mazumder J. Evaluation of defect density, microstructure, residual stress, elastic modulus, hardness and strength of laser-deposited AISI 4340 steel. Acta Materialia 2015;84:172–89. https://doi.org/10.1016/j.actamat.2014.09.028.
[84] Aboulkhair NT, Everitt NM, Ashcroft I, Tuck C. Reducing porosity in AlSi10Mg parts processed by selective laser melting. Additive Manufacturing 2014;1–4:77–86. https://doi.org/10.1016/j.addma.2014.08.001.
[85] Heeling T, Cloots M, Wegener K. Melt pool simulation for the evaluation of process parameters in selective laser melting. Additive Manufacturing 2017;14:116–25. https://doi.org/10.1016/j.addma.2017.02.003.
[86] Wei HL, Mukherjee T, Zhang W, Zuback JS, Knapp GL, De A, et al. Mechanistic models for additive manufacturing of metallic components. Progress in Materials Science 2021;116:100703. https://doi.org/10.1016/j.pmatsci.2020.100703.
[87] Ren K, Chew Y, Zhang YF, Fuh JYH, Bi GJ. Thermal field prediction for laser scanning paths in laser aided additive manufacturing by physics-based machine learning. Computer Methods in Applied Mechanics and Engineering 2020;362:112734. https://doi.org/10.1016/j.cma.2019.112734.
[88] Mondal B, Mukherjee T, DebRoy T. Crack free metal printing using physics informed machine learning. Acta Materialia 2022;226:117612. https://doi.org/10.1016/j.actamat.2021.117612.
[89] Du Y, Mukherjee T, DebRoy T. Physics-informed machine learning and mechanistic modeling of additive manufacturing to reduce defects. Applied Materials Today 2021;24:101123. https://doi.org/10.1016/j.apmt.2021.101123.
[90] DebRoy T, Mukherjee T, Wei HL, Elmer JW, Milewski JO. Metallurgy, mechanistic models and machine learning in metal printing. Nat Rev Mater 2021;6:48–68. https://doi.org/10.1038/s41578-020-00236-1.
[91] Tang L, Landers RG. Melt Pool Temperature Control for Laser Metal Deposition Processes—Part II: Layer-to-Layer Temperature Control. J Manuf Sci Eng 2010;132. https://doi.org/10.1115/1.4000883.
[92] Leung CLA, Marussi S, Atwood RC, Towrie M, Withers PJ, Lee PD. In situ X-ray imaging of defect and molten pool dynamics in laser additive manufacturing. Nat Commun 2018;9:1–9. https://doi.org/10.1038/s41467-018-03734-7.
[93] DePond PJ, Fuller JC, Khairallah SA, Angus JR, Guss G, Matthews MJ, et al. Laser-metal interaction dynamics during additive manufacturing resolved by detection of thermally-induced electron emission. Commun Mater 2020;1:1–10. https://doi.org/10.1038/s43246-020-00094-y.

[94] Wolff SJ, Wang H, Gould B, Parab N, Wu Z, Zhao C, et al. In situ X-ray imaging of pore formation mechanisms and dynamics in laser powder-blown directed energy deposition additive manufacturing. International Journal of Machine Tools and Manufacture 2021;166:103743. https://doi.org/10.1016/j.ijmachtools.2021.103743.
[95] Hojjatzadeh SMH, Parab ND, Guo Q, Qu M, Xiong L, Zhao C, et al. Direct observation of pore formation mechanisms during LPBF additive manufacturing process and high energy density laser welding. International Journal of Machine Tools and Manufacture 2020;153:103555. https://doi.org/10.1016/j.ijmachtools.2020.103555.
[96] Svetlizky D, Das M, Zheng B, Vyatskikh AL, Bose S, Bandyopadhyay A, et al. Directed energy deposition (DED) additive manufacturing: Physical characteristics, defects, challenges and applications. Materials Today 2021. https://doi.org/10.1016/j.mattod.2021.03.020.
[97] Kim FH, Moylan SP. Literature review of metal additive manufacturing defects. Gaithersburg, MD: National Institute of Standards and Technology; 2018. https://doi.org/10.6028/NIST.AMS.100-16.
[98] Wang S, Ning J, Zhu L, Yang Z, Yan W, Dun Y, et al. Role of porosity defects in metal 3D printing: Formation mechanisms, impacts on properties and mitigation strategies. Materials Today 2022. https://doi.org/10.1016/j.mattod.2022.08.014.
[99] du Plessis A. Effects of process parameters on porosity in laser powder bed fusion revealed by X-ray tomography. Additive Manufacturing 2019;30:100871. https://doi.org/10.1016/j.addma.2019.100871.
[100] Snow Z, Scime L, Ziabari A, Fisher B, Paquit V. Observation of spatter-induced stochastic lack-of-fusion in laser powder bed fusion using in situ process monitoring. Additive Manufacturing 2023;61:103298. https://doi.org/10.1016/j.addma.2022.103298.
[101] Wang W, Ning J, Liang SY. Analytical Prediction of Balling, Lack-of-Fusion and Keyholing Thresholds in Powder Bed Fusion. Applied Sciences 2021;11:12053. https://doi.org/10.3390/app112412053.
[102] Tempelman JR, Wachtor AJ, Flynn EB, Depond PJ, Forien J-B, Guss GM, et al. Detection of keyhole pore formations in laser powder-bed fusion using acoustic process monitoring measurements. Additive Manufacturing 2022;55:102735. https://doi.org/10.1016/j.addma.2022.102735.
[103] Wang L, Zhang Y, Chia HY, Yan W. Mechanism of keyhole pore formation in metal additive manufacturing. Npj Comput Mater 2022;8:1–11. https://doi.org/10.1038/s41524-022-00699-6.
[104] Cunningham R, Zhao C, Parab N, Kantzos C, Pauza J, Fezzaa K, et al. Keyhole threshold and morphology in laser melting revealed by ultrahigh-speed x-ray imaging. Science 2019;363:849–52. https://doi.org/10.1126/science.aav4687.
[105] Guo L, Wang H, Liu H, Huang Y, Wei Q, Leung CLA, et al. Understanding keyhole induced-porosities in laser powder bed fusion of aluminum and elimination strategy. International Journal of Machine Tools and Manufacture 2023;184:103977. https://doi.org/10.1016/j.ijmachtools.2022.103977.
[106] Alimardani M, Toyserkani E, Huissoon JP, Paul CP. On the delamination and crack formation in a thin wall fabricated using laser solid freeform fabrication process: An experimental–numerical investigation. Optics and Lasers in Engineering 2009;47:1160–8. https://doi.org/10.1016/j.optlaseng.2009.06.010.
[107] Kou S. Solidification and liquation cracking issues in welding. JOM 2003;55:37–42. https://doi.org/10.1007/s11837-003-0137-4.
[108] Chen Y, Lu F, Zhang K, Nie P, Elmi Hosseini SR, Feng K, et al. Dendritic microstructure and hot cracking of laser additive manufactured Inconel 718 under improved base cooling. Journal of Alloys and Compounds 2016;670:312–21. https://doi.org/10.1016/j.jallcom.2016.01.250.
[109] Lu X, Lin X, Chiumenti M, Cervera M, Hu Y, Ji X, et al. Residual stress and distortion of rectangular and S-shaped Ti-6Al-4V parts by Directed Energy Deposition: Modelling and experimental calibration. Additive Manufacturing 2019;26:166–79. https://doi.org/10.1016/j.addma.2019.02.001.
[110] Hofman JT, de Lange DF, Pathiraj B, Meijer J. FEM modeling and experimental verification for dilution control in laser cladding. Journal of Materials Processing Technology 2011;211:187–96. https://doi.org/10.1016/j.jmatprotec.2010.09.007.
[111] Gonzalez-Val C, Pallas A, Panadeiro V, Rodriguez A. A convolutional approach to quality monitoring for laser manufacturing. J Intell Manuf 2020;31:789–95. https://doi.org/10.1007/s10845-019-01495-8.
[112] Pandiyan V, Drissi-Daoudi R, Shevchik S, Masinelli G, Le-Quang T, Logé R, et al. Semi-supervised Monitoring of Laser powder bed fusion process based on acoustic emissions. Virtual and Physical Prototyping 2021;0:1–17. https://doi.org/10.1080/17452759.2021.1966166.
[113] Wang S, Zhu L, Dun Y, Yang Z, Fuh JYH, Yan W. Multi-physics modeling of direct energy deposition process of thin-walled structures: defect analysis. Comput Mech 2021;67:1229–42. https://doi.org/10.1007/s00466-021-01992-9.
[114] Guo Y, Jia L, Kong B, Wang N, Zhang H. Single track and single layer formation in selective laser melting of

niobium solid solution alloy. Chinese Journal of Aeronautics 2018;31:860–6. https://doi.org/10.1016/j.cja.2017.08.019.

[115] Buchbinder D, Meiners W, Pirch N, Wissenbach K, Schrage J. Investigation on reducing distortion by preheating during manufacture of aluminum components using selective laser melting. Journal of Laser Applications 2014;26:012004. https://doi.org/10.2351/1.4828755.

[116] Barriobero-Vila P, Gussone J, Haubrich J, Sandlöbes S, Da Silva JC, Cloetens P, et al. Inducing Stable α + β Microstructures during Selective Laser Melting of Ti-6Al-4V Using Intensified Intrinsic Heat Treatments. Materials 2017;10:268. https://doi.org/10.3390/ma10030268.

[117] Li R, Liu J, Shi Y, Wang L, Jiang W. Balling behavior of stainless steel and nickel powder during selective laser melting process. Int J Adv Manuf Technol 2012;59:1025–35. https://doi.org/10.1007/s00170-011-3566-1.

[118] Nursyifaulkhair D, Park N, Baek ER, Lee J. Effect of Process Parameters on the Formation of Lack of Fusion in Directed Energy Deposition of Ti-6Al-4V Alloy. Journal of Welding and Joining 2019;37:579–84. https://doi.org/10.5781/JWJ.2019.37.6.7.

[119] Uddin SZ, Murr LE, Terrazas CA, Morton P, Roberson DA, Wicker RB. Processing and characterization of crack-free aluminum 6061 using high-temperature heating in laser powder bed fusion additive manufacturing. Additive Manufacturing 2018;22:405–15. https://doi.org/10.1016/j.addma.2018.05.047.

[120] Szost BA, Terzi S, Martina F, Boisselier D, Prytuliak A, Pirling T, et al. A comparative study of additive manufacturing techniques: Residual stress and microstructural analysis of CLAD and WAAM printed Ti–6Al–4V components. Materials & Design 2016;89:559–67. https://doi.org/10.1016/j.matdes.2015.09.115.

[121] Strantza M, Vrancken B, Prime MB, Truman CE, Rombouts M, Brown DW, et al. Directional and oscillating residual stress on the mesoscale in additively manufactured Ti-6Al-4V. Acta Materialia 2019;168:299–308. https://doi.org/10.1016/j.actamat.2019.01.050.

[122] Bartlett JL, Li X. An overview of residual stresses in metal powder bed fusion. Additive Manufacturing 2019;27:131–49. https://doi.org/10.1016/j.addma.2019.02.020.

[123] Lu QY, Nguyen NV, Hum AJW, Tran T, Wong CH. Identification and evaluation of defects in selective laser melted 316L stainless steel parts via in-situ monitoring and micro computed tomography. Additive Manufacturing 2020;35:101287. https://doi.org/10.1016/j.addma.2020.101287.

[124] Safdar M, Xie J, Lamouche G, Ko H, Lu Y, Zhao YF. TRANSFERABILITY ANALYSIS OF DATA-DRIVEN ADDITIVE MANUFACTURING KNOWLEDGE: A CASE STUDY BETWEEN POWDER BED FUSION AND DIRECTED ENERGY DEPOSITION n.d.

[125] Errico V, Campanelli SL, Angelastro A, Dassisti M, Mazzarisi M, Bonserio C. Coaxial Monitoring of AISI 316L Thin Walls Fabricated by Direct Metal Laser Deposition. Materials 2021;14:673. https://doi.org/10.3390/ma14030673.

[126] Kledwig C, Perfahl H, Reisacher M, Brückner F, Bliedtner J, Leyens C. Analysis of Melt Pool Characteristics and Process Parameters Using a Coaxial Monitoring System during Directed Energy Deposition in Additive Manufacturing. Materials 2019;12:308. https://doi.org/10.3390/ma12020308.

[127] Lane B, Zhirnov I, Mekhontsev S, Grantham S, Ricker R, Rauniyar S, et al. Transient Laser Energy Absorption, Co-axial Melt Pool Monitoring, and Relationship to Melt Pool Morphology. Additive Manufacturing 2020;36:101504. https://doi.org/10.1016/j.addma.2020.101504.

[128] Scime L, Fisher B, Beuth J. Using coordinate transforms to improve the utility of a fixed field of view high speed camera for additive manufacturing applications. Manufacturing Letters 2018;15:104–6. https://doi.org/10.1016/j.mfglet.2018.01.006.

[129] Tang Z, Liu W, Zhu L, Liu Z, Yan Z, Lin D, et al. Investigation on coaxial visual characteristics of molten pool in laser-based directed energy deposition of AISI 316L steel. Journal of Materials Processing Technology 2021;290:116996. https://doi.org/10.1016/j.jmatprotec.2020.116996.

[130] Thiele M, Dillkötter D, Stoppok J, Mönnigmann M, Esen C. Laser metal deposition controlling: melt pool temperature and target / actual height difference monitoring. Procedia CIRP 2020;94:441–4. https://doi.org/10.1016/j.procir.2020.09.161.

[131] Zhang B, Liu S, Shin YC. In-Process monitoring of porosity during laser additive manufacturing process. Additive Manufacturing 2019;28:497–505. https://doi.org/10.1016/j.addma.2019.05.030.

[132] Zhang Y, Hong GS, Ye D, Zhu K, Fuh JYH. Extraction and evaluation of melt pool, plume and spatter information for powder-bed fusion AM process monitoring. Materials & Design 2018;156:458–69. https://doi.org/10.1016/j.matdes.2018.07.002.

[133] Li J, Cao L, Xu J, Wang S, Zhou Q. In situ porosity intelligent classification of selective laser melting based on coaxial monitoring and image processing. Measurement 2022;187:110232. https://doi.org/10.1016/j.measurement.2021.110232.

[134] Liu W, Wang Z, Tian L, Lauria S, Liu X. Melt pool segmentation for additive manufacturing: A generative adversarial network approach. Computers & Electrical Engineering 2021;92:107183. https://doi.org/10.1016/j.compeleceng.2021.107183.
[135] Heigel JC, Lane BM, Levine LE. In Situ Measurements of Melt-Pool Length and Cooling Rate During 3D Builds of the Metal AM-Bench Artifacts. Integr Mater Manuf Innov 2020;9:31–53. https://doi.org/10.1007/s40192-020-00170-8.
[136] Lane B, Heigel J, Ricker R, Zhirnov I, Khromschenko V, Weaver J, et al. Measurements of Melt Pool Geometry and Cooling Rates of Individual Laser Traces on IN625 Bare Plates. Integr Mater Manuf Innov 2020;9:16–30. https://doi.org/10.1007/s40192-020-00169-1.
[137] García-Moreno A-I. A fast method for monitoring molten pool in infrared image streams using gravitational superpixels. J Intell Manuf 2021:1–16. https://doi.org/10.1007/s10845-021-01761-8.
[138] Fang Q, Tan Z, Li H, Shen S, Liu S, Song C, et al. In-situ capture of melt pool signature in selective laser melting using U-Net-based convolutional neural network. Journal of Manufacturing Processes 2021;68:347–55. https://doi.org/10.1016/j.jmapro.2021.05.052.
[139] Hu K, Wang Y, Li W, Wang L. CNN-BiLSTM enabled prediction on molten pool width for thin-walled part fabrication using Laser Directed Energy Deposition. Journal of Manufacturing Processes 2022;78:32–45. https://doi.org/10.1016/j.jmapro.2022.04.010.
[140] Kozjek D, Carter FM, Porter C, Mogonye J-E, Ehmann K, Cao J. Data-driven prediction of next-layer melt pool temperatures in laser powder bed fusion based on co-axial high-resolution Planck thermometry measurements. Journal of Manufacturing Processes 2022;79:81–90. https://doi.org/10.1016/j.jmapro.2022.04.033.
[141] Heigel JC, Lane BM. Measurement of the Melt Pool Length During Single Scan Tracks in a Commercial Laser Powder Bed Fusion Process. Journal of Manufacturing Science and Engineering 2018;140. https://doi.org/10.1115/1.4037571.
[142] Arroud G, Ertveldt J, Guillaume P. A proof-of-concept analysis relating dimensions of a melt pool to its vibrational behavior to control a laser-based additive manufacturing process. Procedia CIRP 2020;94:404–8. https://doi.org/10.1016/j.procir.2020.09.154.
[143] Lison M, Devesse W, de Baere D, Hinderdael M, Guillaume P. Hyperspectral and thermal temperature estimation during laser cladding. Journal of Laser Applications 2019;31:022313. https://doi.org/10.2351/1.5096129.
[144] Ren W, Zhang Z, Lu Y, Wen G, Mazumder J. In-Situ Monitoring of Laser Additive Manufacturing for Al7075 Alloy Using Emission Spectroscopy and Plume Imaging. IEEE Access 2021;9:61671–9. https://doi.org/10.1109/ACCESS.2021.3074703.
[145] Hua T, Jing C, Xin L, Fengying Z, Weidong H. Research on molten pool temperature in the process of laser rapid forming. Journal of Materials Processing Technology 2008;198:454–62. https://doi.org/10.1016/j.jmatprotec.2007.06.090.
[146] Zhirnov I, Mekhontsev S, Lane B, Grantham S, Bura N. Accurate determination of laser spot position during laser powder bed fusion process thermography. Manufacturing Letters 2020;23:49–52. https://doi.org/10.1016/j.mfglet.2019.12.002.
[147] Grasso M, Demir AG, Previtali B, Colosimo BM. In situ monitoring of selective laser melting of zinc powder via infrared imaging of the process plume. Robotics and Computer-Integrated Manufacturing 2018;49:229–39. https://doi.org/10.1016/j.rcim.2017.07.001.
[148] Bappy MM, Liu C, Bian L, Tian W. Morphological Dynamics-Based Anomaly Detection Towards In Situ Layer-Wise Certification for Directed Energy Deposition Processes. Journal of Manufacturing Science and Engineering 2022;144:111007. https://doi.org/10.1115/1.4054805.
[149] Ko H, Lu Y, Yang Z, Ndiaye NY, Witherell P. A framework driven by physics-guided machine learning for process-structure-property causal analytics in additive manufacturing. Journal of Manufacturing Systems 2023;67:213–28. https://doi.org/10.1016/j.jmsy.2022.09.010.
[150] Ko H, Kim J, Lu Y, Shin D, Yang Z, Oh Y. Spatial-Temporal Modeling Using Deep Learning for Real-Time Monitoring of Additive Manufacturing, American Society of Mechanical Engineers Digital Collection; 2022. https://doi.org/10.1115/DETC2022-91021.
[151] Ko H, Witherell P, Lu Y, Kim S, Rosen DW. Machine learning and knowledge graph based design rule construction for additive manufacturing. Additive Manufacturing 2021;37:101620. https://doi.org/10.1016/j.addma.2020.101620.
[152] Bi G, Sun CN, Gasser A. Study on influential factors for process monitoring and control in laser aided additive manufacturing. Journal of Materials Processing Technology 2013;213:463–8. https://doi.org/10.1016/j.jmatprotec.2012.10.006.

[153] Bi G, Gasser A, Wissenbach K, Drenker A, Poprawe R. Identification and qualification of temperature signal for monitoring and control in laser cladding. Optics and Lasers in Engineering 2006;44:1348–59. https://doi.org/10.1016/j.optlaseng.2006.01.009.
[154] Bi G, Gasser A, Wissenbach K, Drenker A, Poprawe R. Characterization of the process control for the direct laser metallic powder deposition. Surface and Coatings Technology 2006;201:2676–83. https://doi.org/10.1016/j.surfcoat.2006.05.006.
[155] Bi G, Gasser A, Wissenbach K, Drenker A, Poprawe R. Investigation on the direct laser metallic powder deposition process via temperature measurement. Applied Surface Science 2006;253:1411–6. https://doi.org/10.1016/j.apsusc.2006.02.025.
[156] Bi G, Schürmann B, Gasser A, Wissenbach K, Poprawe R. Development and qualification of a novel laser-cladding head with integrated sensors. International Journal of Machine Tools and Manufacture 2007;47:555–61. https://doi.org/10.1016/j.ijmachtools.2006.05.010.
[157] Ocylok S, Alexeev E, Mann S, Weisheit A, Wissenbach K, Kelbassa I. Correlations of Melt Pool Geometry and Process Parameters During Laser Metal Deposition by Coaxial Process Monitoring. Physics Procedia 2014;56:228–38. https://doi.org/10.1016/j.phpro.2014.08.167.
[158] Criales LE, Arısoy YM, Lane B, Moylan S, Donmez A, Özel T. Laser powder bed fusion of nickel alloy 625: Experimental investigations of effects of process parameters on melt pool size and shape with spatter analysis. International Journal of Machine Tools and Manufacture 2017;121:22–36. https://doi.org/10.1016/j.ijmachtools.2017.03.004.
[159] Chen M, Lu Y, Wang Z, Lan H, Sun G, Ni Z. Melt pool evolution on inclined NV E690 steel plates during laser direct metal deposition. Optics & Laser Technology 2021;136:106745. https://doi.org/10.1016/j.optlastec.2020.106745.
[160] Chen L, Yao X, Liu K, Tan C, Moon SK. Multisensor fusion-based digital twin in additive manufacturing for in-situ quality monitoring and defect correction 2023. https://doi.org/10.48550/arXiv.2304.05685.
[161] Akbari M, Kovacevic R. An investigation on mechanical and microstructural properties of 316LSi parts fabricated by a robotized laser/wire direct metal deposition system. Additive Manufacturing 2018;23:487–97. https://doi.org/10.1016/j.addma.2018.08.031.
[162] Sampson R, Lancaster R, Sutcliffe M, Carswell D, Hauser C, Barras J. An improved methodology of melt pool monitoring of direct energy deposition processes. Optics & Laser Technology 2020;127:106194. https://doi.org/10.1016/j.optlastec.2020.106194.
[163] Chen L, Yao X, Chew Y, Weng F, Moon SK, Bi G. Data-Driven Adaptive Control for Laser-Based Additive Manufacturing with Automatic Controller Tuning. Applied Sciences 2020;10:7967. https://doi.org/10.3390/app10227967.
[164] Smoqi Z, Gaikwad A, Bevans B, Kobir MH, Craig J, Abul-Haj A, et al. Monitoring and prediction of porosity in laser powder bed fusion using physics-informed meltpool signatures and machine learning. Journal of Materials Processing Technology 2022;304:117550. https://doi.org/10.1016/j.jmatprotec.2022.117550.
[165] Chen L, Yao X, Ng NPH, Moon SK. In-situ Melt Pool Monitoring of Laser Aided Additive Manufacturing using Infrared Thermal Imaging. 2022 IEEE International Conference on Industrial Engineering and Engineering Management (IEEM), 2022, p. 1478–82. https://doi.org/10.1109/IEEM55944.2022.9989715.
[166] Oster S, Breese PP, Ulbricht A, Mohr G, Altenburg SJ. A deep learning framework for defect prediction based on thermographic in-situ monitoring in laser powder bed fusion. J Intell Manuf 2023. https://doi.org/10.1007/s10845-023-02117-0.
[167] Doubenskaia M, Pavlov M, Grigoriev S, Smurov I. Definition of brightness temperature and restoration of true temperature in laser cladding using infrared camera. Surface and Coatings Technology 2013;220:244–7. https://doi.org/10.1016/j.surfcoat.2012.10.044.
[168] Zhu K, Fuh JYH, Lin X. Metal-Based Additive Manufacturing Condition Monitoring: A Review on Machine Learning Based Approaches. IEEE/ASME Transactions on Mechatronics 2021:1–16. https://doi.org/10.1109/TMECH.2021.3110818.
[169] Fu Y, Downey ARJ, Yuan L, Zhang T, Pratt A, Balogun Y. Machine learning algorithms for defect detection in metal laser-based additive manufacturing: A review. Journal of Manufacturing Processes 2022;75:693–710. https://doi.org/10.1016/j.jmapro.2021.12.061.
[170] Goh GD, Sing SL, Yeong WY. A review on machine learning in 3D printing: applications, potential, and challenges. Artif Intell Rev 2021;54:63–94. https://doi.org/10.1007/s10462-020-09876-9.
[171] Qi X, Chen G, Li Y, Cheng X, Li C. Applying Neural-Network-Based Machine Learning to Additive Manufacturing: Current Applications, Challenges, and Future Perspectives. Engineering 2019;5:721–9. https://doi.org/10.1016/j.eng.2019.04.012.
[172] Montazeri M, Nassar AR, Dunbar AJ, Rao P. In-process monitoring of porosity in additive manufacturing

using optical emission spectroscopy. IISE Transactions 2020;52:500–15. https://doi.org/10.1080/24725854.2019.1659525.
[173] Sing SL, Kuo CN, Shih CT, Ho CC, Chua CK. Perspectives of using machine learning in laser powder bed fusion for metal additive manufacturing. Virtual and Physical Prototyping 2021;16:372–86. https://doi.org/10.1080/17452759.2021.1944229.
[174] Atwya M, Panoutsos G. In-situ porosity prediction in metal powder bed fusion additive manufacturing using spectral emissions: a prior-guided machine learning approach. J Intell Manuf 2023. https://doi.org/10.1007/s10845-023-02170-9.
[175] Kwon O, Kim HG, Ham MJ, Kim W, Kim G-H, Cho J-H, et al. A deep neural network for classification of melt-pool images in metal additive manufacturing. J Intell Manuf 2020;31:375–86. https://doi.org/10.1007/s10845-018-1451-6.
[176] Mi J, Zhang Y, Li H, Shen S, Yang Y, Song C, et al. In-situ monitoring laser based directed energy deposition process with deep convolutional neural network. J Intell Manuf 2021. https://doi.org/10.1007/s10845-021-01820-0.
[177] Li X, Siahpour S, Lee J, Wang Y, Shi J. Deep Learning-Based Intelligent Process Monitoring of Directed Energy Deposition in Additive Manufacturing with Thermal Images. Procedia Manufacturing 2020;48:643–9. https://doi.org/10.1016/j.promfg.2020.05.093.
[178] Mahato V, Obeidi MA, Brabazon D, Cunningham P. Detecting voids in 3D printing using melt pool time series data. J Intell Manuf 2020. https://doi.org/10.1007/s10845-020-01694-8.
[179] Taherkhani K, Sheydaeian E, Eischer C, Otto M, Toyserkani E. Development of a defect-detection platform using photodiode signals collected from the melt pool of laser powder-bed fusion. Additive Manufacturing 2021;46:102152. https://doi.org/10.1016/j.addma.2021.102152.
[180] Khanzadeh M, Chowdhury S, Marufuzzaman M, Tschopp MA, Bian L. Porosity prediction: Supervised-learning of thermal history for direct laser deposition. Journal of Manufacturing Systems 2018;47:69–82. https://doi.org/10.1016/j.jmsy.2018.04.001.
[181] Khanzadeh M, Chowdhury S, Tschopp MA, Doude HR, Marufuzzaman M, Bian L. In-situ monitoring of melt pool images for porosity prediction in directed energy deposition processes. IISE Transactions 2019;51:437–55. https://doi.org/10.1080/24725854.2017.1417656.
[182] Lough CS, Liu T, Wang X, Brown B, Landers RG, Bristow DA, et al. Local prediction of Laser Powder Bed Fusion porosity by short-wave infrared imaging thermal feature porosity probability maps. Journal of Materials Processing Technology 2022;302:117473. https://doi.org/10.1016/j.jmatprotec.2021.117473.
[183] McGowan E, Gawade V, Guo W (Grace). A Physics-Informed Convolutional Neural Network with Custom Loss Functions for Porosity Prediction in Laser Metal Deposition. Sensors 2022;22:494. https://doi.org/10.3390/s22020494.
[184] Guo W “Grace,” Tian Q, Guo S, Guo Y. A physics-driven deep learning model for process-porosity causal relationship and porosity prediction with interpretability in laser metal deposition. CIRP Annals 2020;69:205–8. https://doi.org/10.1016/j.cirp.2020.04.049.
[185] Tian Q, Guo S, Melder E, Bian L, Guo W “Grace.” Deep Learning-Based Data Fusion Method for In Situ Porosity Detection in Laser-Based Additive Manufacturing. Journal of Manufacturing Science and Engineering 2020;143. https://doi.org/10.1115/1.4048957.
[186] Scime L, Beuth J. Using machine learning to identify in-situ melt pool signatures indicative of flaw formation in a laser powder bed fusion additive manufacturing process. Additive Manufacturing 2019;25:151–65. https://doi.org/10.1016/j.addma.2018.11.010.
[187] Chen L, Yao X, Feng W, Chew Y, Moon SK. Multimodal sensor fusion for real-time location-dependent defect detection in laser-directed energy deposition 2023. https://doi.org/10.48550/arXiv.2305.13596.
[188] Wang C, Tan XP, Tor SB, Lim CS. Machine learning in additive manufacturing: State-of-the-art and perspectives. Additive Manufacturing 2020;36:101538. https://doi.org/10.1016/j.addma.2020.101538.
[189] Xia C, Pan Z, Li Y, Chen J, Li H. Vision-based melt pool monitoring for wire-arc additive manufacturing using deep learning method. Int J Adv Manuf Technol 2022;120:551–62. https://doi.org/10.1007/s00170-022-08811-2.
[190] Rodríguez-Araújo J, Garcia-Diaz A, Panadeiro V, Knaak C. Uncooled MWIR PbSe technology outperforms CMOS in RT closed-loop control and monitoring of laser processing. Imaging and Applied Optics 2017 (3D, AIO, COSI, IS, MATH, pcAOP) (2017), paper ATh2A.2, Optical Society of America; 2017, p. ATh2A.2. https://doi.org/10.1364/AIO.2017.ATh2A.2.
[191] Knaak C, von Eßen J, Kröger M, Schulze F, Abels P, Gillner A. A Spatio-Temporal Ensemble Deep Learning Architecture for Real-Time Defect Detection during Laser Welding on Low Power Embedded Computing Boards. Sensors 2021;21:4205. https://doi.org/10.3390/s21124205.

[192] Lapido YL, Rodriguez-Araújo J, García-Díaz A, Castro G, Vidal F, Romero P, et al. Cognitive high speed defect detection and classification in MWIR images of laser welding. Industrial Laser Applications Symposium (ILAS 2015), vol. 9657, International Society for Optics and Photonics; 2015, p. 96570B. https://doi.org/10.1117/12.2177890.

[193] Larsen S, Hooper PA. Deep semi-supervised learning of dynamics for anomaly detection in laser powder bed fusion. J Intell Manuf 2022;33:457–71. https://doi.org/10.1007/s10845-021-01842-8.

[194] Lyu J, Manoochehri S. Online Convolutional Neural Network-based anomaly detection and quality control for Fused Filament Fabrication process. Virtual and Physical Prototyping 2021;16:160–77. https://doi.org/10.1080/17452759.2021.1905858.

[195] Cho H-W, Shin S-J, Seo G-J, Kim DB, Lee D-H. Real-time anomaly detection using convolutional neural network in wire arc additive manufacturing: molybdenum material. Journal of Materials Processing Technology 2022:117495. https://doi.org/10.1016/j.jmatprotec.2022.117495.

[196] Yan H, Grasso M, Paynabar K, Colosimo BM. Real-time detection of clustered events in video-imaging data with applications to additive manufacturing. IISE Transactions 2021;0:1–28. https://doi.org/10.1080/24725854.2021.1882013.

[197] Bugatti M, Colosimo BM. Towards real-time in-situ monitoring of hot-spot defects in L-PBF: a new classification-based method for fast video-imaging data analysis. J Intell Manuf 2021:1–17. https://doi.org/10.1007/s10845-021-01787-y.

[198] Schwerz C, Raza A, Lei X, Nyborg L, Hryha E, Wirdelius H. In-situ detection of redeposited spatter and its influence on the formation of internal flaws in laser powder bed fusion. Additive Manufacturing 2021;47:102370. https://doi.org/10.1016/j.addma.2021.102370.

[199] Li J, Zhou Q, Huang X, Li M, Cao L. In situ quality inspection with layer-wise visual images based on deep transfer learning during selective laser melting. J Intell Manuf 2021:1–15. https://doi.org/10.1007/s10845-021-01829-5.

[200] Garland AP, White BC, Jared BH, Heiden M, Donahue E, Boyce BL. Deep Convolutional Neural Networks as a Rapid Screening Tool for Complex Additively Manufactured Structures. Additive Manufacturing 2020;35:101217. https://doi.org/10.1016/j.addma.2020.101217.

[201] Aminzadeh M, Kurfess TR. Online quality inspection using Bayesian classification in powder-bed additive manufacturing from high-resolution visual camera images. J Intell Manuf 2019;30:2505–23. https://doi.org/10.1007/s10845-018-1412-0.

[202] Freeman FSHB, Thomas B, Chechik L, Todd I. Multi-faceted monitoring of powder flow rate variability in directed energy deposition. Additive Manufacturing Letters 2022;2:100024. https://doi.org/10.1016/j.addlet.2021.100024.

[203] Webster S, Giovannini M, Shi Y, Martinez-Prieto N, Fezzaa K, Sun T, et al. High-throughput, in situ imaging of multi-layer powder-blown directed energy deposition with angled nozzle. Review of Scientific Instruments 2022;93:023701. https://doi.org/10.1063/5.0077140.

[204] Tan Phuc L, Seita M. A high-resolution and large field-of-view scanner for in-line characterization of powder bed defects during additive manufacturing. Materials & Design 2019;164:107562. https://doi.org/10.1016/j.matdes.2018.107562.

[205] Li Z, Liu X, Wen S, He P, Zhong K, Wei Q, et al. In Situ 3D Monitoring of Geometric Signatures in the Powder-Bed-Fusion Additive Manufacturing Process via Vision Sensing Methods. Sensors 2018;18:1180. https://doi.org/10.3390/s18041180.

[206] Scime L, Beuth J. A multi-scale convolutional neural network for autonomous anomaly detection and classification in a laser powder bed fusion additive manufacturing process. Additive Manufacturing 2018;24:273–86. https://doi.org/10.1016/j.addma.2018.09.034.

[207] Scime L, Siddel D, Baird S, Paquit V. Layer-wise anomaly detection and classification for powder bed additive manufacturing processes: A machine-agnostic algorithm for real-time pixel-wise semantic segmentation. Additive Manufacturing 2020;36:101453. https://doi.org/10.1016/j.addma.2020.101453.

[208] Shi B, Chen Z. A layer-wise multi-defect detection system for powder bed monitoring: Lighting strategy for imaging, adaptive segmentation and classification. Materials & Design 2021;210:110035. https://doi.org/10.1016/j.matdes.2021.110035.

[209] Xia C, Pan Z, Polden J, Li H, Xu Y, Chen S. Modelling and prediction of surface roughness in wire arc additive manufacturing using machine learning. J Intell Manuf 2021:1–16. https://doi.org/10.1007/s10845-020-01725-4.

[210] Liu C, Wang RR, Ho I, Kong ZJ, Williams C, Babu S, et al. Toward online layer-wise surface morphology measurement in additive manufacturing using a deep learning-based approach. J Intell Manuf 2022:1–17. https://doi.org/10.1007/s10845-022-01933-0.

[211] García-Díaz A, Panadeiro V, Lodeiro B, Rodríguez-Araújo J, Stavridis J, Papacharalampopoulos A, et al. OpenLMD, an open source middleware and toolkit for laser-based additive manufacturing of large metal parts. Robotics and Computer-Integrated Manufacturing 2018;53:153–61. https://doi.org/10.1016/j.rcim.2018.04.006.
[212] Wang R, Cheung CF. CenterNet-based defect detection for additive manufacturing. Expert Systems with Applications 2022;188:116000. https://doi.org/10.1016/j.eswa.2021.116000.
[213] Abdelrahman M, Reutzel EW, Nassar AR, Starr TL. Flaw detection in powder bed fusion using optical imaging. Additive Manufacturing 2017;15:1–11. https://doi.org/10.1016/j.addma.2017.02.001.
[214] Snow Z, Diehl B, Reutzel EW, Nassar A. Toward in-situ flaw detection in laser powder bed fusion additive manufacturing through layerwise imagery and machine learning. Journal of Manufacturing Systems 2021;59:12–26. https://doi.org/10.1016/j.jmsy.2021.01.008.
[215] Boschetto A, Bottini L, Vatanparast S, Veniali F. Part defects identification in selective laser melting via digital image processing of powder bed anomalies. Prod Eng Res Devel 2022. https://doi.org/10.1007/s11740-022-01112-3.
[216] Fischer FG, Zimmermann MG, Praetzsch N, Knaak C. Monitoring of the powder bed quality in metal additive manufacturing using deep transfer learning. Materials & Design 2022;222:111029. https://doi.org/10.1016/j.matdes.2022.111029.
[217] Lu QY, Nguyen NV, Hum AJW, Tran T, Wong CH. Optical in-situ monitoring and correlation of density and mechanical properties of stainless steel parts produced by selective laser melting process based on varied energy density. Journal of Materials Processing Technology 2019;271:520–31. https://doi.org/10.1016/j.jmatprotec.2019.04.026.
[218] Iravani-Tabrizipour M, Toyserkani E. An image-based feature tracking algorithm for real-time measurement of clad height. Machine Vision and Applications 2007;18:343–54. https://doi.org/10.1007/s00138-006-0066-7.
[219] Donadello S, Motta M, Demir AG, Previtali B. Monitoring of laser metal deposition height by means of coaxial laser triangulation. Optics and Lasers in Engineering 2019;112:136–44. https://doi.org/10.1016/j.optlaseng.2018.09.012.
[220] Jamnikar ND, Liu S, Brice C, Zhang X. In situ microstructure property prediction by modeling molten pool-quality relations for wire-feed laser additive manufacturing. Journal of Manufacturing Processes 2022;79:803–14. https://doi.org/10.1016/j.jmapro.2022.05.013.
[221] Jamnikar ND, Liu S, Brice C, Zhang X. In-process comprehensive prediction of bead geometry for laser wire-feed DED system using molten pool sensing data and multi-modality CNN. Int J Adv Manuf Technol 2022;121:903–17. https://doi.org/10.1007/s00170-022-09248-3.
[222] Estalaki SM, Lough CS, Landers RG, Kinzel EC, Luo T. Predicting defects in laser powder bed fusion using in-situ thermal imaging data and machine learning. Additive Manufacturing 2022;58:103008. https://doi.org/10.1016/j.addma.2022.103008.
[223] Xie X, Bennett J, Saha S, Lu Y, Cao J, Liu WK, et al. Mechanistic data-driven prediction of as-built mechanical properties in metal additive manufacturing. Npj Comput Mater 2021;7:1–12. https://doi.org/10.1038/s41524-021-00555-z.
[224] Altenburg SJ, Straße A, Gumenyuk A, Maierhofer C. In-situ monitoring of a laser metal deposition (LMD) process: comparison of MWIR, SWIR and high-speed NIR thermography. Quantitative InfraRed Thermography Journal 2020;0:1–18. https://doi.org/10.1080/17686733.2020.1829889.
[225] Lane B, Jacquemetton L, Piltch M, Beckett D. Thermal calibration of commercial melt pool monitoring sensors on a laser powder bed fusion system. Gaithersburg, MD: National Institute of Standards and Technology; 2020. https://doi.org/10.6028/NIST.AMS.100-35.
[226] Devesse W, De Baere D, Guillaume P. High Resolution Temperature Measurement of Liquid Stainless Steel Using Hyperspectral Imaging. Sensors 2017;17:91. https://doi.org/10.3390/s17010091.
[227] Touloukian YS, DeWitt DP. Thermophysical Properties of Matter - The TPRC Data Series. Volume 7. Thermal Radiative Properties - Metallic Elements and Alloys. THERMOPHYSICAL AND ELECTRONIC PROPERTIES INFORMATION ANALYSIS CENTER LAFAYETTE IN; 1970.
[228] Yeung H, Lane BM, Donmez MA, Moylan S. In-situ calibration of laser/galvo scanning system using dimensional reference artefacts. CIRP Annals 2020;69:441–4. https://doi.org/10.1016/j.cirp.2020.03.016.
[229] Pandiyan V, Drissi-Daoudi R, Shevchik S, Masinelli G, Logé R, Wasmer K. Analysis of time, frequency and time-frequency domain features from acoustic emissions during Laser Powder-Bed fusion process. Procedia CIRP 2020;94:392–7. https://doi.org/10.1016/j.procir.2020.09.152.
[230] Gutknecht K, Cloots M, Sommerhuber R, Wegener K. Mutual comparison of acoustic, pyrometric and thermographic laser powder bed fusion monitoring. Materials & Design 2021;210:110036.

https://doi.org/10.1016/j.matdes.2021.110036.
[231] Song S, Chen H, Lin T, Wu D, Chen S. Penetration state recognition based on the double-sound-sources characteristic of VPPAW and hidden Markov Model. Journal of Materials Processing Technology 2016;234:33–44. https://doi.org/10.1016/j.jmatprotec.2016.03.002.
[232] Lv N, Xu Y, Li S, Yu X, Chen S. Automated control of welding penetration based on audio sensing technology. Journal of Materials Processing Technology 2017;250:81–98. https://doi.org/10.1016/j.jmatprotec.2017.07.005.
[233] Jinachandran S, Ning Y, Wu B, Li H, Xi J, Prusty BG, et al. Cold Crack Monitoring and Localization in Welding Using Fiber Bragg Grating Sensors. IEEE Transactions on Instrumentation and Measurement 2020;69:9228–36. https://doi.org/10.1109/TIM.2020.3001367.
[234] Zhang Z, Wen G, Chen S. Audible Sound-Based Intelligent Evaluation for Aluminum Alloy in Robotic Pulsed GTAW: Mechanism, Feature Selection, and Defect Detection. IEEE Transactions on Industrial Informatics 2018;14:2973–83. https://doi.org/10.1109/TII.2017.2775218.
[235] Shevchik SA, Le-Quang T, Farahani FV, Faivre N, Meylan B, Zanoli S, et al. Laser Welding Quality Monitoring via Graph Support Vector Machine With Data Adaptive Kernel. IEEE Access 2019;7:93108–22. https://doi.org/10.1109/ACCESS.2019.2927661.
[236] Shevchik S, Le-Quang T, Meylan B, Farahani FV, Olbinado MP, Rack A, et al. Supervised deep learning for real-time quality monitoring of laser welding with X-ray radiographic guidance. Sci Rep 2020;10:1–12. https://doi.org/10.1038/s41598-020-60294-x.
[237] Cai W, Wang J, Jiang P, Cao L, Mi G, Zhou Q. Application of sensing techniques and artificial intelligence-based methods to laser welding real-time monitoring: A critical review of recent literature. Journal of Manufacturing Systems 2020;57:1–18. https://doi.org/10.1016/j.jmsy.2020.07.021.
[238] Asif K, Zhang L, Derrible S, Indacochea JE, Ozevin D, Ziebart B. Machine learning model to predict welding quality using air-coupled acoustic emission and weld inputs. J Intell Manuf 2020:1–15. https://doi.org/10.1007/s10845-020-01667-x.
[239] Yaacoubi S, Dahmene F, El Mountassir M, Bouzenad AE. A novel AE algorithm-based approach for the detection of cracks in spot welding in view of online monitoring: case study. Int J Adv Manuf Technol 2021;117:1807–24. https://doi.org/10.1007/s00170-021-07848-z.
[240] Sejdić E, Djurović I, Jiang J. Time–frequency feature representation using energy concentration: An overview of recent advances. Digital Signal Processing 2009;19:153–83. https://doi.org/10.1016/j.dsp.2007.12.004.
[241] Allen J. Short term spectral analysis, synthesis, and modification by discrete Fourier transform. IEEE Transactions on Acoustics, Speech, and Signal Processing 1977;25:235–8. https://doi.org/10.1109/TASSP.1977.1162950.
[242] Zhang D. Wavelet Transform. In: Zhang D, editor. Fundamentals of Image Data Mining: Analysis, Features, Classification and Retrieval, Cham: Springer International Publishing; 2019, p. 35–44. https://doi.org/10.1007/978-3-030-17989-2_3.
[243] Muda L, Begam M, Elamvazuthi I. Voice Recognition Algorithms using Mel Frequency Cepstral Coefficient (MFCC) and Dynamic Time Warping (DTW) Techniques 2010. https://doi.org/10.48550/arXiv.1003.4083.
[244] Klapuri A, Davy M, editors. Signal processing methods for music transcription. New York: Springer; 2006.
[245] Scheirer E, Slaney M. Construction and evaluation of a robust multifeature speech/music discriminator. 1997 IEEE International Conference on Acoustics, Speech, and Signal Processing, vol. 2, 1997, p. 1331–4 vol.2. https://doi.org/10.1109/ICASSP.1997.596192.
[246] Dubnov S. Generalization of spectral flatness measure for non-Gaussian linear processes. IEEE Signal Processing Letters 2004;11:698–701. https://doi.org/10.1109/LSP.2004.831663.
[247] Jiang D-N, Lu L, Zhang H-J, Tao J-H, Cai L-H. Music type classification by spectral contrast feature. Proceedings. IEEE International Conference on Multimedia and Expo, vol. 1, 2002, p. 113–6 vol.1. https://doi.org/10.1109/ICME.2002.1035731.
[248] Geoffroy P. A Large Set of Audio Features for Sound Description (Similarity and Classification) in the CUIDADO Project. n.d.
[249] Misra H, Ikbal S, Bourlard H, Hermansky H. Spectral entropy based feature for robust ASR. 2004 IEEE International Conference on Acoustics, Speech, and Signal Processing, vol. 1, 2004, p. I–193. https://doi.org/10.1109/ICASSP.2004.1325955.
[250] Dixon S. Onset detection revisited. Proceedings of the 9th international conference on digital audio effects, Montreal, Canada: 2006, p. 133–7.
[251] Pandiyan V, Drissi-Daoudi R, Shevchik S, Masinelli G, Le-Quang T, Logé R, et al. Deep Transfer Learning of Additive Manufacturing Mechanisms Across Materials in Metal-Based Laser Powder Bed Fusion Process. Journal of Materials Processing Technology 2022:117531. https://doi.org/10.1016/j.jmatprotec.2022.117531.

[252] Khadkevich M, Omologo M. Reassigned spectrum-based feature extraction for GMM-based automatic chord recognition. EURASIP Journal on Audio, Speech, and Music Processing 2013;2013:15. https://doi.org/10.1186/1687-4722-2013-15.
[253] Wang H, Li B, Xuan F-Z. Acoustic emission for in situ process monitoring of selective laser melting additive manufacturing based on machine learning and improved variational modal decomposition. Int J Adv Manuf Technol 2022:1–16. https://doi.org/10.1007/s00170-022-10032-6.
[254] Kononenko DY, Nikonova V, Seleznev M, van den Brink J, Chernyavsky D. An in situ crack detection approach in additive manufacturing based on acoustic emission and machine learning. Additive Manufacturing Letters 2023;5:100130. https://doi.org/10.1016/j.addlet.2023.100130.
[255] Chen Z, Wang D, Zhang Y. Microphone signal specialities in laser powder bed fusion: single-track scan and multi-track scan. Journal of Materials Research and Technology 2023;24:1344–62. https://doi.org/10.1016/j.jmrt.2023.03.091.
[256] Ito K, Kusano M, Demura M, Watanabe M. Detection and location of microdefects during selective laser melting by wireless acoustic emission measurement. Additive Manufacturing 2021;40:101915. https://doi.org/10.1016/j.addma.2021.101915.
[257] Shevchik SA, Kenel C, Leinenbach C, Wasmer K. Acoustic emission for in situ quality monitoring in additive manufacturing using spectral convolutional neural networks. Additive Manufacturing 2018;21:598–604. https://doi.org/10.1016/j.addma.2017.11.012.
[258] Wasmer K, Kenel C, Leinenbach C, Shevchik SA. In Situ and Real-Time Monitoring of Powder-Bed AM by Combining Acoustic Emission and Artificial Intelligence. In: Meboldt M, Klahn C, editors. Industrializing Additive Manufacturing - Proceedings of Additive Manufacturing in Products and Applications - AMPA2017, Cham: Springer International Publishing; 2018, p. 200–9. https://doi.org/10.1007/978-3-319-66866-6_20.
[259] Shevchik SA, Masinelli G, Kenel C, Leinenbach C, Wasmer K. Deep Learning for In Situ and Real-Time Quality Monitoring in Additive Manufacturing Using Acoustic Emission. IEEE Transactions on Industrial Informatics 2019;15:5194–203. https://doi.org/10.1109/TII.2019.2910524.
[260] Wasmer K, Le-Quang T, Meylan B, Shevchik SA. In Situ Quality Monitoring in AM Using Acoustic Emission: A Reinforcement Learning Approach. J of Materi Eng and Perform 2019;28:666–72. https://doi.org/10.1007/s11665-018-3690-2.
[261] Ye D, Hong GS, Zhang Y, Zhu K, Fuh JYH. Defect detection in selective laser melting technology by acoustic signals with deep belief networks. Int J Adv Manuf Technol 2018;96:2791–801. https://doi.org/10.1007/s00170-018-1728-0.
[262] Ghayoomi Mohammadi M, Mahmoud D, Elbestawi M. On the application of machine learning for defect detection in L-PBF additive manufacturing. Optics & Laser Technology 2021;143:107338. https://doi.org/10.1016/j.optlastec.2021.107338.
[263] Luo S, Ma X, Xu J, Li M, Cao L. Deep Learning Based Monitoring of Spatter Behavior by the Acoustic Signal in Selective Laser Melting. Sensors 2021;21:7179. https://doi.org/10.3390/s21217179.
[264] Zhang W, Abranovic B, Hanson-Regalado J. Flaw Detection in Metal Additive Manufacturing Using Deep Learned Acoustic Features, n.d., p. 8.
[265] Koester LW, Taheri H, Bigelow TA, Bond LJ, Faierson EJ. In-situ acoustic signature monitoring in additive manufacturing processes. AIP Conference Proceedings 2018;1949:020006. https://doi.org/10.1063/1.5031503.
[266] Taheri H, Koester LW, Bigelow TA, Faierson EJ, Bond LJ. In Situ Additive Manufacturing Process Monitoring With an Acoustic Technique: Clustering Performance Evaluation Using K-Means Algorithm. Journal of Manufacturing Science and Engineering 2019;141. https://doi.org/10.1115/1.4042786.
[267] Hossain MS, Taheri H. In-situ process monitoring for metal additive manufacturing through acoustic techniques using wavelet and convolutional neural network (CNN). Int J Adv Manuf Technol 2021:1–16. https://doi.org/10.1007/s00170-021-07721-z.
[268] Chen L, Yao X, Moon SK. In-situ acoustic monitoring of direct energy deposition process with deep learning-assisted signal denoising. Materials Today: Proceedings 2022. https://doi.org/10.1016/j.matpr.2022.09.008.
[269] Bevans B, Ramalho A, Smoqi Z, Gaikwad A, Santos TG, Rao P, et al. Monitoring and flaw detection during wire-based directed energy deposition using in-situ acoustic sensing and wavelet graph signal analysis. Materials & Design 2023;225:111480. https://doi.org/10.1016/j.matdes.2022.111480.
[270] Ramalho A, Santos TG, Bevans B, Smoqi Z, Rao P, Oliveira JP. Effect of contaminations on the acoustic emissions during wire and arc additive manufacturing of 316L stainless steel. Additive Manufacturing 2022;51:102585. https://doi.org/10.1016/j.addma.2021.102585.
[271] Xu Y, Weaver JB, Healy DM, Lu J. Wavelet transform domain filters: a spatially selective noise filtration technique. IEEE Transactions on Image Processing 1994;3:747–58. https://doi.org/10.1109/83.336245.

[272] Surovi NA, Soh GS. Acoustic feature based geometric defect identification in wire arc additive manufacturing. Virtual and Physical Prototyping 2023;18:e2210553. https://doi.org/10.1080/17452759.2023.2210553.
[273] Gaja H, Liou F. Defects monitoring of laser metal deposition using acoustic emission sensor. Int J Adv Manuf Technol 2017;90:561–74. https://doi.org/10.1007/s00170-016-9366-x.
[274] Whiting J, Springer A, Sciammarella F. Real-time acoustic emission monitoring of powder mass flow rate for directed energy deposition. Additive Manufacturing 2018;23:312–8. https://doi.org/10.1016/j.addma.2018.08.015.
[275] Lu QY, Wong CH. Additive manufacturing process monitoring and control by non-destructive testing techniques: challenges and in-process monitoring. Virtual and Physical Prototyping 2018;13:39–48. https://doi.org/10.1080/17452759.2017.1351201.
[276] Huang C, Wang G, Song H, Li R, Zhang H. Rapid surface defects detection in wire and arc additive manufacturing based on laser profilometer. Measurement 2021:110503. https://doi.org/10.1016/j.measurement.2021.110503.
[277] Binega E, Yang L, Sohn H, Cheng JCP. Online geometry monitoring during directed energy deposition additive manufacturing using laser line scanning. Precision Engineering 2022;73:104–14. https://doi.org/10.1016/j.precisioneng.2021.09.005.
[278] Kaji F, Nguyen-Huu H, Budhwani A, Narayanan JA, Zimny M, Toyserkani E. A deep-learning-based in-situ surface anomaly detection methodology for laser directed energy deposition via powder feeding. Journal of Manufacturing Processes 2022;81:624–37. https://doi.org/10.1016/j.jmapro.2022.06.046.
[279] Li Y, Li X, Zhang G, Horváth I, Han Q. Interlayer closed-loop control of forming geometries for wire and arc additive manufacturing based on fuzzy-logic inference. Journal of Manufacturing Processes 2021;63:35–47. https://doi.org/10.1016/j.jmapro.2020.04.009.
[280] Wang K. Contrastive learning-based semantic segmentation for In-situ stratified defect detection in additive manufacturing. Journal of Manufacturing Systems 2023;68:465–76. https://doi.org/10.1016/j.jmsy.2023.05.001.
[281] Ye Z, Liu C, Tian W, Kan C. A Deep Learning Approach for the Identification of Small Process Shifts in Additive Manufacturing using 3D Point Clouds. Procedia Manufacturing 2020;48:770–5. https://doi.org/10.1016/j.promfg.2020.05.112.
[282] Ye Z, Liu C, Tian W, Kan C. In-situ point cloud fusion for layer-wise monitoring of additive manufacturing. Journal of Manufacturing Systems 2021;61:210–22. https://doi.org/10.1016/j.jmsy.2021.09.002.
[283] Bernauer C, Leitner P, Zapata A, Garkusha P, Grabmann S, Schmoeller M, et al. Segmentation-based closed-loop layer height control for enhancing stability and dimensional accuracy in wire-based laser metal deposition. Robotics and Computer-Integrated Manufacturing 2024;86:102683. https://doi.org/10.1016/j.rcim.2023.102683.
[284] Tang S, Wang G, Zhang H. In situ 3D monitoring and control of geometric signatures in wire and arc additive manufacturing. Surf Topogr: Metrol Prop 2019;7:025013. https://doi.org/10.1088/2051-672X/ab1c98.
[285] Ding D, Zhao Z, Huang R, Dai C, Zhang X, Xu T, et al. Error Modeling and Path Planning for Freeform Surfaces by Laser Triangulation On-Machine Measurement. IEEE Transactions on Instrumentation and Measurement 2021;70:1–11. https://doi.org/10.1109/TIM.2021.3063751.
[286] Ding Y, Huang W, Kovacevic R. An on-line shape-matching weld seam tracking system. Robotics and Computer-Integrated Manufacturing 2016;42:103–12. https://doi.org/10.1016/j.rcim.2016.05.012.
[287] Li X, Liu B, Mei X, Wang W, Wang X, Li X. Development of an In-Situ Laser Machining System Using a Three-Dimensional Galvanometer Scanner. Engineering 2020;6:68–76. https://doi.org/10.1016/j.eng.2019.07.024.
[288] Liska J, Vanicek O, Chalus M. Hand-Eye Calibration of a Laser Profile Scanner in Robotic Welding. 2018 IEEE/ASME International Conference on Advanced Intelligent Mechatronics (AIM), 2018, p. 316–21. https://doi.org/10.1109/AIM.2018.8452270.
[289] Li M, Du Z, Ma X, Dong W, Gao Y. A robot hand-eye calibration method of line laser sensor based on 3D reconstruction. Robotics and Computer-Integrated Manufacturing 2021;71:102136. https://doi.org/10.1016/j.rcim.2021.102136.
[290] Lyu J, Akhavan Taheri Boroujeni J, Manoochehri S. In-Situ Laser-Based Process Monitoring and In-Plane Surface Anomaly Identification for Additive Manufacturing Using Point Cloud and Machine Learning, American Society of Mechanical Engineers Digital Collection; 2021. https://doi.org/10.1115/DETC2021-69436.
[291] Singh A, Yadav A, Rana A. K-means with Three different Distance Metrics. International Journal of Computer Applications 2013;67.
[292] Cantzler H. Random Sample Consensus (RANSAC) n.d.:4.

[293] Chen L, Yao X, Xu P, Moon SK, Bi G. Surface Monitoring for Additive Manufacturing with in-situ Point Cloud Processing. 2020 6th International Conference on Control, Automation and Robotics (ICCAR), 2020, p. 196–201. https://doi.org/10.1109/ICCAR49639.2020.9108092.

[294] Wu Z, Tang G, Clark SJ, Meshkov A, Roychowdhury S, Gould B, et al. High frequency beam oscillation keyhole dynamics in laser melting revealed by in-situ x-ray imaging. Commun Mater 2023;4:1–10. https://doi.org/10.1038/s43246-023-00332-z.

[295] Hocine S, Van Petegem S, Frommherz U, Tinti G, Casati N, Grolimund D, et al. A miniaturized selective laser melting device for operando X-ray diffraction studies. Additive Manufacturing 2020;34:101194. https://doi.org/10.1016/j.addma.2020.101194.

[296] Hocine S, Van Swygenhoven H, Van Petegem S, Chang CST, Maimaitiyili T, Tinti G, et al. Operando X-ray diffraction during laser 3D printing. Materials Today 2020;34:30–40. https://doi.org/10.1016/j.mattod.2019.10.001.

[297] Zhao C, Parab ND, Li X, Fezzaa K, Tan W, Rollett AD, et al. Critical instability at moving keyhole tip generates porosity in laser melting. Science 2020;370:1080–6. https://doi.org/10.1126/science.abd1587.

[298] Soundarapandiyan G, Leung CLA, Johnston C, Chen B, Khan RHU, McNutt P, et al. In situ monitoring the effects of Ti6Al4V powder oxidation during laser powder bed fusion additive manufacturing. International Journal of Machine Tools and Manufacture 2023:104049. https://doi.org/10.1016/j.ijmachtools.2023.104049.

[299] Wang A, Wei Q, Tang Z, Ren P, Zhang X, Wu Y, et al. Effects of processing parameters on pore defects in blue laser directed energy deposition of aluminum by in and ex situ observation. Journal of Materials Processing Technology 2023:118068. https://doi.org/10.1016/j.jmatprotec.2023.118068.

[300] Hamidi Nasab M, Masinelli G, de Formanoir C, Schlenger L, Van Petegem S, Esmaeilzadeh R, et al. Harmonizing sound and light: X-ray imaging unveils acoustic signatures of stochastic inter-regime instabilities during laser melting. Nat Commun 2023;14:8008. https://doi.org/10.1038/s41467-023-43371-3.

[301] Ren Z, Gao L, Clark SJ, Fezzaa K, Shevchenko P, Choi A, et al. Machine learning–aided real-time detection of keyhole pore generation in laser powder bed fusion. Science 2023;379:89–94. https://doi.org/10.1126/science.add4667.

[302] Pandiyan V, Masinelli G, Claire N, Le-Quang T, Hamidi-Nasab M, de Formanoir C, et al. Deep learning-based monitoring of laser powder bed fusion process on variable time-scales using heterogeneous sensing and operando X-ray radiography guidance. Additive Manufacturing 2022;58:103007. https://doi.org/10.1016/j.addma.2022.103007.

[303] Wolff SJ, Wu H, Parab N, Zhao C, Ehmann KF, Sun T, et al. In-situ high-speed X-ray imaging of piezo-driven directed energy deposition additive manufacturing. Sci Rep 2019;9:962. https://doi.org/10.1038/s41598-018-36678-5.

[304] Fleming TG, Rees DT, Marussi S, Connolley T, Atwood RC, Jones MA, et al. In situ Correlative Observation of Humping-Induced Cracking in Directed Energy Deposition of Nickel-Based Superalloys. Additive Manufacturing 2023:103579. https://doi.org/10.1016/j.addma.2023.103579.

[305] Kogel-Hollacher M, Strebel M, Staudenmaier C, Schneider H-I, Regulin D. OCT sensor for layer height control in DED using SINUMERIK® controller. Laser 3D Manufacturing VII, vol. 11271, International Society for Optics and Photonics; 2020, p. 112710O. https://doi.org/10.1117/12.2540167.

[306] Kong L, Peng X, Chen Y, Wang P, Xu M. Multi-sensor measurement and data fusion technology for manufacturing process monitoring: a literature review. Int J Extrem Manuf 2020;2:022001. https://doi.org/10.1088/2631-7990/ab7ae6.

[307] Marshall GJ, Thompson SM, Shamsaei N. Data indicating temperature response of Ti–6Al–4V thin-walled structure during its additive manufacture via Laser Engineered Net Shaping. Data in Brief 2016;7:697–703. https://doi.org/10.1016/j.dib.2016.02.084.

[308] Bevans B, Barrett C, Spears T, Gaikwad A, Riensche A, Smoqi Z, et al. Heterogeneous sensor data fusion for multiscale, shape agnostic flaw detection in laser powder bed fusion additive manufacturing. Virtual and Physical Prototyping 2023;18:e2196266. https://doi.org/10.1080/17452759.2023.2196266.

[309] Petrich J, Snow Z, Corbin D, Reutzel EW. Multi-modal sensor fusion with machine learning for data-driven process monitoring for additive manufacturing. Additive Manufacturing 2021;48:102364. https://doi.org/10.1016/j.addma.2021.102364.

[310] Gaikwad A, Williams RJ, de Winton H, Bevans BD, Smoqi Z, Rao P, et al. Multi phenomena melt pool sensor data fusion for enhanced process monitoring of laser powder bed fusion additive manufacturing. Materials & Design 2022;221:110919. https://doi.org/10.1016/j.matdes.2022.110919.

[311] Liu H, Gobert C, Ferguson K, Abranovic B, Chen H, Beuth JL, et al. Inference of highly time-resolved melt pool visual characteristics and spatially-dependent lack-of-fusion defects in laser powder bed fusion using acoustic and thermal emission data 2023. https://doi.org/10.48550/arXiv.2310.05289.

[312] Vandone A, Baraldo S, Valente A. Multisensor Data Fusion for Additive Manufacturing Process Control. IEEE Robotics and Automation Letters 2018;3:3279–84. https://doi.org/10.1109/LRA.2018.2851792.
[313] Feng S, Lu Y, Jones A, Yang Z. Additive Manufacturing In-situ and Ex-Situ Geometric Data Registration. Journal of Computing and Information Science in Engineering 2022:1–13. https://doi.org/10.1115/1.4054202.
[314] message_filters - ROS Wiki n.d. http://wiki.ros.org/message_filters (accessed January 16, 2023).
[315] Kim J, Yang Z, Ko H, Cho H, Lu Y. Deep learning-based data registration of melt-pool-monitoring images for laser powder bed fusion additive manufacturing. Journal of Manufacturing Systems 2023;68:117–29. https://doi.org/10.1016/j.jmsy.2023.03.006.
[316] Xie T, Huang X, Choi S-K. Intelligent Mechanical Fault Diagnosis Using Multisensor Fusion and Convolution Neural Network. IEEE Transactions on Industrial Informatics 2022;18:3213–23. https://doi.org/10.1109/TII.2021.3102017.
[317] Yeong DJ, Velasco-Hernandez G, Barry J, Walsh J. Sensor and Sensor Fusion Technology in Autonomous Vehicles: A Review. Sensors 2021;21:2140. https://doi.org/10.3390/s21062140.
[318] Li J, Zhang X, Zhou Q, Chan FTS, Hu Z. A feature-level multi-sensor fusion approach for in-situ quality monitoring of selective laser melting. Journal of Manufacturing Processes 2022;84:913–26. https://doi.org/10.1016/j.jmapro.2022.10.050.
[319] Li J, Zhou Q, Cao L, Wang Y, Hu J. A convolutional neural network-based multi-sensor fusion approach for in-situ quality monitoring of selective laser melting. Journal of Manufacturing Systems 2022;64:429–42. https://doi.org/10.1016/j.jmsy.2022.07.007.
[320] Perani M, Baraldo S, Decker M, Vandone A, Valente A, Paoli B. Track geometry prediction for Laser Metal Deposition based on on-line artificial vision and deep neural networks. Robotics and Computer-Integrated Manufacturing 2023;79:102445. https://doi.org/10.1016/j.rcim.2022.102445.
[321] Wang R, Standfield B, Dou C, Law AC, Kong ZJ. Real-time process monitoring and closed-loop control on laser power via a customized laser powder bed fusion platform. Additive Manufacturing 2023;66:103449. https://doi.org/10.1016/j.addma.2023.103449.
[322] Bremer J, Walderich P, Pirch N, Schleifenbaum JH, Gasser A, Schopphoven T. Effects of path accuracy on additively manufactured specimens by laser material deposition using six-axis robots. Journal of Laser Applications 2021;33:012045. https://doi.org/10.2351/7.0000308.
[323] Sammons PM, Gegel ML, Bristow DA, Landers RG. Repetitive Process Control of Additive Manufacturing With Application to Laser Metal Deposition. IEEE Transactions on Control Systems Technology 2019;27:566–75. https://doi.org/10.1109/TCST.2017.2781653.
[324] Moralejo S, Penaranda X, Nieto S, Barrios A, Arrizubieta I, Tabernero I, et al. A feedforward controller for tuning laser cladding melt pool geometry in real time. Int J Adv Manuf Technol 2017;89:821–31. https://doi.org/10.1007/s00170-016-9138-7.
[325] Chechik L, Goodall AD, Christofidou KA, Todd I. Controlling grain structure in metallic additive manufacturing using a versatile, inexpensive process control system. Sci Rep 2023;13:10003. https://doi.org/10.1038/s41598-023-37089-x.
[326] Liao S, Webster S, Huang D, Council R, Ehmann K, Cao J. Simulation-guided variable laser power design for melt pool depth control in directed energy deposition. Additive Manufacturing 2022;56:102912. https://doi.org/10.1016/j.addma.2022.102912.
[327] Ogoke F, Farimani AB. Thermal control of laser powder bed fusion using deep reinforcement learning. Additive Manufacturing 2021;46:102033. https://doi.org/10.1016/j.addma.2021.102033.
[328] Freeman F, Chechik L, Thomas B, Todd I. Calibrated closed-loop control to reduce the effect of geometry on mechanical behaviour in directed energy deposition. Journal of Materials Processing Technology 2023;311:117823. https://doi.org/10.1016/j.jmatprotec.2022.117823.
[329] Su Y, Wang Z, Xu X, Luo K, Lu J. Effect of closed-loop controlled melt pool width on microstructure and tensile property for Fe-Ni-Cr alloy in directed energy deposition. Journal of Manufacturing Processes 2022;82:708–21. https://doi.org/10.1016/j.jmapro.2022.08.049.
[330] Gibson BT, Bandari YK, Richardson BS, Henry WC, Vetland EJ, Sundermann TW, et al. Melt pool size control through multiple closed-loop modalities in laser-wire directed energy deposition of Ti-6Al-4V. Additive Manufacturing 2020;32:100993. https://doi.org/10.1016/j.addma.2019.100993.
[331] Liu Y, Wang L, Brandt M. Model predictive control of laser metal deposition. Int J Adv Manuf Technol 2019;105:1055–67. https://doi.org/10.1007/s00170-019-04279-9.
[332] Yeung H, Lane B, Fox J. Part geometry and conduction-based laser power control for powder bed fusion additive manufacturing. Additive Manufacturing 2019;30:100844. https://doi.org/10.1016/j.addma.2019.100844.
[333] Shi T, Lu B, Shen T, Zhang R, Shi S, Fu G. Closed-loop control of variable width deposition in laser metal

deposition. Int J Adv Manuf Technol 2018;97:4167–78. https://doi.org/10.1007/s00170-018-1895-z.
[334] Hofman JT, Pathiraj B, van Dijk J, de Lange DF, Meijer J. A camera based feedback control strategy for the laser cladding process. Journal of Materials Processing Technology 2012;212:2455–62. https://doi.org/10.1016/j.jmatprotec.2012.06.027.
[335] Song L, Mazumder J. Feedback Control of Melt Pool Temperature During Laser Cladding Process. IEEE Trans Contr Syst Technol 2011;19:1349–56. https://doi.org/10.1109/TCST.2010.2093901.
[336] Song L, Bagavath-Singh V, Dutta B, Mazumder J. Control of melt pool temperature and deposition height during direct metal deposition process. Int J Adv Manuf Technol 2012;58:247–56. https://doi.org/10.1007/s00170-011-3395-2.
[337] Tang L, Landers RG. Melt Pool Temperature Control for Laser Metal Deposition Processes—Part I: Online Temperature Control. J Manuf Sci Eng 2010;132. https://doi.org/10.1115/1.4000882.
[338] Xu P, Yao X, Chen L, Liu K, Bi G. Heuristic Kinematics of a Redundant Robot-Positioner System for Additive Manufacturing. 2020 6th International Conference on Control, Automation and Robotics (ICCAR), 2020, p. 119–23. https://doi.org/10.1109/ICCAR49639.2020.9108047.
[339] Chen L, Yao X, Xu P, Moon SK, Zhou W, Bi G. In-Process Sensing, Monitoring and Adaptive Control for Intelligent Laser-Aided Additive Manufacturing. Transactions on Intelligent Welding Manufacturing, Springer, Singapore; 2023, p. 3–30. https://doi.org/10.1007/978-981-19-6149-6_1.
[340] Snow Z, Scime L, Ziabari A, Fisher B, Paquit V. Scalable In Situ Non-Destructive Evaluation of Additively Manufactured Components Using Process Monitoring, Sensor Fusion, and Machine Learning. Additive Manufacturing 2023:103817. https://doi.org/10.1016/j.addma.2023.103817.
[341] Jafari-Marandi R, Khanzadeh M, Tian W, Smith B, Bian L. From in-situ monitoring toward high-throughput process control: cost-driven decision-making framework for laser-based additive manufacturing. Journal of Manufacturing Systems 2019;51:29–41. https://doi.org/10.1016/j.jmsy.2019.02.005.
[342] Chung J, Shen B, Law ACC, Kong Z (James). Reinforcement learning-based defect mitigation for quality assurance of additive manufacturing. Journal of Manufacturing Systems 2022;65:822–35. https://doi.org/10.1016/j.jmsy.2022.11.008.
[343] Wang Y, Hu K, Li W, Wang L. Prediction of melt pool width and layer height for Laser Directed Energy Deposition enabled by physics-driven temporal convolutional network. Journal of Manufacturing Systems 2023;69:1–17. https://doi.org/10.1016/j.jmsy.2023.06.002.
[344] Sharma R, Raissi M, Guo Y. Physics-informed deep learning of gas flow-melt pool multi-physical dynamics during powder bed fusion. CIRP Annals 2023. https://doi.org/10.1016/j.cirp.2023.04.005.
[345] Li B, Zhang Y, Lei Y, Wei H, Chen C, Liu F, et al. A Single-Sensor Multi-Scale Quality Monitoring Methodology for Laser-Directed Energy Deposition: Example with Height Instability and Porosity Monitoring in Additive Manufacturing of Ceramic Thin-Walled Parts. Additive Manufacturing 2023:103923. https://doi.org/10.1016/j.addma.2023.103923.
[346] Zhang W, Wang J, Tang M, Ma H, Wang L, Zhang Q, et al. 2-D Transformer-Based Approach for Process Monitoring of Metal 3-D Printing via Coaxial High-Speed Imaging. IEEE Transactions on Industrial Informatics 2023:1–11. https://doi.org/10.1109/TII.2023.3314071.
[347] Yin M, Zhuo S, Xie L, Chen L, Wang M, Liu G. Online monitoring of local defects in robotic laser additive manufacturing process based on a dynamic mapping strategy and multibranch fusion convolutional neural network. Journal of Manufacturing Systems 2023;71:494–503. https://doi.org/10.1016/j.jmsy.2023.10.005.
[348] Kim S, Jeon I, Sohn H. Infrared thermographic imaging based real-time layer height estimation during directed energy deposition. Optics and Lasers in Engineering 2023;168:107661. https://doi.org/10.1016/j.optlaseng.2023.107661.
[349] Cai W, Jiang P, Shu L, Geng S, Zhou Q. Real-time laser keyhole welding penetration state monitoring based on adaptive fusion images using convolutional neural networks. J Intell Manuf 2021:1–15. https://doi.org/10.1007/s10845-021-01848-2.
[350] Zhang Y, Soon HG, Ye D, Fuh JYH, Zhu K. Powder-Bed Fusion Process Monitoring by Machine Vision With Hybrid Convolutional Neural Networks. IEEE Transactions on Industrial Informatics 2020;16:5769–79. https://doi.org/10.1109/TII.2019.2956078.
[351] Drissi-Daoudi R, Masinelli G, de Formanoir C, Wasmer K, Jhabvala J, Logé RE. Acoustic emission for the prediction of processing regimes in Laser Powder Bed Fusion, and the generation of processing maps. Additive Manufacturing 2023;67:103484. https://doi.org/10.1016/j.addma.2023.103484.